\documentstyle[amstex,amssymb,rmp,aps,array,epsf,epsfig]{revtex}

\begin{document}
\title{Proximity effects in superconductor-ferromagnet heterostructures }
\author{A.I.Buzdin}
\address{Institut Universitaire de France and  Universit\'{e} Bordeaux 1, CPMOH, UMR
5798, \\
F-33405 Talence Cedex, France}
\date{\today}
\maketitle

\begin{abstract}
The very special characteristic of the proximity effect in
superconductor-ferromagnet systems is the damped oscillatory behavior of the
Cooper pair wave function in a ferromagnet. In some sense, this is analogous
to the inhomogeneous superconductivity, predicted long time ago by Larkin
and Ovchinnikov (1964), and Fulde and Ferrell (1964), and constantly
searched since that. After the qualitative analysis of the peculiarities of
the proximity effect in the presence of the exchange field, the author
provides a unified description of the properties of the
superconductor-ferromagnet heterostructures. Special attention is paid to
the striking non-monotonous dependance of the critical temperature of the
multilayers and bilayers on the ferromagnetic layer thickness and conditions
of the realization of the ''$\pi $''- Josephson junctions. The recent
progress in the preparation of the high quality hybrid systems permitted to
observe on experiments many interesting effects, which are also discussed in
the article. Finally, the author analyzes the phenomenon of the domain-wall
superconductivity and the influence of superconductivity on the magnetic
structure in superconductor-ferromagnet bilayers.
\end{abstract}

\pacs{XXXX}

\tableofcontents%
%

\section{Introduction}

Due to their antagonistic characters, singlet superconductivity and
ferromagnetic order cannot coexist in bulk samples with realistic physical
parameters. Ginzburg (1956) was the first to set up theoretically the
problem of magnetism and superconductivity coexistence taking into account
the orbital mechanism of superconductivity destruction (interaction of the
superconducting order parameter with a vector-potential ${\bf A}$ of the
magnetic field). After the creation of BCS theory, it became clear that
superconductivity (in the singlet state) can be also destroyed by the
exchange mechanism. The exchange field, in the magnetically ordered state,
tends to align spins of Cooper pairs in the same direction, thus preventing
a pairing effect. This is the so-called paramagnetic effect (Saint-James 
{\it et al.}, 1969). Anderson and Suhl (1959) demonstrated that
ferromagnetic ordering is unlikely to appear in the superconducting phase.
The main reason for that is the suppression of the zero wave-vector
component of the electronic paramagnetic susceptibility in the presence of
superconductivity. In such situation the gain of energy for the
ferromagnetic ordering decreases and instead of the ferromagnetic order the
non-uniform magnetic ordering should appear. Anderson and Suhl (1959) called
this state cryptoferromagnetic.

The 1977 discovery of ternary rare earth (RE) compounds (RE)Rh$_{4}$B$_{4}$
and (RE)Mo$_{6}$X$_{8}$ (X=S, Se) (as a review see, for example, Maple and
Fisher, 1982) provided the first experimental evidence of magnetism and
superconductivity coexistence in stoichiometrical compounds. It turned out
that in many of these systems, superconductivity (with the critical
temperature $T_{c}$) coexists rather easily with antiferromagnetic order
(with the N\'{e}el temperature $T_{N}$), and usually the situation with $%
T_{N}<T_{c}$ \ is realized.

The more recent discovery of superconductivity in the quaternary
intermetallic compounds (RE)Ni$_{2}$B$_{2}$C\ (as a review see, for example,
M\"{u}ller and Narozhnyi, 2001) gives another example of antiferromagnetism
and superconductivity coexistence.

Indeed, superconductivity and antiferromagnetism can coexist quite
peacefully because, on average, at distances of the order of the Cooper pair
size (superconducting coherence length) the exchange and orbital fields are
zero. Much more interesting \ a re-entrant behavior of the superconductivity
was observed in ErRh$_{4}$B$_{4}$ and HoMo$_{6}$S$_{8}$ (Maple and
Fisher,1982). For example, ErRh$_{4}$B$_{4}$ becomes superconductor below $%
T_{c}=8.7$ K. When it is cooled to the Curie temperature $\Theta \approx 0.8$
K an inhomogeneous magnetic order appears in the superconducting state. With
further cooling the superconductivity is destroyed by the onset of a
first-order ferromagnetic transition at the second critical temperature $%
T_{c2}\approx 0.7$ K. HoMo$_{6}$S$_{8}$ gives another example of the
re-entrant superconductivity with $T_{c}=1.8$ K, \ $\Theta \approx 0.74$ K,
and $T_{c2}\approx 0.7$ K.

In these compounds at Curie temperature, following the prediction of
Anderson and Suhl (1959) a non-uniform magnetic order appears. Its presence
was confirmed by neutron scattering experiments. The period of this magnetic
structure is smaller than the superconducting coherence length, but larger
than the interatomic distance. In some sense this structure is a realization
of the compromise between superconductivity and ferromagnetism : for the
superconductivity it is seen as an antiferromagnetism, but for the magnetism
it looks like a ferromagnetism. Theoretical analysis, taking into account
both orbital and exchange mechanisms and magnetic anisotropy (as a review
see Bulaevskii {\it et al.}, 1985), revealed that the coexistence phase is a
domain-like structure with very small period. The region of magnetism and
superconductivity coexistence in ErRh$_{4}$B$_{4}$ and HoMo$_{6}$S$_{8}$ is
narrow, but in HoMo$_{6}$Se$_{8}$ the domain coexistence phase survives till 
$T=0$ K.

The first truly ferromagnetic superconductors $UGe_{2}$ (Saxena {\it et al.}%
, 2000) and $URhGe$ (Aoki {\it et al.}, 2001) have\ been discovered only
recently, and apparently the coexistence of superconductivity with
ferromagnetism is possible due to the triplet character of the
superconducting pairing. Indeed, the superconductivity in $URhGe$ (Aoki {\it %
et al.}, 2001) appears below 0.3 K in the ferromagnetic phase which has the
Curie temperature $\Theta =9.5$ K$;$ this makes the singlet scenario of
superconductivity rather improbable.

Though the coexistence of\ singlet superconductivity with ferromagnetism is
very unlikely in bulk compounds, it may be easily achieved in artificially
fabricated layered ferromagnet/superconductors (F/S) systems. Due to the
proximity effect, the Cooper pairs can penetrate into the F layer and induce
superconductivity there. In such case we have the unique possibility to
study the properties of superconducting electrons under the influence of a
huge exchange field acting on the electron spins. In addition, it is
possible to study the interplay between superconductivity and magnetism in a
controlled manner, since varying the layer thicknesses we change the
relative strength of two competing orderings. The behavior of the
superconducting condensate under these conditions is quite peculiar.

Long time ago Larkin and Ovchinnikov (1964), and Fulde and Ferrell (1964)
demonstrated that in a pure ferromagnetic superconductor at low temperature
the superconductivity may be non-uniform. Due to the incompatibility of
ferromagnetism and superconductivity it is not easy to verify this
prediction on experiment. It occurs that in S/F systems there exists some
analogy with the non-uniform superconducting state. The Cooper pair wave
function has damped oscillatory behavior in a ferromagnet in contact with a
superconductor. It results in many new effects that we discuss in this
article : the spacial oscillations of the electron's density of states, the
non-monotonous dependance of the critical temperature of S/F multilayers and
bilayers on the ferromagnet layer thickness, the realization of the
Josephson ''$\pi $''- junctions in S/F/S systems. The spin-walve effect in
the complex S/F structures gives another example of the interesting
interplay between magnetism and superconductivity, promising for the
potential applications. We discuss also the issues of the localized
domain-wall superconductivity in S/F bilayers and the inverse influence of
superconductivity on ferromagnetism, which favors the non-uniform magnetic
structures. An interesting example of atomic thickness S/F multilayers is
provided by the layered superconductors like Sm$_{1.85}$Ce$_{0.15}$CuO$_{4}$
and RuSr$_{2}$GdCu$_{2}$O$_{8}$. For such systems the exchange field in F
layer also favors the ''$\pi $''-phase behavior, with an alternating order
parameter in adjacent superconducting layers.

Note that practically all interesting effects related with the interplay
between the superconductivity and the magnetism in S/F structures occurs at
the nanoscopic range of layers thicknesses. The observation of these effects
became possible only recently due to the great progress in the preparation
of high-quality hybrid F/S systems. The experimental progress and the
possibility of potential applications in its turn stimulated a revival of
the interest to the superconductivity and ferromagnetism interplay in
heterostructures. It seems to be timely to review the present state of the
research in this domain and outline the perspectives.

\section{Paramagnetic limit and qualitative explanation of the non-uniform
phase formation}

\subsection{The (H, T) phase diagram}

For a pure paramagnetic effect, the critical field$\ $of a superconductor $%
H_{p}$ at $T=0$ may be found from the comparison of the energy gain $\Delta
E_{n\text{ }}$due to the electron spin polarization in the normal state and
the superconducting condensation energy $\Delta E_{s}.$ Really, in the
normal state, the polarization of the electron gas changes its energy in the
magnetic field by 
\begin{equation}
\Delta E_{n\text{ }}=-\chi _{n}\frac{H^{2}}{2}{\bf ,}
\end{equation}
where $\chi _{n}=2\mu _{B}^{2}N(0)$ is the spin susceptibility of the normal
metal, $\mu _{B}$ is the Bohr magneton, $2N(0)$ is the density of electron
states at Fermi level (per two spin projections), and the electron $g$
factor is supposed to be equal to 2.

On the other hand, in a superconductor the polarization is absent, but the
BCS pairing decreases its energy by 
\begin{equation}
\Delta E_{s\text{ }}=-N(0)\frac{\Delta _{0}^{2}}{2}{\bf ,}  \label{Es}
\end{equation}
where $\Delta _{0}=1.76T_{c}$ is the superconducting gap at $T=0$. From the
condition $\Delta E_{n\text{ }}=\Delta E_{s},$ we find the Chandrasekhar
(1962) - Clogston (1962) limit (the paramagnetic limit at $T=0$) 
\begin{equation}
H_{p}(0{\bf )}=\frac{\Delta _{0}}{\sqrt{2}\mu _{B}}{\bf \,.}
\end{equation}
Note that it is the field of the first-order phase transition from a normal
to a superconducting state. The complete analysis (Saint-James {\it et al.},
1969) demonstrates that at $T=0$ this critical field is higher than the
field of the second order phase \ transition $H_{p}^{II}(0{\bf )}=\Delta
_{0}/2\mu _{B}$, and the transition from a normal to a uniform
superconducting state is of the second-order at $T^{\ast }<T<T_{c}$ only,
where $T^{\ast }=0.56T_{c},$ $H^{\ast }=H(T^{\ast })=0.61\Delta _{0}/\mu
_{B}=1.05T_{c}/\mu _{B}.$ However, Larkin and Ovchinnikov (1964), and Fulde
and Ferrell (1964) predicted in the framework of the model of pure
paramagnetic effect the appearance of the non-uniform superconducting state
with a sinusoidal modulation of the superconducting order parameter at the
scale of the superconducting coherence length $\xi _{s}$ (the FFLO state).
In this FFLO state, the Cooper pairs have a finite momentum, compared with
zero momentum in conventional superconductors. Recently Casalbuoni and
Nardulli (2004) reviewed the theory of the inhomogeneous superconductivity
applied to the condensed matter and quantum chromodynamics at high density
and low temperature.

The critical field of the second-order transition into FFLO state goes
somewhere above the first-order transition line into a uniform
superconducting state (Saint-James {\it et al.}, 1969). At $T=0,$ it is $\
H^{FFLO}(0)=0.755\Delta _{0}/\mu _{B}$ (whereas $H_{p}=0.7\Delta _{0}/\mu
_{B}).$ This FFLO\ state only appears in the temperature interval $%
0<T<T^{\ast }$,\ and is sensitive to impurities (Aslamazov, 1968). In a
dirty limit it is suppressed, and the first-order transition into the
uniform superconducting state takes place instead. The phase diagram for the
3D superconductors in the model of pure paramagnetic effect is presented in
Fig. 1 (Saint-James {\it et al.}, 1969). Up to now, there were no
unambiguous experimental proofs of this state observation. Note however
that, recently, the magnetic-field-induced superconductivity has been
observed in the quasi \ two-dimensional organic conductor $%
(BETS)_{2}FeCl_{4} $ (Uji {\it et al.,} 2001) which is an excellent
candidate for the FFLO state formation (Balicas {\it et al.}, 2001 and
Houzet {\it et al.,} 2002).

\subsection{Exchange field in the ferromagnet}

In a ferromagnet the exchange interaction between the electrons and the
magnetic moments may be considered as some effective Zeeman field. In the
case of magnetic moments with spin ${\bf S}_{i},$ localized in the sites $%
{\bf r}_{i}$, their interaction with electron spins is described by the
exchange Hamiltonian 
\begin{equation}
H_{int}=%
\mathrel{\mathop{\int d^{3}r{\bf \Psi }^{+}{\bf (r)}\left\{ %
\mathrel{\mathop{\sum }\limits_{i}}J({\bf r-r}_{i}){\bf S}_{i}{\bf \sigma }\right\} {\bf \Psi (r),}}\limits_{}}%
\label{exch Ham}
\end{equation}
where ${\bf \Psi (r)}$ is the electron's spinor operator, ${\bf \sigma }%
=\left\{ \sigma _{x},\sigma _{y},\sigma _{z}\right\} $ are the Pauli
matrices, and $J({\bf r})$ is the exchange integral. Below the Curie
temperature $\Theta $, the average value of the localized spins $%
\left\langle {\bf S}_{i}\right\rangle $ is non-zero, and the exchange
interaction may be considered as some effective Zeeman field $\ H^{eff}=%
\frac{\left\langle S_{i}^{z}\right\rangle n}{\mu _{B}}\int J({\bf r})d^{3}r$%
, where $n$ is the concentration of localized moments, and the spin
quantization $z$-axis is chosen along the ferromagnetic moment. It is
convenient to introduce the exchange field $h$ as 
\begin{equation}
h=\mu _{B}H^{eff}=\left\langle S_{i}^{z}\right\rangle n\int J({\bf r}%
)d^{3}r=s(T)h_{0},
\end{equation}
where $s(T)=\left\langle S_{i}^{z}\right\rangle /\left\langle
S_{i}^{z}\right\rangle _{T=0}$ is the dimensionless magnetization and $h_{0}$
is the maximum value of an exchange field at $T=0$ . The exchange field $h$
describes the spin-dependent part of the electron's energy and the exchange
Hamiltonian Eq. (\ref{exch Ham}) is then simply written as 
\begin{equation}
H_{int}=%
\mathrel{\mathop{\int d^{3}r{\bf \Psi }^{+}{\bf (r)}h{\bf \sigma _{z}\Psi (r).}}\limits_{}}%
\label{Hint}
\end{equation}
If we also want to take into account the proper Zeeman field of
magnetization $M$, then we may simply replace $h$ in Eq. (\ref{Hint}) by \ $%
h $ $+4\pi M\mu _{B}$ . The reader is warned that in principle, if the
exchange integral is negative, the exchange field may have the direction
opposite to the magnetic moments and the interesting compensation
Jaccarino-Peter (1962) effect is possible. However, in the ferromagnetic
metals, the contribution of the magnetic induction to the spin splitting is
several order of magnitude smaller than that of the exchange interaction and
may be neglected$.$ In the case of the Ruderman-Kittel-Kasuya-Yosida (RKKY)
mechanism of the ferromagnetic ordering, the Curie temperature $\Theta \sim
h_{0}^{2}/E_{F}$ and in all real systems the exchange field $h_{0}>>\Theta ,$
$T_{c}$. \ This explains that the conditions of singlet superconductivity
and ferromagnetism coexistence are very stringent. Indeed, if $\Theta >T_{c}$
the exchange field in a ferromagnet $h>>T_{c}$, which strongly exceeds the
paramagnetic limit. On the other hand, if $\Theta <T_{c}$ then, instead of
the ferromagnetic transition the inhomogeneous magnetic ordering appears
(Maple and Fisher, 1982; \ Bulaevskii {\it et al.}, 1985). The very high
value of the exchange field in ferromagnet permits us to concentrate on the
paramagnetic effect and neglect the orbital one (note that well below the
Curie temperature the magnetic induction $4\pi M$ in ferromagnets is of the
order of several $koe$ only).

\subsection{Why does the Fulde-Ferrell-Larkin-Ovchinnikov state appear?}

What is the physical origin of the superconducting order parameter
modulation in the FFLO state ? The appearance of modulation of the
superconducting order parameter is related to the Zeeman's splitting of the
electron's level under a magnetic field acting on electron spins. To
demonstrate this, we consider the simplest case of the 1D superconductor.

In the absence of the field, a Cooper pair is formed by two electrons with
opposite momenta $+k_{F}$ and $-k_{F}$ and opposite spins ($\uparrow $) and (%
$\downarrow $) respectively. The resulting momentum of the Cooper pair $%
k_{F}+(-k_{F})=0$. Under a magnetic field, because of the Zeeman's
splitting, the Fermi momentum of the electron with spin ($\uparrow $) will
shift from $k_{F}$ to $k_{1}=k_{F}+\delta k_{F}$ ,\ where $\delta k_{F}=\mu
_{B}H/v_{F\text{ }}$ and $v_{F\text{ }}$ is the Fermi velocity. Similarly,
the Fermi momentum of an electron with spin ($\downarrow $) will shift from $%
-k_{F}$ \ to $\ k_{2}=-k_{F}+\delta k_{F}$ (see Fig. 2) . Then, the
resulting momentum of the Cooper pair will be $k_{1}+k_{2}=2\delta k_{F}\neq
0,$ which just implies the space modulation of the superconducting order
parameter with a resulting wave-vector $2\delta k_{F}.$ Such type of
reasoning explains the origin of the non-uniform superconducting state
formation in the presence of the field acting on electron spins, and, at the
same time, demonstrates the absence of a paramagnetic limit (at T$%
\rightarrow 0)$ for the 1D superconductor (Buzdin and Polonskii, 1987). For
3D (Larkin and Ovchinnikov, 1964 and Fulde and Ferrell, 1964) or 2D
(Bulaevskii, 1973) superconductors, it is not possible to choose the single
wave vector $\delta k_{F}$ which compensates the Zeeman splitting for all
electrons on the Fermi surface (as $\delta k_{F}$ depends on direction of $%
v_{F}),$ and the paramagnetic limit is preserved. However, the critical
field for a non-uniform state \ at $T=0$ is always higher than for a uniform
one. However, the critical field for a non-uniform state \ at $T=0$ is
always higher than for a uniform one. At finite temperature (when $T\gtrsim
\mu _{B}H$ ), the smearing of the electrons distribution function near the
Fermi energy decreases the difference of energies between the non-uniform
and uniform states. As it follows from the microscopical calculations, at $%
T>T^{\ast }=0.56T_{c}$ the uniform superconducting phase is always more
favorable (Saint-James {\it et al.}, 1969).

\subsection{Generalized Ginzburg-Landau functional}

Qualitatively, the phenomenon of the FFLO phase formation and the
particularities of the proximity effect in S/F systems may be described in
the framework of the generalized Ginzburg-Landau expansion. Let us first
recall the form of the standard Ginzburg-Landau functional (see, for
example, De Gennes, 1966) 
\begin{equation}
F=a\left| \psi \right| ^{2}+\gamma \left| \overrightarrow{{\bf %
\bigtriangledown }}\psi \right| ^{2}+\frac{b}{2}\left| \psi \right| ^{4},
\label{G-L}
\end{equation}
where $\psi $ is the superconducting order parameter, and the coefficient $a$
vanishes at the transition temperature $T_{c}.$ At $T<T_{c},$ the
coefficient $a$ is negative and the minimum of $F$ in Eq. (\ref{G-L}) is
achieved for a uniform superconducting state with $\left| \psi \right| ^{2}=-%
\frac{a}{b}.$ If we consider also the paramagnetic effect of the magnetic
field, all the coefficients in Eq. (\ref{G-L}) will depend on the energy of
the Zeeman splitting $\mu _{B}H$, i. e. an exchange field $h$ in the
ferromagnet. Note that we neglect the orbital effect, so there is no
vector-potential ${\bf A}$ in Eq. (\ref{G-L}). To take into account the
orbital effect in the Ginzburg-Landau functional, we may substitute the
gradient by its gauge-invariant form $\overrightarrow{{\bf \bigtriangledown }%
}\rightarrow \overrightarrow{{\bf \bigtriangledown }}-\frac{2ie}{c}{\bf A}$.
Usually, the orbital effect is much more important for the superconductivity
destruction than the paramagnetic one. It explains why in the standard
Ginzburg-Landau theory there is no need to take into account the field and
temperature dependence of the coefficients $\gamma $ and \ $b$. However,
when the paramagnetic effect becomes predominant, this approximation fails.
What are the consequences ? If it was simply some renormalization of the
coefficients in Ginzburg-Landau functional, the general superconducting
properties of the system would basically be the same. However, the
qualitatively new physics emerges due to the fact that the coefficient $%
\gamma $ changes its sign at the point $(H^{\ast },T^{\ast })$ of the phase
diagram, see Fig. 1. The negative sign of $\gamma $ means that the minimum
of the functional does not correspond to an uniform state anymore, and a
spatial variation of the order parameter decreases the energy of the system.
To describe such a situation it is necessary to add a higher order
derivative term in the expansion (\ref{G-L}), and the generalized
Ginzburg-Landau expansion will be: 
\begin{eqnarray}
F_{G} &=&a(H,T)\left| \psi \right| ^{2}+\gamma (H,T)\left| \overrightarrow{%
{\bf \bigtriangledown }}\psi \right| ^{2}+  \label{gen G-L} \\
&&+\frac{\eta (H,T)}{2}\left| \overrightarrow{{\bf \bigtriangledown }}%
^{2}\psi \right| ^{2}+\frac{b(H,T)}{2}\left| \psi \right| ^{4}\ .  \nonumber
\end{eqnarray}
The critical temperature of the second order phase transition into a
superconducting state may be found from the solution of the linear equation
for the superconducting order parameter 
\begin{equation}
a\psi -\gamma \Delta \psi +\frac{\eta }{2}\Delta ^{2}\psi =0.
\label{eq for psi}
\end{equation}
If we seek for a non-uniform solution $\psi =\psi _{_{0}}\exp (i{\bf qr})$,
the corresponding critical temperature depends on the wave-vector ${\bf q}$
and is given by the expression 
\begin{equation}
a=-\gamma q^{2}-\frac{\eta }{2}q^{4}.
\end{equation}
Note that the coefficient $a$ may be written as $a=\alpha (T-T_{cu}(H)),$
where $T_{cu}(H)$ is the critical temperature of the transition into the
uniform superconducting state. In a standard situation, the gradient term in
the Ginzburg-Landau functional is positive, $\gamma >0$, and the highest
transition temperature coincides with $T_{cu}(H);$ it is realized for the
uniform state with $q=0$. However, in the case $\gamma <0,$ the maximum
critical temperature corresponds to the finite value of the modulation
vector $q_{0}^{2}=-\gamma /\eta $ and the corresponding transition
temperature into the non-uniform FFLO state $T_{ci}(H)$ is given by 
\begin{equation}
a=\alpha (T_{ci}-T_{cu})=\frac{\gamma ^{2}}{2\eta }.
\end{equation}
It is higher than the critical temperature $T_{cu}$ of the uniform state.
Therefore, we see that the FFLO state appearance may simply be interpreted
as a change of the sign of the gradient term in the Ginzburg-Landau
functional. A more detailed analysis of the FFLO state in the framework of
the generalized Ginzburg-Landau functional shows that it is not an
exponential but a one dimensional sinusoidal modulation of the order
parameter which gives the minimum energy (Buzdin and Kachkachi, 1997; Houzet 
{\it et al.}, 1999). In fact, the generalized Ginzburg-Landau functional
describes new type of superconductors with very different properties, and
the whole theory of superconductivity must be redone on the basis of this
functional. The orbital effect in the framework of the generalized
Ginzburg-Landau functional may be introduced by the usual gauge-invariant
procedure $\overrightarrow{{\bf \bigtriangledown }}\rightarrow 
\overrightarrow{{\bf \bigtriangledown }}-\frac{2ie}{c}{\bf A}$. The
resulting expression for the superconducting current is quite a special one
and the critical field may correspond to the higher Landau level solutions
as well as new types of vortex lattices may exist (Houzet and Buzdin, 2000;
Houzet and Buzdin, 2001).

\section{Proximity effect in ferromagnets}

\subsection{Some generalities about superconducting proximity effect}

The contact of materials with different long-range ordering modifies their
properties near the interface. In the case of a superconductor-normal metal
interface, the Cooper pairs can penetrate the normal metal at some distance.
If the electrons motion is diffusive, this distance is of the order of the
thermal diffusion length scale $L_{T}\thicksim \sqrt{D/T},$ where $D$ is the
diffusion constant. In the case of pure normal metal the coresponding
characteristic distance is $\xi _{T}\thicksim v_{F}/T$. Therefore the
superconducting-like properties may be induced in the normal metal, and
usually this phenomenon is called the proximity effect. At the same time the
leakage of the Coopers pairs weakens the superconductivity near the
interface with a normal metal. Sometime this effect is called the ''inverse
proximity effect'', and it results in the decrease of the superconducting
transition temperature in thin superconducting layer in contact with a
normal metal. If the thickness of a superconducting layer is smaller than
some critical one, the proximity effect totally suppresses the
superconducting transition. All these phenomena and the earlier experimental
and theoretical works on the proximity effect were reviewed by Deutscher and
de Gennes (1969).

Note that the proximity effect is a rather general phenomenon not limited by
the superconducting phase transition. For example, in the case of the
surface magnetism (White and Geballe, 1979) the critical temperature at the
surface can be higher then the bulk one. In the result the magnetic
transition at the surface induces the magnetisation nearby. On the other
hand, the volume strongly affects the surface transition characteristics.

However, the unique and very important characteristic of the\
superconducting proximity effect is the Andreev reflection revealed at the
microscopical level. Andreev (1964) demonstrated how the single electron
states of the normal metal are converted into Cooper pairs and explained the
mechanism of the transformation at the interface of the dissipative
electrical current into the dissipationless supercurrent. An electron with
an energy below the superconducting gap is reflected at the interface as a
hole. The corresponding charge $2e$ is transferred to the Cooper pair which
appears on the superconducting side of the interface. The manifestation of
this double charge transfer is that for a perfect contact the sub-gap
conductance occurs to be twice the normal state conductance. The classical
work by Blonder, Tinkham and Klapwijk (1982) gives the detailed theory of
this phenomenon.

Andreev reflection plays a primary role for the understanding of quantum
transport properties of superconductor/normal metal systems. The interplay
between Andreev reflection and proximity effect was reviewed by Pannetier
and Courtois (2000). The reader can find a detailed description of the
Andreev reflection in the normal metal-superconductor junctions in the
framework of the scattering theory formalism in the review by Beenakker
(1997). Recent review by Deutscher (2005) is devoted to the Andreev
reflection spectroscopy of the superconductors.

\subsection{Damped oscillatory dependence of the Cooper pair wave function
in the ferromagnets}

The physics of the oscillating Cooper pair wave function in a ferromagnet is
similar to the physics of the superconducting order parameter modulation in
the FFLO state - see section II.C. Qualitative picture of this effect has
been well presented by Demler, Arnold, and Beasley (1997). When a
superconductor is in a contact with a normal metal the Cooper pairs
penetrate across the interface at some distance inside the metal. A Cooper
pair in a superconductor comprises two electrons with opposite spins and
momenta. In a ferromagnet the up spin electron (with the spin orientation
along the exchange field) decreases its energy by $h$ , while the down spin
electron increases its energy by the same value. To compensate this energy
variation, the up spin electron increases its kinetic energy, while the down
spin electron decreases its. In the result the Cooper pair acquires a center
of mass momentum $2\delta k_{F}=2h/v_{F\text{ }}$, which implies the
modulation of the order parameter with the period $\pi v_{F\text{ }}/h$. The
direction of the modulation wave vector must be perpendicular to the
interface, because only this orientation is compatible with the uniform
order parameter in the superconductor.

To get some idea about the peculiarity of the proximity effect in S/F
structures, we may start\ also from the description based on the generalized
Ginzburg-Landau functional Eq. (\ref{gen G-L}). Such approach is adequate
for a small wave-vector modulation case, i. e. in the vicinity of the ($%
H^{\ast },T^{\ast })$ point of the ($H,T)$ phase diagram, otherwise the
microscopical theory must be used. This situation corresponds to a very weak
ferromagnet with an extremely small exchange field $h\approx \mu _{B}H^{\ast
}=1.05T_{c}$ , which is non realistic as\ usually $h>>T_{c}$. However, we
will discuss this case to get a preliminary understanding of the phenomenon.
We address the question of the proximity effect for a weak ferromagnet
described by the generalized Ginzburg-Landau functional Eq. (\ref{gen G-L}%
).\ More precisely, we consider the decay of the order parameter in the
normal phase, i. e. at $T>T_{ci}$ assuming that our system is in contact
with another superconductor with a higher critical temperature, and the $x$
axis is choosen perpendicular to the interface (see Fig. 3).

The induced superconductivity is weak and to deal with it, we may use the
linearized equation for the order parameter (\ref{eq for psi}), which is
written for our geometry as 
\begin{equation}
a\psi -\gamma \frac{\partial ^{2}\psi }{\partial x^{2}}+\frac{\eta }{2}\frac{%
\partial ^{4}\psi }{\partial x^{4}}=0.
\end{equation}
The solutions of this equation in the normal phase are of the type $\psi
=\psi _{_{0}}\exp (kx)$, with a {\it complex wave-vector} $k=k_{1}+ik_{2}$,
and 
\begin{eqnarray}
k_{1}^{2} &=&\frac{\left| \gamma \right| }{2\eta }\left( \sqrt{1+\frac{%
T-T_{ci}}{T_{ci}-T_{cu}}}-1\right) , \\
k_{2}^{2} &=&\frac{\left| \gamma \right| }{2\eta }\left( 1+\sqrt{1+\frac{%
T-T_{ci}}{T_{ci}-T_{cu}}}\right) .
\end{eqnarray}
If we choose the gauge with the real order parameter in the superconductor,
then the solution for the decaying order parameter\ in the ferromagnet is
also real 
\begin{equation}
\psi (x)=\psi _{_{1}}\exp (-k_{1}x)\cos (k_{2}x),  \label{psi(x) mGL}
\end{equation}
where the choice of the root for $k$ is the condition $k_{1}>0$. So the
decay of the order parameter is accompanied by its oscillation (Fig. 3b),
which is the characteristic feature of the proximity effect in the
considered system. When we approach the critical temperature $T_{ci}$ the
decaying wave-vector vanishes, $k_{1}\rightarrow 0,$ while the oscillating
wave-vector $k_{2\text{ }}$goes to the FFLO wave-vector, $k_{2\text{ }%
}\rightarrow \sqrt{\frac{\left| \gamma \right| }{\eta }}$, so a FFLO phase
emerges. Let us compare this behavior with the standard proximity effect
(Deutscher and De Gennes, 1969) described by the linearized Ginzburg-Landau
equation for the order parameter 
\begin{equation}
a\psi -\gamma \frac{\partial ^{2}\psi }{\partial x^{2}}=0,
\end{equation}
with $\gamma >0$. In such case $T_{c}$ simply coincides with $T_{cu},$ and
the decaying solution is $\psi =\psi _{_{0}}\exp (-x/\xi (T)),$ where the
coherence length $\xi (T)=\sqrt{\gamma /a}$ (Fig. 3a). This simple analysis
brings in evidence the appearance of the oscillations of the order parameter
in the presence of an exchange field. This is a fundamental difference
between the proximity effect in S/F and S/N systems, and it is at the origin
of many peculiar characteristics of S/F heterostructures.

In real ferromagnets, the exchange field is very large compared with
superconducting temperature and energy scales, so the gradients of the
superconducting order parameter variations are large too, and can not be
treated in the framework of the generalized Ginzburg-Landau functional. To
describe the relevant experimental situation we need to use a microscopical
approach. The most convenient scheme to do this (see Appendix A and B) is
the use of the Boboliubov-de Gennes equations or the Green's functions in
the framework of the quasiclassical Eilenberger (Eilenberger, 1968) or
Usadel (Usadel, 1970) equations.

If the electron scattering mean free path $l$ is small (which is usually the
case in S/F systems), the most natural approach is to use the Usadel
equations for the Green's functions averaged over the Fermi surface
(Appendix). Linearized over the pair potential $\Delta (x)$, the Usadel
equation for the anomalous function $F(x,\omega )$ depending only on one
coordinate $x$ is 
\begin{equation}
\left( \left| \omega \right| +ih\cdot sgn(\omega )-\frac{D}{2}\frac{\partial
^{2}}{\partial x^{2}}\right) F(x,\omega )=\Delta (x),
\label{Usadel linearized}
\end{equation}
where ${\bf \omega =}\left( 2n+1\right) \pi T$ are the Matsubara
frequencies, and $D=\frac{1}{3}v_{F}l$ \ is the diffusion coefficient. In
the F region, we may neglect the Matsubara frequencies compared to the large
exchange field ($h>>T_{c}$), and the pairing potential $\Delta $ is absent
(we assume that the BCS coupling constant $\lambda $ is zero there).This
results in a very simple form of the Usadel equation for the anomalous
function $F_{f}$ \ in the ferromagnet 
\begin{equation}
ihsgn\left( \omega \right) F_{f}-\frac{D_{f}}{2}\frac{\partial ^{2}F_{f}}{%
\partial x^{2}}=0,  \label{Usadel F}
\end{equation}
where $D_{f}$ is the diffusion coefficient in the ferromagnet. For the
geometry in Fig. 3\ and $\omega >0,$ the decaying solution for $F_{f}$ is 
\begin{equation}
F_{f}\left( x,\omega >0\right) =A\exp \left( -\frac{i+1}{\xi _{f}}x\right) ,
\label{F in F}
\end{equation}
where $\xi _{f}=\sqrt{\frac{D_{f}}{h}}$ is the characteristic length of the
superconducting correlations decay (with oscillations) in F- layer (see
Table I). Due to the condition $h>>T_{c}$, this length is much smaller than
the superconducting coherence length $\xi _{s}=\sqrt{\frac{D_{s}}{2\pi T_{c}}%
}$, i.e. $\xi _{f}<<\xi _{s}$. The constant $A$ is determined by the
boundary conditions at the S/F interface. For example, in the case of a low
resistivity of a ferromagnet, at first approximation the anomalous function
in a superconductor $F_{s}$ is independent on coordinate and practically the
same as in the absence of the ferromagnet, i.e. $F_{s}=\Delta /\sqrt{\Delta
^{2}+\omega ^{2}}$. If, in addition, the interface is transparent then the
continuity of the function $F$ at the F/S boundary gives $A=\Delta /\sqrt{%
\Delta ^{2}+\omega ^{2}}$. For $\omega <0$, we simply have $F_{f}\left(
x,\omega <0\right) =F_{f}^{\ast }\left( x,\omega >0\right) $. In a
ferromagnet, the role of the Cooper pair wave function is played by $\Psi $
than decays as 
\begin{equation}
\Psi \sim \sum_{\omega }F(x,\omega )\sim \Delta \exp (-\frac{x}{\xi _{f}}%
)\cos (\frac{x}{\xi _{f}}).
\end{equation}
We retrieve the damping oscillatory behavior of the order parameter Eq. (\ref
{psi(x) mGL}), Fig. 3b. The important conclusion we obtain from the
microscopic approach is that in the dirty limit the scale for the
oscillation and decay of the Cooper pair wave function in a ferromagnet is
the same.

In the case of a clean ferromagnet the damped oscillatory behavior of the
Cooper pair wave function remains, though at zero temperature the damping is
non-exponential and much weaker $\left( \sim \frac{1}{x}\right) .$ Indeed,
the decaying solution of the Eilenberger equation in the clean limit (see
Appendix B) is

\begin{equation}
f(x,\theta ,\omega )\sim \exp \left( -\frac{2(\omega +ih)x}{v_{Ff}\cos
\theta }\right) ,  \label{F clean1}
\end{equation}

where $\theta $ is the angle between $\ x$-axis and Fermi velocity in a
ferromagnet, and $v_{Ff}$ is its modulus. After averaging over the angle $%
\theta $ and summation over the Matsubara frequencies $\omega $ we obtain

\begin{equation}
\Psi \sim \sum_{\omega }\stackrel{\pi }{%
\mathrel{\mathop{\int }\limits_{0}}%
}f(x,\theta ,\omega )\sin \theta d\theta \sim \frac{1}{x}\exp (-\frac{x}{\xi
_{1f}})\sin (\frac{x}{\xi _{2f}}).  \label{F clean2}
\end{equation}

Here the decaying length $\xi _{1f}=\frac{v_{Ff}}{2\pi T}$, and the
oscillating length $\xi _{2f}=\frac{v_{Ff}}{2h}$ (see Table I)$.$ At low
temperature $\xi _{1f}\longrightarrow 0$ and the Cooper pair wave function
decays very slowly $\sim \frac{1}{x}\sin (\frac{x}{\xi _{2f}})$. An
important difference with the proximity effect for the normal metal is the
presence of the short-ranged oscillations of the order parameter with the
temperature independent period 2$\pi \xi _{2f}$. In contrast with the dirty
limit in a clean ferromagnet the characteristic lengths of the
superconducting correlations' decay and oscillations are not the same.
Halterman and Valls (2001) performed the studies of the
ferromagnet-superconductor interfaces on the basis of the self-consistent
numerical solution of the microscopical Bogoliubov-de Gennes equations. They
clearly observed the damped oscillatory behavior of the Cooper pair wave
function of the type $\Psi \sim \frac{1}{x}\sin (\frac{x}{\xi _{2f}}).$

We may conclude that at low temperatures the proximity effect in clean
ferromagnet metals is long-ranged. On the other hand,\ in the dirty limit
the use of the Usadel equations gives the exponential decay of $\Psi $. This
is due to the fact that the Usadel equations are obtained by averaging over
the impurities configurations. Zyuzin {\it et al.} (2003) pointed out that
at distances $x>>\xi _{f}$ the anomalous Green's function $F$ (as well as
the Cooper pair wave function) has a random sample-specific sign, while the
modulus does not decay exponentially. This circumstance leads to the
survival of the proximity effect in the dirty ferromagnet at distances $%
x>>\xi _{f}$. The use of the Usadel equations at such distances may be
misleading. However, from the practical point of view the range of interest
is $x<5\xi _{f}$, because at larger distances it is difficult to observe the
oscillating phenomena on experiment. In this range the use of the Usadel
equation is adequate.

The characteristic length of the induced superconductivity variation in a
ferromagnet is small compared with a superconducting length, and it implies
the use of the microscopic theory of the superconductivity to describe the
proximity effect in S/F structures. In this context, the calculations of the
free energy of S/F structures in the framework of the standard
Ginzburg-Landau functional (Ryazanov {\it et al.}, 2001a; Ryazanov {\it et
al.}, 2001b) can not be justified. Indeed, the possibility to neglect the
higher gradient terms in the Ginzburg-Landau functional implies that the
length scale of the variation of the order parameter must be larger than the
correlation length. In the ferromagnet the correlation length is $\xi _{f}=%
\sqrt{\frac{D_{f}}{h}}$ in the dirty limit and $\xi _{f}^{0}=\frac{v_{Ff}}{h}
$ in the clean limit. We see that they coincide with the characteristic
lengths of the order parameter variation in a ferromagnet. Therefore the
higher gradient terms in the Ginzburg-Landau functional will be of the same
order of magnitude as the term with the first derivative.

\subsection{Density of states oscillations}

Superconductivity creates a gap in the electronic density of states (DOS)
near the Fermi energy $E_{F}$, i. e. the DOS is zero for a energy $E$ in the
interval $E_{F}-\Delta <E<E_{F}+\Delta $. So, it is natural, that the
induced superconductivity in S/N structures decreases DOS at $E_{F}$ near
the interface. Detailed experimental studies of this phenomenon have been
performed by Moussy {\it et al.} (2001). Damped oscillatory dependence of
the Cooper pair wave function in ferromagnet hints that a similar damped
oscillatory behavior may be expected for the variation of the DOS due to the
proximity effect. Indeed, the DOS $N(\varepsilon )$, where $\varepsilon
=E-E_{F}$ is the energy calculated from the Fermi energy, is directly
related to the normal Green function in the ferromagnet $G_{f}(x,\omega )$
(Abrikosov {\it et al.}, 1975) 
\begin{equation}
N_{f}(\varepsilon )=N(0)%
\mathop{\rm Re}%
G_{f}(x,\omega \rightarrow i\varepsilon ),
\end{equation}
where $N(0)$ is the DOS of the ferromagnetic metal. In a dirty limit taking
into account the relation between the normal and anomalous Green functions $%
G_{f}^{2}+F_{f}^{2}=1$ (Usadel, 1970), and using for $F_{f}=\frac{\Delta }{%
\sqrt{\Delta ^{2}+\omega ^{2}}}\exp \left( -\frac{i+1}{\xi _{f}}x\right) $,
we directly obtain the DOS at the Fermi energy ($\varepsilon =0$) in a
ferromagnet (Buzdin, 2000) at the distance $x>>\xi _{f}$%
\begin{equation}
N_{f}(\varepsilon =0)\approx N(0)\left( 1-\frac{1}{2}\exp (-\frac{2x}{\xi
_{f}})\cos (\frac{2x}{\xi _{f}})\right) .  \label{DOS}
\end{equation}
This simple calculation implies $\Delta <<T_{c}$ . An interesting conclusion
is that at certain distances the DOS at the Fermi energy may be higher than
in the absence of superconductor. This contrasts with the proximity effect
in the S/N systems. Such behavior has been observed experimentally by Kontos 
{\it et al.} (2001) in the measurements of the DOS by planar-tunneling
spectroscopy in Al/Al$_{2}$O$_{3}$/PdNi/Nb junctions, see Fig. 4.

For the PdNi layer thickness 50 \AA\ we are at the distance when the term $%
\cos (\frac{2x}{\xi _{f}})$ in Eq. (\ref{DOS}) is positive and we have the
normal decrease of the DOS inside the gap due to the proximity effect.
However, for PdNi layer thickness 75 \AA\ the $\cos (\frac{2x}{\xi _{f}})$
term changes its sign and the DOS becomes a little bit larger than its value
in the normal effect. Such inversion of the DOS permits us to roughly
estimate $\xi _{f}$ for the PdNi alloy used by Kontos {\it et al.} (2001) as
60 \AA .

At the moment, there exist only one experimental work on the DOS in S/F
systems, while several theoretical papers treat this subject more in
details. In a series of papers Halterman and Valls (2001, 2002, 2003)
performed extensive theoretical studies of the local DOS behavior in S/F
systems in a clean limit in the framework of the self-consistent
Bogoliubov-De Gennes approach. They calculated the DOS spectra on both S and
F sides and took into account the Fermi wave vectors mismatch, interfacial
barrier and sample size.

Fazio and Lucheroni (1999) performed numerical self-consistent calculations
of the local DOS in S/F system in the framework of the Usadel equation. The
influence of the impurity scattering on the DOS oscillations has been
studied by Baladi\'{e} and Buzdin, (2001) and Bergeret {\it et al. }(2002).
An interesting conclusion is that the oscillations disappear in the clean
limit. In this context it is quite understandable, that the calculations of
the DOS oscillations made in the ballistic regime for the ferromagnetic film
on the top of the superconductor (Zareyan {\it et al.}, 2001, Zareyan {\it %
et al.}, 2002 ) depend essentially on the boundary conditions at the
ferromagnet-vacuum interface. Sun {\it et al.} (2002) used the
quasiclassical version of the Bogoliubov-De Gennes equations for the
numerical calculations of the DOS in the S/F system with semi-infinite
ferromagnet. They obtained in the clean limit the oscillations of the DOS
and presented a quantitative fit of the experimental data of Kontos {\it et
al.} (2001). Astonishingly, in the another quasiclassical approach on the
basis of Eilenberger equations the oscillations of DOS are absent in the
case of an infinite electron mean free path (Baladie and Buzdin, 2001 and
Bergeret {\it et al.}, 2002).

DOS oscillations in ferromagnets hint on the similar oscillatory behavior of
the local magnetic moment of the electrons. The corresponding magnetic
moment induced by the proximity effect may be written as

\begin{equation}
\delta M=i\mu _{B}N(0)\pi T\sum_{\omega }\left( G_{f}(x,\omega
,h)-G_{f}(x,\omega ,-h)\right) .
\end{equation}
Assuming the low resistivity of a ferromagnet in the dirty limit at
temperature near $T_{c}$, the magnetic moment is

\begin{equation}
\delta M=-\mu _{B}N(0)\pi \frac{\Delta ^{2}}{2T_{c}}\exp (-\frac{2x}{\xi _{f}%
})\sin (\frac{2x}{\xi _{f}}).
\end{equation}
Note that the total electron's magnetic moment in a ferromagnet being

\begin{equation}
M=\delta M+\mu _{B}N(0)h.
\end{equation}
Similarly to the DOS the local magnetic moment oscillates, and curiously in
some regions it may be higher than in the absence of superconductivity.
Proximity effect also induces the local magnetic moment in a superconductor
near the S/F interface at the distance of the order of superconducting
coherence length $\xi _{s}.$

The proximity induced magnetism was studied on the basis of the Usadel
equations by Bergeret{\it \ al.} (2004a, 2004b) and Krivoruchko and Koshina
(2002). Numerical calculations of Krivoruchko and Koshina (2002) revealed
the damped oscillatory behavior of the local magnetic moment in a
superconductor at the scale of $\xi _{s}$ with positive magnetization at the
interface. On the other hand Bergeret{\it \ al.} (2004a) argued that the
induced magnetic moment in a superconductor must be negative. This is
related to the Cooper pairs located in space in such a way that one electron
of the pair is in superconductor, while the other is in the ferromagnet. The
direction along the magnetic moment in the ferromagnet is preferable for the
electron of the pair located there and this makes the spin of the other
electron of the pair (located in superconductor) to be antiparallel.

The microscopic calculations of the local magnetic moment in the pure limit
in the framework of Bogoliubov-de Gennes equations (Halterman and Valls,
2004) also revealed the damped oscillatory behavior of the local magnetic
moment but at the atomic length scale. Probably in the quasiclassical
approach the oscillations of the local magnetic moments disappear in the
clean limit, similarly to the case of DOS oscillations. The magnitude of the
proximity induced magnetic moment is very small, and at present time there
are no manifestations of this phenomena on experiment.

\subsection{Andreev reflection at the S/F interface}

The spin effects play an important role in the Andreev reflection at the S/F
interface. Indeed, an incident spin up electron in ferromagnet is reflected
by the interface as a spin down hole, and in the result a Cooper pair of
electrons with opposite spins appears in a superconductor. Therefore the
both spin up and spin down bands of electrons in ferromagnet are involved in
this process. De Jong and Beenakker (1995) were the first to demonstrate the
major influence of spin polarization in ferromagnet on the subgap
conductance of the S/F interface. Indeed, in the fully spin-polarized metal
all carriers have the same spin and Andreev reflection is totally
suppressed. In general, with the increase of the spin polarization the
subgap conductance drops from the double of the normal state conductance to
a small value for the highly polarized metals. Following de Jong and
Beenakker (1995) let us consider a simple intuitive picture of the
conductance through a ballistic S/F point contact. Using the language of the
scattering channels (subbands which cross the Fermi level), the conductance
at $T=0$ of a ferromagnet-normal metal contact is given by the Landauer
formula

\begin{equation}
G_{FN}=\frac{e^{2}}{h}N.  \label{Gfn}
\end{equation}

The total number of scattering channels $N$ is the sum of the spin up $%
N_{\uparrow \text{ }}$ and spin down $N_{\downarrow \text{ }}$ channels $%
N=N_{\uparrow \text{ }}+N_{\downarrow \text{ }}$, and the spin polarization
implies that $N_{\uparrow \text{ }}>N_{\downarrow \text{ }}$. In the case of
the contact of the superconductor with the non-polarized metal all electrons
are reflected as the holes, which doubles the number of scattering channels
and the conductance itself. For the spin-polarized metal where $N_{\uparrow 
\text{ }}>N_{\downarrow \text{ }},$ all the spin down electrons will be
reflected as the spin up holes. However, only the part $N_{\downarrow \text{ 
}}/N_{\uparrow \text{ }}<1$ of the spin up electrons can be Andreev
reflected. The subgap conductance of the S/F contact is then

\begin{equation}
G_{FS}=\frac{e^{2}}{h}\left( 2N_{\downarrow \text{ }}+2N_{\uparrow \text{ }}%
\frac{N_{\downarrow \text{ }}}{N_{\uparrow \text{ }}}\right) =4\frac{e^{2}}{h%
}N_{\downarrow \text{ }}.  \label{Gfs}
\end{equation}

Comparing this expression with Eq. (\ref{Gfn}) we see that $%
G_{FS}/G_{FN}=4N_{\uparrow \text{ }}/(N_{\downarrow \text{ }}+N_{\uparrow 
\text{ }})<2$ and $G_{FS}=0$ for the full-polarized ferromagnet with $%
N_{\downarrow \text{ }}=0.$ If the spin polarization is defined as $P=\left(
N_{\uparrow \text{ }}-N_{\downarrow \text{ }}\right) /(N_{\downarrow \text{ }%
}+N_{\uparrow \text{ }})$, then the suppression of the normalized zero-bias
conductance gives the direct access to the value of $P$ :

\begin{equation}
\frac{G_{FS}}{G_{FN}}=2\left( 1-P\right) .  \label{Polarization}
\end{equation}

The subsequent experimental measurements of the spin polarization with
Andreev reflection (Upadhyay {\it et al., }1998;{\it \ }and Soulen {\it et
al., }1998{\it ) }fully confirmed the efficiency of this method to probe the
ferromagnets. The Andreev point contact spectroscopy permits to measure the
spin polarization in a much wider range of materials (Zutic, Fabian and Das
Sarma, 2004) comparing with the spin-polarized electron tunneling (Meservey
and Tedrow, 1994).

However, the interpretation of the Andreev reflection data on the
conductance of the S/F interfaces and the comparison of the spin
polarization with the one obtained from the tunneling data, may be
complicated by the band structure effects (Mazin, 1999). Zutic and Valls
(1999, 2000), Zutic and Das Sarma (1999) generalized the results of the
theoretical analysis of Blonder, Tinkham and Klapwijk (1982) to the case of
the S/F interface. An interesting striking result is that in the absence of
the potential barrier at the S/F interface, the spin polarization could
increase the subgap conductance. The condition of perfect transparency of
the interface is $v_{F\uparrow }v_{F\downarrow }=v_{s}^{2}$, where $%
v_{F\uparrow }$\ and $v_{F\downarrow }$ are the Fermi velocities for two
spin polarizations in ferromagnet, and $v_{s}$\ is the Fermi velocity in
superconductor. Vodopyanov and Tagirov (2003a) proposed a quasiclassical
theory of Andreev reflection in F/S nanocontacts and analyzed the spin
polarization calculated from the conductance and tunneling measurements.

Note that a rather high spin polarization has been measured in CrO$_{2}$
films $P=90\%$ and in La$_{0.7}$Sr$_{0.3}$MnO$_{3}$ films $P=78\%$ (Soulen 
{\it et al., }1998). The spin-polarized tunneling data for these systems is
lacking.

Another interesting effect related with the crossed Andreev reflection has
been predicted by Deutsher and Feinberg (2000) (see also Deutsher, 2004 and
Yamashita, Takahashi and Maekawa, 2003). The electric current between two
ferromagnetic leads attached to the superconductor strongly depends on the
relative orientation of the magnetization in these leads. If we assume that
the leads are fully polarized, then the electron coming from one lead cannot
experience the Andreev reflection in the same lead. However, this reflection
is possible in the second lead, provided its polarization is opposite, and
the distance between the leads is smaller than the superconducting coherence
length. The resistance between the leads will be high for the parallel
orientation of the magnetizations and low for the antiparallel orientation.

\section{Oscillatory superconducting transition temperature in S/F
multilayers and bilayers}

\subsection{First experimental evidences of the anomalous proximity effect
in S/F systems}

The damped oscillatory behavior of the superconducting order parameter in
ferromagnets may produce the commensurability effects between the period of
the order parameter oscillation (which is of the order of $\xi _{f}$) and
the thickness of a F layer. This results in the striking non-monotonous
superconducting transition temperature dependence on the F layer thickness
in S/F multilayers and bilayers. Indeed, for a F layer thickness smaller
than $\xi _{f}$ , the pair wave function in the F layer changes a little and
the superconducting order parameter in the adjacent S layers must be the
same. The phase difference between the superconducting order parameters in
the S layers is absent and we call this state the ''$0$''-phase. On the
other hand, if the F layer thickness becomes of the order of $\xi _{f}$, the
pair wave function may go trough zero at the center of F layer providing the
state with the opposite sign (or $\pi $ shift of the phase) of the
superconducting order parameter in the adjacent S layers, which we call the
''$\pi $''-phase. The increase of the thickness of the F layers may provoke
the subsequent transitions from ''$0$''- to ''$\pi $''-phases, what
superpose on the commensurability effect and result in a very special
dependence of the critical temperature on the F layer thickness. For the S/F
bilayers, the transitions between ''$0$'' and ''$\pi $''-phases are
impossible; the commensurability effect between $\xi _{f}$\ and F layer
thickness nevertheless leads to the non-monotonous dependence of $T_{c}$\ on
the F layer thickness.

The predicted oscillatory type dependence of the critical temperature
(Buzdin and Kuprianov, 1990; Radovic {\it et al.}, 1991) was subsequently
observed experimentally in Nb/Gd (Jiang {\it et al.}, 1995), Nb/CuMn
(Mercaldo {\it et al.}, 1996) and Nb/Co and V/Co (Obi {\it et al.}, 1999)
multilayers, as well as in bilayers Nb/Ni (Sidorenko {\it et al.}, 2003),
trilayers Fe/V/Fe (Garifullin {\it et al.}, 2002),\ Fe/Nb/Fe (M\"{u}hge {\it %
et al.}, 1996), Nb/[Fe/Cu] layers (V\'{e}lez {\it et al.}, 1999) and
Fe/Pb/Fe (Lazar {\it et al.}, 2000).

The strong pair-breaking influence of the ferromagnet and the nanoscopic
range of the oscillation period complicate the observation of this effect.
Advances in thin film processing techniques were crucial for the study of
this subtle phenomenon. The first indications on the non-monotonous
variation of $T_{c}$ versus the thickness of the F layer was obtained by
Wong {\it et al. }(1986) for V/Fe superlattices. However, in the subsequent
experiments of Koorevaar {\it et al.} (1994), no oscillatory behavior of $%
T_{c}$ was found, while the recent studies by Garifullin {\it et al.} (2002)
of the superconducting properties of Fe/V/Fe trilayers even revealed the
re-entrant $T_{c}$ behavior as a function of the F layer thickness.
Bourgeois and Dynes (2002) studied amorphous Pb/Ni bilayer quench-condensed
films and observed only monotonic depairing effect with the increase of the
Ni layer thickness. In the work of Sidorenko {\it et al.} (2003), the
comparative analysis of different techniques of the sample preparation was
made and the conclusion is, that the molecular beam epitaxy (MBE) grown
samples do not reveal $T_{c}$ oscillations, whereas magnetron sputtered
samples do. This difference is attributed to the appearance of magnetically
''dead'' interdiffused layer at the S/F interface which plays an important
role for the MBE grown samples. The transition metal ferromagnets, such as
Fe, have a strongly itinerant character of the magnetic moment which is very
sensitive to the local coordination. In thin Fe layers, the magnetism may be
strongly decreased and even vanished. Probably the best choice is to use the
rare-earth ferromagnetic metal with localized magnetic moments. This has
been done by Jiang {\it et al.} (1995) who prepared the magnetron sputtered
Nb/Gd multilayers, which clearly revealed the $T_{c}$ oscillations, Fig. 5.

The curves show a pronounced non-monotoneous dependence of $T_{c}$ on the Gd
layer thickness. The increase of $T_{c}$ implies the transition from the ''$%
0 $''-phase to the ''$\pi $''-phase. Note that the previous experiments on
the MBE grown Nb/Gd samples (Strunk {\it et al.}, 1994) only revealed the
step-like decrease of $T_{c}$ with increasing Gd layer thickness. The
comprehensive analysis of different problems related to the samples quality
was made by Chien and Reich (1999). Aarts {\it et al. }(1997), studied in
detail the proximity effect in the system consisting of the superconducting
V and ferromagnetic$\ $V$_{1-x}$Fe$_{x}$ alloys and demonstrated the
important role of the interface transparency for the understanding of the
pair-breaking mechanism.

\subsection{Theoretical description of the S/F multilayers}

To provide the theoretical description of the non-monotoneous dependence of $%
T_{c}$, we consider the S/F multilayered system with a thickness of the F
layer $2d_{f}$ and the S layer $2d_{s}$, see Fig. 6.

The $x$-axis is chosen perpendicular to the layers with $x=0$ at the center
of the S layer. The ''$0$''-phase case corresponds to the same
superconducting order parameter sign in all S layers (Fig. 6a) while in the
''$\pi $''-phase the sign of the superconducting order parameter in adjacent
S layers is opposite (Fig. 6b). In the case of a S/F bilayer, the anomalous
Green function $F(x)$ has zero derivative at the boundary with vacuum, see
Eq. (\ref{boundary conds}) below. It is just the case for the function $F(x)$
in the ''$0$''-phase at the centers of\ the S and F layers. So the
superconducting characteristics of a S/F bilayer with thicknesses $d_{s}$
and $d_{f}$ of the S and F layers respectively are equivalent to that of the
S/F multilayer with double layer thicknesses ($2d_{s}$ and $2d_{f}$).

The approach based on the quasiclassical Eilenberger (1968) or Usadel (1970)
equations is very convenient to deal with S/F systems (see Appendix B). In
fact, it is much simpler than the complete microscopical theory, it does not
need the detailed knowledge of all the characteristics of the S and F
metals, and is applicable for scales larger than the atomic one. Then, it
must work for thicknesses of the layers in the range $20-200$ \AA\ , which
is of primary interest for S/F systems.

In the dirty limit, if the electron elastic scattering time $\tau =l/v_{f}$
is small, more precisely $T_{c}\tau \ll 1$ and $h\tau \ll 1,$ the use of the
Usadel equations is justified. The second condition, however is much more
restrictive due to a large value of the exchange field ($h\gg T_{c}$). The
Usadel equations deal only with the Green's functions $G(x,\omega )$ and $%
F(x,\omega )$ averaged over the Fermi surface. Moreover, to calculate the
critical temperature of the second-order superconducting transition in S/F
systems, it is enough to deal with the limit of the small superconducting
order parameter ($\Delta \rightarrow 0)$ in the Usadel equations. This
linearization permits to put $G=sgn(\omega )$ and in the form linearized
over $\Delta $, the Usadel equation for the anomalous function $F_{s}$ in
the S region is written as 
\begin{equation}
\left( \left| \omega \right| -\frac{D_{s}}{2}\frac{\partial ^{2}}{\partial
x^{2}}\right) F_{s}=\Delta (x),  \label{Us eq for Fs}
\end{equation}
where $D_{s}$ is the diffusion coefficient in the S layer. In the F region,
the exchange field is present while the pairing potential $\Delta $ is
absent, and the corresponding Usadel equation for the anomalous function $%
F_{f}$ is just the Eq. (\ref{Usadel F}).

The equations for $F_{s}$ and $F_{f}$ must be supplemented by the boundary
conditions. At the superconductor-vacuum interface, the boundary condition
is simply a zero derivative of the anomalous Green function, which implies
the absence of the superconducting current through the interface. The
general boundary conditions for the Usadel equations at the
superconductor-normal metal interface have been derived by Kupriyanov and
Lukichev (1988) and near the critical temperature they read 
\[
\left( \frac{\partial F_{s}}{\partial x}\right) _{x=0}=\frac{\sigma _{f}}{%
\sigma _{s}}\left( \frac{\partial F_{f}}{\partial x}\right) _{x=0}, 
\]
\begin{equation}
F_{s}\left( 0\right) =F_{f}\left( 0\right) -\xi _{n}\gamma _{B}\left( \frac{%
\partial F_{f}}{\partial x}\right) _{x=0},  \label{boundary conds}
\end{equation}
where $\sigma _{f}$ $\left( \sigma _{s}\right) $ is the conductivity of the
F-layer $\left( \text{S-layer above }T_{c}\right) .$ The parameter $\gamma
_{B}$ characterizes the interface transparency $T=\frac{1}{1+\gamma _{B}}$
and is related to the S/F boundary resistance per unit area $R_{b}$ via the
following simple relationship $\gamma _{B}=\frac{R_{b}\sigma _{f}}{\xi _{n}}$
(Kupriyanov and Lukichev, 1988). By analogy with the superconducting
coherence length $\xi _{s}=\sqrt{\frac{D_{s}}{2\pi T_{c}}}$, we introduce
the normal metal coherence length $\xi _{n}=\sqrt{\frac{D_{f}}{2\pi T_{c}}}.$
The presented form of the boundary conditions corresponds to the S/F
interface $x=0$ and the positive direction of the $x$ axis chosen along the
outer normal to the S surface (i. e. the $x$ axis is directed from the S to
the F metal). It is worth notify that the boundary conditions for the Usadel
equations (Kupriyanov and Lukichev, 1988) have been obtained for
superconductor/normal metal interfaces, and their applicability for S/F
interfaces is justified, provided that the exchange field in the ferromagnet
is much smaller than the Fermi energy, i. e. $h<<$ $E_{F}$. For a
ferromagnet with localized moments, such as $Gd$, this condition is always
fulfilled, while it becomes more stringent for transition metals and
violated for half-metals. Recently Vodopyanov and Tagirov (2003b) obtained
the boundary conditions for Eilenberger equations in the case of a strong
ferromagnet. They used them to study the critical temperature of a S/F
bilayer when ferromagnet is in the clean limit. Nevertheless the important
question about the boundary conditions for Usadel equations at the interface
superconductor/strong ferromagnet is still open.

Provided the solutions of Usadel equations in the F and S layers are known,
the critical temperature $T_{c}^{\ast }$ may be found from the
self-consistency equation for the pair potential $\Delta (x)$ in a
superconducting layer 
\begin{equation}
\Delta (x)=\pi T_{c}^{\ast }\lambda \sum_{\omega }F_{s}(x,\omega ),
\label{sel cons}
\end{equation}
where $\lambda $ is BCS coupling constant in S layer (while in F layer it is
supposed to be equal to zero). This equation is more convenient to write in
the following form 
\begin{equation}
\Delta (x)\ln \frac{T_{c}^{\ast }}{T_{c}}+\pi T_{c}^{\ast }\sum_{\omega
}\left( \frac{\Delta (x)}{\left| \omega \right| }-F_{s}(x,\omega )\right) =0,
\label{self consit*}
\end{equation}
where $T_{c}$ is the bare transition temperature of the superconducting
layer in the absence of the proximity effect.

The Usadel equations provide a good basis for the complete numerical
solution of the problem of the transition temperature of S/F superlattices.
Firstly such a solution has been obtained for a S/F system with no interface
barrier by Radovic {\it et al.} (1988, 1991), using the Fourier transform
method, and this case was treated analytically by Buzdin and Kuprianov
(1990) and Buzdin {\it et al.} (1992). The role of the\ S/F interface
transparency has been elucidated by Proshin and Khusainov (1997), (for more
references see also the review by Izyumov {\it et al.} 2002) and Tagirov
(1998). Recently Fominov {\it et al.} (2002), performed a detailed analysis
of the non-monotonous critical temperature dependence of S/F bilayers for
arbitrary interface transparency and compared the results of different
approximations with exact numerical calculations.

Below we illustrate the appearance of the non-monotonous superconducting
transition temperature dependence for the case of a thin S-layer, which has
a simple analytical solution. More precisely, we consider the case $d_{s}\ll
\xi _{s}$, which implies that the variations of the superconducting order
parameter and anomalous Green's function in the S layer are small. We may
write the following expansion up to the $x^{2}$ order term for $F_{s}$ in
the S layer centered at $x=0:$\ 
\begin{equation}
F_{s}(x,\omega )=F_{0}\left( 1-\frac{\beta _{\omega }}{2}x^{2}\right) ,
\label{exp F}
\end{equation}
where $F_{0}$ is the value of the anomalous Green's functions at the center
of the S-layer, and the linear over$\ x$ term is absent due to the symmetry
of the problem in both ''$0$''- and ''$\pi $''-phases (see Fig. 4). Putting
this form of $F_{s}$ into the Usadel equation (\ref{Us eq for Fs}), we
readily find 
\begin{equation}
F_{0}=\frac{\Delta }{\omega +\tau _{s}^{-1}},  \label{F0}
\end{equation}
where we have introduced the complex pair-breaking parameter $\tau _{s}^{-1}=%
\frac{D_{s}}{2}\beta _{\omega }$ and in the first approximation over $%
d_{s}/\xi _{s}\ll 1$, the pair potential $\Delta $ may be considered as
spatially independent. The pair-breaking parameter $\tau _{s}^{-1}$, is
directly related to the logarithmic derivative of $F_{s}$ at $x=d_{s}$%
\begin{equation}
\frac{F_{s}^{^{\prime }}(d_{s})}{F_{s}(d_{s})}\simeq -d_{s}\beta _{\omega }=-%
\frac{2d_{s}\tau _{s}^{-1}}{D_{s}}.
\end{equation}

The boundary conditions Eq. (\ref{boundary conds}) permit us to calculate
the parameter $\tau _{s}^{-1}$, provided the anomalous Green function in the
F layer is known: 
\begin{equation}
\tau _{s}^{-1}=-\frac{D_{s}}{2d_{s}}\frac{\sigma _{f}}{\sigma _{s}}\frac{%
F_{f}^{^{\prime }}(d_{s})/F_{f}(d_{s})}{1-\xi _{n}\gamma _{B}F_{f}^{^{\prime
}}(d_{s})/F_{f}(d_{s})}.
\end{equation}

\subsection{''$0$''- and ''$\protect\pi $''-phases}

The solution of the Usadel equation (\ref{Usadel F}) in the F layer is
straightforward but different for ''$0$''- and ''$\pi $''-phases. Let us
start first with a ''$0$''-phase. In such a case (see Fig. 6a), we must take
as a solution for $F_{f}(x)$ at $\omega >0$ in the interval $%
d_{s}<x<d_{s}+2d_{f}$ the function symmetrical relative to the plane $%
x=d_{s}+d_{f}$, i. e. 
\begin{equation}
F_{f}(x,\omega >0)=A\cosh \left[ \frac{i+1}{\xi _{f}}\left(
x-d_{s}-d_{f}\right) \right] .
\end{equation}
Therefore the pair-breaking parameter $\tau _{s0}^{-1}$ for ''$0$''-phase at 
$\omega >0$ is 
\begin{eqnarray}
\tau _{s,0}^{-1}(\omega &>&0)=\frac{D_{s}}{2d_{s}}\frac{\sigma _{f}}{\sigma
_{s}}\frac{i+1}{\xi _{f}}\frac{\tanh \left( \frac{i+1}{\xi _{f}}d_{f}\right) 
}{1+\frac{i+1}{\xi _{f}}\xi _{n}\gamma _{B}\tanh \left( \frac{i+1}{\xi _{f}}%
d_{f}\right) },  \nonumber \\
&&  \label{tau-0}
\end{eqnarray}
and does not depend on the Matsubara frequencies $\omega .$ For a negative $%
\omega $ we simply have $\tau _{s,0}^{-1}(\omega <0)=\left( \tau
_{s,0}^{-1}(\omega >0)\right) ^{\ast }$.

Now, let us address the case of \ the ''$\pi $''-phase. The only difference
is that in such case we must choose the asymmetrical solution for $F_{f}(x)$%
\begin{equation}
F_{f}(x,\omega >0)=B\sinh \left[ \frac{i+1}{\xi _{f}}\left(
x-d_{s}-d_{f}\right) \right] ,
\end{equation}
and the corresponding pair-breaking parameter $\tau _{s,\pi }^{-1}$ is given
by the expression 
\begin{eqnarray}
\tau _{s,\pi }^{-1}(\omega &>&0)=\left( \tau _{s,\pi }^{-1}(\omega
<0)\right) ^{\ast }=  \label{tau-Pi} \\
&=&\frac{D_{s}}{2d_{s}}\frac{\sigma _{f}}{\sigma _{s}}\frac{i+1}{\xi _{f}}%
\frac{\coth \left( \frac{i+1}{\xi _{f}}d_{f}\right) }{1+\frac{i+1}{\xi _{f}}%
\xi _{n}\gamma _{B}\coth \left( \frac{i+1}{\xi _{f}}d_{f}\right) }. 
\nonumber
\end{eqnarray}
We see that in all cases the pair-breaking parameter $\tau _{s}^{-1}$ is
complex and depends on the sign of the Matsubara frequency only but not on
its value. As a result, with the help of the self-consistency equation (\ref
{self consit*}), we obtain the following expression for the critical
temperature $T_{c}^{\ast }$ of the S/F multilayer 
\begin{equation}
\ln \frac{T_{c}^{\ast }}{T_{c}}=\Psi \left( \frac{1}{2}\right) -%
\mathop{\rm Re}%
\Psi \left\{ \frac{1}{2}+\frac{1}{2\pi T_{c}^{\ast }\tau _{s}}\right\} ,
\label{temperature critique}
\end{equation}
where $\Psi $ is the Digamma function, and the pair-breaking parameter $\tau
_{s}^{-1}$ is given by Eqs. (\ref{tau-0}) and (\ref{tau-Pi}) for the ''$0$%
''- and ''$\pi $''-phases respectively. This type of expression for $%
T_{c}^{\ast }$ reminds the corresponding formula for the critical
temperature of a superconductor with magnetic impurities $($Abrikosov and
Gor'kov, 1960), though the ''magnetic scattering time'' $\tau _{s}$ is
complex in our system. If the critical temperature variation is small ($%
\frac{T_{c}-T_{c}^{\ast }}{T_{c}}<<1$), the formula for the critical
temperature shift Eq. (\ref{temperature critique}) may be simplified 
\begin{equation}
\frac{T_{c}-T_{c}^{\ast }}{T_{c}}=\frac{\pi }{4T_{c}}%
\mathop{\rm Re}%
\left( \tau _{s}^{-1}\right) .  \label{tau}
\end{equation}

\subsection{Oscillating critical temperature}

To illustrate the oscillatory behavior of the critical temperature, we
consider the case of a transparent S/F interface $\gamma _{B}=0$. The
critical temperatures $T_{c}^{\ast 0}$ and $T_{c}^{\ast \pi }$ for the ''$0$%
''- and ''$\pi $''-phases respectively, are 
\begin{eqnarray}
\frac{T_{c}-T_{c}^{\ast 0}}{T_{c}} &=&\frac{\pi }{4T_{c}\tau _{0}}\left( 
\frac{\sinh (2y)-\sin (2y)}{\cosh (2y)+\cos (2y)}\right) ,
\label{Tc 0 and PI} \\
\frac{T_{c}-T_{c}^{\ast \pi }}{T_{c}} &=&\frac{\pi }{4T_{c}\tau _{0}}\left( 
\frac{\sinh (2y)+\sin (2y)}{\cosh (2y)-\cos (2y)}\right) ,
\end{eqnarray}
where $\tau _{0}^{-1}=\frac{D_{s}}{2d_{s}\xi _{f}}\frac{\sigma _{f}}{\sigma
_{s}}$ and $2y=2d_{f}/\xi _{f}$ is the dimensionless thickness of the F
layer. The critical temperature variation versus the F layer thickness is
presented in Fig. 7.

We see that for the small F layer thicknesses, the ''$0$''-phase has a
higher transition temperature. The first crossing of the curves $T_{c}^{\ast
0}(y)$ and $T_{c}^{\ast \pi }(y)$ occurs at $2y_{c}\thickapprox 2.36$ and in
the interval of thickness $2.36\xi _{f}<2d_{f}<5.5\xi _{f}$\ ,\ the ''$\pi $%
''-phase has a higher critical temperature. The oscillations of the critical
temperature rapidly decay with the increase of $y,$ and it is not realistic
to observe on experiment more than two periods of oscillations.

In the general case, the F-layer thickness dependence of the critical
temperature Eq. (\ref{temperature critique}) may be written\ for ''$0$%
''-phase in the following form convenient for numerical calculations 
\begin{eqnarray}
\ln \frac{T_{c}^{\ast 0}}{T_{c}} &=&\Psi \left( \frac{1}{2}\right) -
\label{Tc 0 gen} \\
&&-%
\mathop{\rm Re}%
\Psi \left\{ \frac{1}{2}+\frac{2T_{c}}{T_{c}^{\ast 0}\tilde{\tau}_{0}}\frac{1%
}{\tilde{\gamma}+\frac{1-i}{2}\coth \left[ (1+i)y\right] }\right\} , 
\nonumber
\end{eqnarray}
where the dimensionless parameter $\tilde{\tau}_{0}^{-1}=1/\left( 4\pi
T_{c}\tau _{0}\right) $ and $\tilde{\gamma}=\gamma _{B}\left( \xi _{n}/\xi
_{f}\right) .$ The corresponding formula for the critical temperature for
the ''$\pi $''-phase is simply obtained from Eq. (\ref{Tc 0 gen}) by the
substitution $\coth \rightarrow \tanh .$

In Fig. 8, we present the examples of calculations of the thickness
dependence of the critical temperature for S/F multilayers for different
interface transparencies.

The oscillations of the critical temperature are most pronounced for
transparent interface $\tilde{\gamma}=0,$ and they rapidly decrease with the
increase of the boundary barrier (at $\tilde{\gamma}\gtrsim 2$ the
oscillations are hardly observable). Note that, for certain values of the
parameters $\tilde{\tau}_{0}$ and $\tilde{\gamma},$ the $T_{c}^{\ast
0}(d_{f})$ dependence may show the infinite derivative, which indicates the
change of the order of the superconducting transition from second-order to
the first-order one. This\ question was studied in detail by Tollis (2004).
The increase of the boundary barrier not only decreases the amplitude of the
critical temperature oscillations, but also it decreases the critical
thickness of F layer $y_{c}$, corresponding to the ''$0$''- ''$\pi $''-phase
transition. The limit $\tilde{\gamma}=\gamma _{B}\left( \xi _{n}/\xi
_{f}\right) >>1$ is rather special one. In such case the S/F interface
barrier becomes a tunnel barrier, and the critical thickness $y_{c}$ may be
much smaller than 1. Indeed, if the critical temperature variation is small,
(more precisely if $\tilde{\gamma}\tau _{0}>>1$), the condition $%
\mathop{\rm Re}%
(\tau _{s,0}^{-1})=%
\mathop{\rm Re}%
(\tau _{s,\pi }^{-1})$ is realized at 
\begin{equation}
d_{f}^{c}=\frac{\xi _{f}}{2}\left( \frac{3}{\tilde{\gamma}}\right) ^{1/3},
\label{dfc}
\end{equation}
and the mechanism of the ''$0$''- ''$\pi $''-phase transition is now related
to the peculiarity of tunneling through the F layer. This is very different
from the case of low interface transparency, when the transition occurs due
to the spatial oscillations of the anomalous Green's function. It must be
very difficult to observe the low transparency regime of\ the ''$0$''- ''$%
\pi $''transition with the help of the critical temperature measurements due
to the fact that at $\tilde{\gamma}>>1$ the oscillations of $T_{c}^{\ast
}(d_{f})$ become very small. On the other hand, the measurements of the
critical current in S/F/S Josephson junctions may be the adequate technique
to reveal the ''$0$''- ''$\pi $''transition in this regime (see next
Section).

It is interesting to note that for small thicknesses of F layer ($d_{f}<\xi
_{f}$) the critical temperature decreases with the increase of the interface
barrier (provided the condition $\tilde{\gamma}(d_{f}/\xi _{f})<1$ is
fulfilled) - see Fig. 8. Such a counterintiutive behavior may be explained
in the following way. The low penetration of the barrier prevents the quick
return of the Cooper pair from thin F layer. Therefore, the Cooper pair
stays for a relatively long time in the F layer before going back to the S
layer. In the results, the pair-breaking role of the exchange field in the F
layer occurs to be strongly enhanced.

The cases of S/F bilayers or F/S/F threelayers with parallel magnetization
are equivalent to the ''$0$''-phase case for the multilayers (with double F
layers thickness) and the corresponding $T_{c,0}^{\ast }(d_{f})$ dependence
reveals a rather weak non-monotonous behavior in the case of finite
transparency of the S/F interface (see Fig. 8). The comparison of the
experimental data of Ryazanov {\it et al. }(2003) for the critical
temperature of the bilayer Nb/Cu$_{0.43}$Ni$_{0.57}$ vs the thickness of the
ferromagnetic layer with the theoretical fit (Fominov {\it et al.}, 2002) is
presented in Fig. 9.

Now let us address a question, if it is possible to have a transition into a
state with the phase difference another than $\ 0$ and $\pi $ ? For example
the state with the intermediate phase difference $0<\varphi _{0}<\pi $ may
be expected at F layer thicknesses near $d_{f}^{c}$. The numerical
calculations of Radovic {\it et al. }(1991) indeed revealed the presence of
the intermediate phase. However, the relative width of the region of its
existence near $d_{f}^{c}$ was very small - around several percents only. On
the other hand, the analytical calculations show that for the thin S layer
case the states without current (corresponding to the highest $T_{c}^{\ast }$%
) are possible only for the phase difference $0$ or $\pi $ . Also, in the
S/F/S junctions the transitions between ''$0$''- and '' $\pi $''-states are
discontinuous - see discussion in the next section. Probably the narrow
region of the ''$\varphi _{0}$''- phase existence obtained by the numerical
calculations (Radovic {\it et al. }, 1991) is simply related with its
accuracy $\thicksim 1\%,$ and the width of this region may decrease with the
increase of the accuracy. Nevertheless there is another mechanism of the
realization of the ''$\varphi _{0}$''- phase due to the fluctuations of the
thickness of F layer. In such case near the critical F layer thickness $%
d_{f}^{c}$ the regions of ''$0$''- and '' $\pi $''-phases would coexist. If
the characteristic dimensions of these regions are smaller than the
Josephson length in S/F structure, then the average phase difference would
be different from $0$ and $\pi $\ (Buzdin and Koshelev, 2003).

The quasiclassical Eilenberger and Usadel equations are not adequate for
treating the strong ferromagnets with $h\thicksim E_{F}$ because the period
of Green's function oscillations becomes comparable with the interatomic
distance. On the other hand, the approach based on the Bogoliubov-de Gennes
equations in clean limit is universal. Halterman and Valls (2003, 2004a)
applied it to study the properties of clean S/F multilayers, at low
temperature. They obtained the excitation spectrum through numerical
solution of the self-consistent Bogoliubov-De Gennes equations and discussed
the influence of the interface barrier and Fermi energy mismatch on the
local density of states. Comparing the energy of the ''$0$'' and ''$\pi $''
phases Halterman and Valls confirmed the existence of the transitions
between them with the increase of F layer thickness. It is of interest that
the local density of states is quite different in the ''$0$'' and ''$\pi $''
phases, and its measurements could permit to trace the ''$0$'' - ''$\pi $''
transition. In the more recent work Halterman and Valls (2004b) showed that
a lot of different order parameter configurations may correspond to the
local energy minima in S/F heterostructures.

The calculations of the energy spectrum in the S/F/S system in ''$0$'' and ''%
$\pi $'' phases on the basis of Eilenberger equations were performed by
Dobrosavljevic-Grujic, Zikic and Radovic (2000) for s-wave superconductivity
and d-wave superconductivity (Zikic {\it et al.,} 1999). The large peaks in
the density of states were attributed to the spin-split bound states
appearing due to the special case of the Andreev reflection at the
ferromagnetic barrier.

In the previous analysis the spin-orbit and magnetic scattering were
ignored. Demler, Arnold, and Beasley (1997) theoretically studied the
influence of the spin-orbit scattering on the properties of S/F systems and
demonstrated that it is quite harmful for the observation of the oscillatory
effects. A similar effect is produced by the magnetic scattering which at
some extend is always present in S/F systems due to the non-stoichiometry of
the F layers (and it may be rather large when the magnetic alloy is used as
F layer). The calculations of the critical temperature of the S/F
multilayers in the presence of the magnetic scattering were firstly
performed by Tagirov (1998). In the framework of the formalism presented in
this section it is very easy to take into account the magnetic diffusion
with the spin-flip scattering time $\tau _{m}$ - it is enough to substitute
the exchange field $h\ $in the linearized Usadel equation (\ref{Usadel
linearized}) by $h-isgn(\omega )\tau _{m}^{-1}.$ This renormalization leads
to the decrease of the damping length and the increase of the oscillation
period, which makes the $T_{c}^{\ast }(d_{f})$ oscillations less pronounced
(Tagirov, 1998).

\section{Superconductor-ferromagnet-superconductor ''$\protect\pi $%
''-junction}

\subsection{General characteristics of the ''$\protect\pi $''-junction}

A Josephson junction at equilibrium has usually a zero phase difference $%
\varphi $ between two superconductors. The energy $E$ of Josephson junction
may be written as (see for example De Gennes, 1966a) 
\begin{equation}
E=\frac{\Phi _{0}I_{c}}{2\pi c}\left( 1-\cos \varphi \right) ,  \label{E(fi)}
\end{equation}
where $I_{c}$ is the Josephson critical current, and the current-phase
relation is $I_{s}(\varphi )=\frac{2e}{\text{%
h\hskip-.2em\llap{\protect\rule[1.1ex]{.325em}{.1ex}}\hskip.2em%
}}\frac{\partial E}{\partial \varphi }=I_{c}\sin \varphi .$ At the standard
situation, the constant $I_{c}>0,$ and the minimum energy of a Josephson
junction is achieved at $\varphi =0.$ However, in the previous section it
has been demonstrated that in the S/F multilayers the transition into the ''$%
\pi $''-phase may occur. This means that for the Josephson S/F/S junction
(with the same thickness of F layer which corresponds to the ''$\pi $%
''-phase in the multilayered system) the equilibrium phase difference would
be equal to $\pi ,$ and it is natural to call such a junction the ''$\pi $%
''-junction. For the ''$\pi $''-junction, the constant $I_{c}$ in the
equation (\ref{E(fi)}) is negative, and the transition from ''$0$''- to ''$%
\pi $''-state may be considered as a change of the sign of the critical
current, though the experimentally measured critical current is always
positive and equals to $\left| I_{c}\right| .$ The S/F/S junctions would
reveal the striking non-monotonous behavior of the critical current as a
function of F layer thickness. The vanishing of the critical current signals
the transition from ''$0$''- to ''$\pi $''-state.

The possibility of the negative Josephson coupling was firstly noted by
Kulik (1966), who discussed the spin-flip tunneling through an insulator
with magnetic impurities. Bulaevskii {\it et al. }(1977) put forward the
arguments that under certain conditions such a spin-flip tunneling could
dominate the direct tunneling and lead to the ''$\pi $''-junction
appearance. Up to now there are no experimental evidences of the ''$\pi $%
''-coupling in the Josephson junctions with magnetic impurities. On the
other hand, Buzdin {\it et al.} (1982) showed that in the ballistic S/F/S
weak link $I_{c}$ displays damped oscillations as a function of the
thickness of the F layer and its exchange field. Later, Buzdin and Kuprianov
(1991) demonstrated that these oscillations remain in the diffusive regime
and so, the ''$\pi $''-coupling is the inherent property of the S/F/S
junctions. The characteristic thickness of F layer corresponding to the
transition from the ''$0$''- to ''$\pi $''-phase is $\xi _{f}=\sqrt{\frac{%
D_{f}}{h},}$ and it is rather small $\left( 10-50\right) $ \AA\ in the
typical ferromagnets because of the large value of the exchange field $%
\left( h\gtrsim 1000K\right) $. So, the experimental verification of the ''$%
\pi $''-coupling in S/F/S junction was not easy, due to the needed very
careful control of the F layer thickness. Finally the first experimental
evidence for a ''$\pi $''-junction was obtained by Ryazanov {\it et al. }%
(2001a) for the Josephson junction with a weakly ferromagnetic interlayer of
a $Cu_{x}Ni_{1-x}$ alloy. Such choice of F layer permitted to have a
ferromagnet with a relatively weak exchange field $\left( h\thicksim 100-500%
\text{ }K\right) $ and, therefore the relatively large $\xi _{f}$ length.

\subsection{Theory of ''$\protect\pi $''-junction}

The complete qualitative analysis of the S/F/S junctions is rather
complicated, because the ferromagnetic layer may strongly modify
superconductivity near the S/F interface. In addition, the boundary
transparence and electron mean free path, as well as magnetic and spin-orbit
scattering, are important parameters affecting the critical current.

To introduce the physics of ''$\pi $''-coupling, we prefer to concentrate on
the rather simple approach based on the Usadel equation and consider the
S/F/S junction with a F-layer of thickness $2d_{f}$, (see Fig. 10) and
identical S/F interfaces. In the case of small conductivity of F layer or
small interface transparency $\sigma _{f}\xi _{s}$/$\sigma _{s}\xi _{f}$ $%
<<max(1,\gamma _{_{B}})$ we may use\ the ''rigid boundary'' conditions
(Golubov {\it et al.}, 2004) with $F_{s}\left( -d_{f}\right) =\Delta
e^{-i\varphi /2}/\sqrt{\omega ^{2}+\Delta ^{2}}$ and $F_{s}\left(
d_{f}\right) =\Delta e^{i\varphi /2}/\sqrt{\omega ^{2}+\Delta ^{2}}.$

The solution of Eq. $(\ref{Usadel F})$ in a ferromagnet satisfying the
corresponding boundary conditions is written as 
\begin{eqnarray}
F(x) &=&\frac{\Delta }{\sqrt{\omega ^{2}+\Delta ^{2}}}  \nonumber \\
&&\left\{ \frac{\cos \left( \varphi /2\right) \cosh \left( kx\right) }{%
\left( \cosh \left( kd_{f}\right) +k\gamma _{_{B}}\xi _{n}\sinh \left(
kd_{f}\right) \right) }+\right.  \label{F(x)} \\
&&\left. \frac{i\sin \left( \varphi /2\right) \sinh \left( kx\right) }{%
\left( \sinh \left( kd_{f}\right) +k\gamma _{_{B}}\xi _{n}\cosh \left(
kd_{f}\right) \right) }\right\} ,  \nonumber
\end{eqnarray}
where the complex wave-vector $k=\sqrt{2\left( \left| \omega \right|
+isign\left( \omega \right) h\right) /D_{f}.}$ This solution describes the $%
F(x)$ behavior near the critical temperature. Note, that in principle, at
arbitrary temperature, the boundary conditions are different from the Eq. ($%
\ref{boundary conds}$), see for example (Golubov {\it et al.}, 2004).
However, in the limit of low S/F interface transparency ($\gamma _{_{B}}>>1)$%
, when the amplitude of the $F$ function in F- layer is small, we may use
the linearized Usadel equation ($\ref{Usadel F}$) at all temperatures. The
only modification in the boundary conditions Eq. $\left( \ref{boundary conds}%
\right) $ is that $F_{s}$ must be substituted by $F_{s}/\left| G_{s}\right| $
and $\gamma _{_{B}}$ by $\gamma _{_{B}}/\left| G_{s}\right| $, where the
normal Green function in superconducting electrode $G_{s}=\omega /\sqrt{%
\omega ^{2}+\Delta ^{2}}$. Taking this renormalization into account\ in the
explicit form Eq. $\left( \ref{F(x)}\right) ,$ we may use it in the formula
for the supercurrent 
\begin{equation}
I_{s}\left( \varphi \right) =ieN(0)D_{f}\pi T%
\mathrel{\mathop{S\stackrel{\infty }{\sum }}\limits_{-\infty }}%
\left( F\frac{d}{dx}\stackrel{\symbol{126}}{F}-\stackrel{\symbol{126}}{F}%
\frac{d}{dx}F\right) ,  \label{current}
\end{equation}
where $\stackrel{\symbol{126}}{F}(x,h)=F^{\ast }(x,-h),$ $S$ is the area of
the cross section of the junction and $N(0)$ is the electron density of
state for a one spin projection. This expression gives the usual sinusoidal
current-phase dependence $I_{s}\left( \varphi \right) =I_{c}\sin (\varphi )$
with the critical current 
\begin{eqnarray}
I_{c} &=&eSN(0)D_{f}\pi T  \nonumber \\
&&\stackrel{\infty }{%
\mathrel{\mathop{\sum }\limits_{-\infty }}%
}\frac{\Delta ^{2}}{\omega ^{2}}\frac{2k/\cosh \left( 2kd_{f}\right) }{\tanh
\left( 2kd_{f}\right) \left( 1+\Gamma _{\omega }^{2}k^{2}\right) +2k\Gamma
_{\omega }},  \label{Ic}
\end{eqnarray}
where $\Gamma _{\omega }=\gamma _{_{B}}\xi _{n}/\left| G_{s}\right| $. This
expression may be easily generalized to take into account the different
interface transparencies $\gamma _{_{B1}}$, $\gamma _{_{B2}}>>1$, it is
enough to substitute in Eq. (\ref{Ic}) $\Gamma _{\omega }^{2}\rightarrow
\gamma _{_{B1}}\gamma _{_{B2}}\left( \xi _{n}/\left| G_{s}\right| \right)
^{2}$ and $2\Gamma _{\omega }\rightarrow \left( \gamma _{_{B1}}+\gamma
_{_{B2}}\right) \xi _{n}/\left| G_{s}\right| $. Near $T_{c}$ and in the case
of transparent interface $\gamma _{_{B}}\rightarrow 0$ (Buzdin and
Kuprianov, 1991) 
\begin{eqnarray}
I_{c} &=&eSN(0)D_{f}\frac{\pi \Delta ^{2}}{2T_{c}}\left| 
\mathop{\rm Re}%
\left[ \frac{k}{\sinh \left( 2kd_{f}\right) }\right] \right| =
\label{Ic near Tc} \\
&=&\frac{V_{0}}{R_{n}}4y\left| \frac{\cos \left( 2y\right) \sinh \left(
2y\right) +\sin \left( 2y\right) \cosh \left( 2y\right) }{\cosh \left(
4y\right) -\cos \left( 4y\right) }\right| ,  \nonumber
\end{eqnarray}
where $2y=2d_{f}/\xi _{f}$ is the dimensionless thickness of the F layer, $%
R_{n}=2d_{f}/\left( \sigma _{f}S\right) $ is the resistance of the junction (%
$\sigma _{f}=2e^{2}N(0)D_{f}$ is the conductivity of the F layer), and $%
V_{0}=\frac{\pi \Delta ^{2}}{4eT_{c}}.$

The dependence $I_{c}R_{n}/V_{0}$ vs. $2y$ is presented in Fig. 11.\ The
first vanishing of the critical current\ signals the transition from ''$0$%
''- to ''$\pi $''-state. It occurs at $2y_{c}\thickapprox 2.36$ which is
exactly the critical value of F layer thickness in S/F multilayer system
corresponding to the ''$0$''$-$ ''$\pi $''-state transition, i. e. to the
condition $T_{c}^{\ast 0}=$ $T_{c}^{\ast \pi }$ in the Eqs.( \ref{Tc 0 and
PI}). The theoretical description of the S/F/S junctions with arbitrary
interface transparencies near the critical temperature was proposed by
Buzdin and Baladie (2003).

At low temperature or low S/F barrier the amplitude of the anomalous Green's
function $F_{f}(x)$ is not small and we need to use the complete
(non-linearized) Usadel equation. In the limit of large thickness of F layer 
$d_{f}>>\xi _{f}$ \ and $\gamma _{_{B}}=0,$ the analytical solution was
obtained by Buzdin and Kuprianov (1991), and the critical current is 
\begin{equation}
I_{c}R_{n}=64\sqrt{2}\frac{\left| \Delta \right| }{e}{\cal F}\left( \frac{%
\left| \Delta \right| }{T}\right) 2y\exp (-2y)\left| \sin \left( 2y+\frac{%
\pi }{4}\right) \right| ,  \label{Ic(2df) low temp}
\end{equation}
with the function 
\begin{equation}
{\cal F}\left( \frac{\left| \Delta \right| }{T}\right) =\pi T\stackrel{%
\infty }{%
\mathrel{\mathop{\sum }\limits_{\omega >0}}%
}\frac{\left| \Delta \right| }{\left( \Omega +\omega \right) \left[ \sqrt{%
2\Omega }+\sqrt{\Omega +\omega }\right] ^{2}},
\end{equation}
where $\Omega =\sqrt{\omega ^{2}+\left| \Delta \right| ^{2}}$, and ${\cal F}%
\left( \frac{\left| \Delta \right| }{T}\right) \thickapprox \frac{\pi }{128}%
\frac{\left| \Delta \right| }{T_{c}}$ at $T\thickapprox T_{c}$ while at low
temperature $T<<T_{c},$ the function ${\cal F}\left( \frac{\left| \Delta
\right| }{T}\right) \thickapprox 0.071.$

Note that in the clean limit ($\tau h>>1$) the thickness dependence of the
critical current is very different (Buzdin {\it et al.,} 1982) and near $%
T_{c}$ it is 
\begin{equation}
I_{c}R_{n}=\frac{\pi \Delta ^{2}}{4e}\frac{\left| \sin (\frac{4hd_{f}}{v_{F}}%
)\right| }{(\frac{4hd_{f}}{v_{F}})},  \label{Ic-clean}
\end{equation}
i. e. the critical current decreases $\backsim 1/d_{f}$ and not
exponentially like in the dirty limit case. In general, in the clean limit
the S/F proximity effect is not exponential, but a power low one.

The expression (\ref{Ic-clean}) was obtained on the basis of Eilenberger
equations. In the case of a strong ferromagnet $h\lesssim E_{F\text{ }},$
the period of the oscillations of the Green's functions becomes of the order
of the interatomic distance, and this approach does not work anymore. Using
the technique of the Bogoliubov-de Gennes equations, Cayssol and Montambaux
(2004) demonstrated that the quasiclassical result (\ref{Ic-clean}), where
the only relevant parameter for the critical current oscillations being $%
hd_{f}/v_{F}$, is not applicable for the strong ferromagnets. This is
related to the progressive suppression of the Andreev reflection channels
with the increase of the exchange energy.

In the framework of the Bogoliubov-de Gennes equations Radovic {\it et al. }%
(2003) studied the general case of the ballistic S/F/S junction for a strong
exchange field, arbitrary interfacial transparency and Fermi wave vectors
mismatch. The characteristic feature of such ballistic junction is the
short-period geometrical oscillations of the supercurrent as the function of 
$d_{f}$ due to the quasiparticle transmission resonances. In the case of
strong ferromagnet, the period of ''$0$''$-$ ''$\pi $'' oscillations becomes
comparable with the period of geometrical oscillations, and their interplay
provides very special $I_{c}(d_{f})$ dependences. Also Radovic {\it et al. }%
(2003) demonstrated that the current-phase relationship may strongly deviate
from the simple sinusoidal one, and studied how it depends on the junction
parameters. While the temperature variation of $I_{c}$ is usually a
monotonic decay with increasing temperature, near the critical thickness $%
d_{f}$ corresponding to ''$0$''$-$ ''$\pi $'' transition, a nonmonotonic
dependence $I_{c}$ on temperature was obtained. Radovic {\it et al. }(2001)
showed that at low temperature the characteristic multimode anharmonicity of
the current-phase relation in clean S/F/S junctions implies the coexistence
of stable and metastable ''$0-$''$\ $and ''$\pi -$'' states. As a
consequence, the coexistence of integer and half-integer flixoid
configuration of SQUID was predicted. Note that for strong ferromagnets the
details of the electrons energy bands become important for the description
of the properties of S/F/S junction.

The weak link between $d-$wave superconductors may also produce the $\pi $
shift effect (as a review, see for example Van Harlingen, 1995). The
situation of the Josephson coupling in a ferromagnetic weak link between $d-$%
wave superconductors was studied in the clean limit theoretically by Radovic 
{\it et al. }(1999).

It is interesting that in the limit $kd<<1$ (i.e. $d_{f}<<\xi _{f}$ ) the
oscillations of the\ anomalous function in the F layer are absent, but as it
has been noted previously, for the case of the low transparency of the
barrier $\gamma _{_{B}}>>1,$ the critical current can nevertheless change
its sign. Indeed, in this limit, the expression for the critical current Eq.
(\ref{Ic}) reads 
\begin{eqnarray}
I_{c} &=&eN(0)D_{f}\pi T%
\mathrel{\mathop{S\stackrel{\infty }{%
\mathrel{\mathop{\sum }\limits_{-\infty }}}\frac{2\left| \Delta \right| ^{2}}{\omega ^{2}+\left| \Delta \right| ^{2}}}\limits_{}}%
\frac{1}{\gamma _{_{B}}^{2}\xi _{n}^{2}2d_{f}}\left( \frac{1}{k^{2}}-\right.
\nonumber \\
&&\left. -\frac{2d_{f}{}^{2}}{3}-\frac{1}{\gamma _{_{B}}\xi _{n}d_{f}k^{4}}%
\frac{\left| \omega \right| }{\sqrt{\omega ^{2}+\left| \Delta \right| ^{2}}}%
\right) .  \label{Crit current}
\end{eqnarray}
Usually at experiment, the Curie temperature $\Theta $ of ferromagnet is
higher than the superconducting critical temperature $T_{c}$. For RKKY
mechanism of ferromagnetic transition $\Theta \sim h^{2}/E_{F}$ and \ so the
exchange field $h$ occurs to be much larger than the superconducting
critical temperature $T_{c}$. In the case of the itinerant ferromagnetism,
the exchange field is usually several times higher than the Curie
temperature and also the limit $h$ $>>T_{c}$ holds. Taking this into account
and performing the summation over Matsubara frequencies of the first two
terms in the brackets of the Eq. (\ref{Crit current}), we finally obtain 
\begin{eqnarray}
I_{c} &=&\frac{eN(0)SD_{f}\Delta \xi _{f}^{2}}{4\gamma _{_{B}}^{2}d_{f}\xi
_{n}^{2}}  \nonumber \\
&&\left\{ \frac{\Delta }{h}\left[ \Psi \left( \frac{1}{2}+i\frac{h}{2\pi T}%
\right) -\Psi \left( \frac{1}{2}+i\frac{\Delta }{2\pi T}\right) +c.c\right]
+\right.  \label{Crit current1} \\
&&\left. \frac{2\pi T\Delta \xi _{f}^{2}}{\gamma _{_{B}}\xi _{n}d_{f}}%
\stackrel{}{%
\mathrel{\mathop{\sum }\limits_{\omega >0}}%
}\frac{\omega }{\left( \omega ^{2}+\Delta ^{2}\right) ^{3/2}}-\frac{4\pi }{3}%
\left( \frac{d_{f}^{2}}{\xi _{f}^{2}}\right) \tanh \left( \frac{\Delta }{2T}%
\right) \right\} .  \nonumber
\end{eqnarray}
We start with the analysis of $I_{c}$ over $d_{f}$ dependence in the limit
of very large $\gamma _{_{B}}$(more precisely when $\gamma _{_{B}}>>\left( 
\frac{h}{T_{c}}\right) $). In such case we may neglect the term proportional
to $1/\gamma _{_{B}}$ in the brackets of Eq. (\ref{Crit current1}), and then
we obtain that at $T\rightarrow 0$ the transition into the $\pi -$phase
occurs ( $I_{c}$\ changes its sign) at 
\begin{equation}
d_{f}^{c}\approx \xi _{f}\sqrt{\frac{2\Delta (0)}{h}ln\left( \frac{h}{\Delta
(0)}\right) }.
\end{equation}
Indeed the condition $d_{f}<<\xi _{f}$ is satisfied. In the case of very low
boundary transparencies, the relevant formula obtained in (Buzdin and
Baladie, 2003) near the critical temperature in the limit $\left(
T_{c}/h\right) \longrightarrow 0$ also reveals the crossover between $"0"-$
and $"\pi "-$phase. On the other hand, no transition into $\pi -$phase was
obtained in the analysis of S/F/S system by Golubov {\it et al.}\ (2002b),
which is apparently related to the use of the gradient expansion of the
anomalous function in ferromagnet when only the first term has been retained.

It is interesting to note that the critical F-layer thickness $d_{f}^{c}$,
when the transition from $"0"-$ to $"\pi "-$phase occurs, depends on the
temperature. The corresponding temperature dependences are presented in Fig.
12 for different value of $\left( T_{c}/h\right) $ ratios. We see that $%
d_{f}^{c}(T)$ decreases when the temperature decreases. This is a very
general feature and it is true also for the subsequent $"0"-"\pi "$
transitions occurring at higher F layer thickness. So for some range of
F-layer thicknesses the transition from $"0"-$ to ''$\pi "-$phase is
possible when the temperature lowers.

For the case of moderately large $\gamma _{_{B}}$ , i.e. when $1<<\gamma
_{_{B}}<<\frac{h}{T_{c}},$ the terms with $\Psi $ functions in Eq.\ (\ref
{Crit current1}) can be neglected, and at $T=T_{c\text{ }}$the critical
thickness $d_{f}^{c}\ $is 
\begin{equation}
d_{f}^{c}\left( T=T_{c}\right) =\frac{\xi _{f}}{2}\left( \frac{3\xi _{f}}{%
\gamma _{_{B}}\xi _{n}}\right) ^{1/3},  \label{dc}
\end{equation}
while at $T\rightarrow 0$ the critical thickness is somewhat smaller $%
d_{f}^{c}\left( T=0\right) =\frac{\xi _{f}}{2}\left( \frac{6\xi _{f}}{\pi
\gamma _{_{B}}\xi _{n}}\right) ^{1/3}$. The critical F layer thickness,
given by Eq. (\ref{dc}), naturally coincides with the corresponding
expression Eq. (\ref{dfc}) obtained for S/F multilayers in the limit $%
h>>T_{c}$. The examples of different non-monotonous $I_{c}(T)$ dependences
for low barrier transparency limit $\gamma _{_{B}}>>\left( \frac{h}{T_{c}}%
\right) $\ are presented in Fig. 13. In fact, in the limit of low barrier
transparency and thin F layer, we deal with the superconducting electrons
tunneling through ferromagnetically ordered atoms. The situation is in some
sense reminiscent the tunneling through magnetic impurities, considered by
Kulik (1966) and Bulaevskii {\it et al. }(1977). What may be more relevant
is the analogy with the mechanism of the ''$\pi "-$phase realization due to
the tunneling through a ferromagnetic layer in the atomic S/F multilayer
structure, which we consider in the section 7.

Fogelstr\"{o}m (2000) considered the ferromagnetic layer as a partially
transparent barrier with different transmission for two spin projections. In
some sense this work may be considered as a further development of
Bulaevskii {\it et al. }(1977) approach . The Andreev bound states appearing
near the spin-active interface within the superconducting gap are tunable
with the magnetic properties of the interface. This can result to the switch
of the junction from ''$0"-$ to ''$\pi "-$state with changing the
transmission characteristics of the interface. This approach was also
applied by Andersson, Cuevas and Fogelstr\"{o}m (2002) to study the coupling
of two superconductors through a ferromagnetic dot. They demonstrated that
the realization of the ''$\pi "-$junction is possible in this case as well.
In the framework of the Bogoliubov-de Gennes approach Tanaka and Kashiwaya
(1997) analyzed the system consisting of two superconductors separated by $%
\delta -$functional barrier with the spin-orientation dependent height.

Similarly to the case of S/F multilayers we may discuss the question of the
existence of the S/F/S junction with arbitrary equilibrium phase difference $%
\varphi _{0}.$ Naturally, the form Eq. (\ref{E(fi)}) for the energy of the
junction may give the minima at $\varphi =0$ and $\varphi =\pi $ only. A
more general expression for the Josephson junction energy takes into account
the higher order terms over the critical current which leads to the
appearance of the higher harmonics over $\varphi $ in the current-phase
relationship. Up to the second harmonic, the energy is 
\begin{equation}
E=\frac{\Phi _{0}I_{c}}{2\pi c}\left( 1-\cos \varphi \right) -\frac{\Phi _{0}%
}{2\pi c}\frac{I_{2}}{2}\cos 2\varphi ,  \label{Ewith harmonics}
\end{equation}
and the current is 
\begin{equation}
j\left( \varphi \right) =I_{c}\sin \varphi +I_{2}\sin 2\varphi .
\label{j with harm}
\end{equation}
If the sign of the second harmonic term is negative $I_{2}<0,$ then the
transition from ''$0"-$ to ''$\pi "-$phase will be continuous, and the
realization of the ''$\varphi _{0}"-$junction becomes possible. In general,
the ''$\varphi _{0}"-$junction may exist if $j\left( \varphi _{0}\right) =0$
and $\left( \partial j/\partial \varphi \right) _{\varphi _{0}}>0$. The
calculations of the current-phase relationships for different types of S/F/S
junctions (Golubov {\it et al., }2004, Radovic {\it et al., }2003 and
Cayssol and Montambaux, 2004) show that $\left( \partial j/\partial \varphi
\right) <0$, and therefore the transition between ''$0"-$ and ''$\pi "-$%
states occurs to be discontinuous.\ 

The presence of the higher harmonics in the $j\left( \varphi \right) $
relationship prevents the vanishing of the critical current at the
transition from ''$0"-$ to ''$\pi "-$state. This is always the case when the
transition occurs at low temperature. Theoretical studies of the properties
of clean S/F/S junctions at $T<T_{c}$ (Buzdin {\it et al., }1982,
Chtchelkatchev {\it et al., }2001, and Radovic {\it et al., }2003) confirm
this conclusion.

Zyuzin and Spivak (2000) argued that the mesoscopic fluctuations of the
critical current may produce the ''$\pi /2"-$ superconducting Josephson
junction. Such situation is possible when the thickness of F layer is close
to $2d_{f}^{c}$ . The spatial variations of the thickness of F layer lead to
the appearance of the second harmonic term in Eq. (\ref{j with harm}) with $%
I_{2}<0$ (Buzdin and Koshelev, 2003), and thus the realization of the ''$%
\varphi _{0}"-$junction becomes possible at $2d_{f}\thickapprox 2d_{f}^{c}$ .

\subsection{Experiments with ''$\protect\pi $''-junctions}

The temperature dependence of the critical thickness $d_{f}^{c}$ is at the
origin of the observed by Ryazanov {\it et al.\ }(2001a) very specific
temperature dependence of the critical current $I_{c}(T)$ (see Fig. 14).
With decreasing temperature for specific thicknesses of the F layer (around
27 nm), a maximum of $I_{c}$ is followed by a strong decrease down to zero,
after which $I_{c}$ rises again.

This was the first unambiguous experimental confirmation of the $"0-\pi "$
transition via the critical current measurements. Ryazanov {\it et al.\ }%
(2001a) explained their results by a model with a small exchange field $%
h\sim T_{c}$. The Cu$_{x}$Ni$_{1-x}$ alloy used in their experiments has the
Curie temperature $\Theta \sim 20-30K$ and this implies that the exchange
field must be higher $100K$. In consequence, it seems more probable that the
thickness of the F layer was in the range $d_{f}^{c}\left( 0\right)
<d_{f}<d_{f}^{c}\left( T_{c}\right) $, which provides the strong
non-monotonous temperature dependence of $I_{c}$. Also, the experimental
estimate of $\xi _{f}\thicksim 10$ $nm$ is too large for expected value of
the exchange field.

Recent systematic studies of the thickness dependence of the critical
current in junctions with Cu$_{x}$Ni$_{1-x}$ alloy as a F layer (Ryazanov 
{\it et al.,\ }2004), have revealed very strong variation of $I_{c}$ with
the F layer thickness. Indeed, the five orders change of the critical
current was observed in the thickness interval $(12-26)$ $nm$. The natural
explanation of such a strong thickness dependence is the magnetic scattering
effect which is inherent to the ferromagnetic alloys. The presence of rather
strong magnetic scattering in Cu$_{x}$Ni$_{1-x}$ alloy S/F/S junctions was
noted also by Sellier {\it et al. }(2003). The magnetic scattering
strengthens the decrease of the critical current with the increase of the F
layer thickness, and at the same time it increases the period of $%
I_{c}(2d_{f})$ oscillations. The general expression for the $I_{c}(2d_{f})$
dependence, taking into account the magnetic scattering is given in Appendix
B, \ Eq. (\ref{Ic general}). The attempts to describe the experimental data
of Ryazanov {\it et al.\ (}2004) on the $I_{c}(2d_{f})$ dependence with the
help of this expression provided hints on the existence of the another
minimum $I_{c}(2d_{f})$ at smaller F layer thickness - around $10$ $nm$. The
very recent experiments with the junctions with the F layer thicknesses up
to $7$ $nm$ have confirmed this prediction (Ryazanov {\it et al.,\ }2005) -
see Fig.15. The existence of the first $"0-\pi "$ transition at $%
2d_{f}\thickapprox 11$ $nm$ means that previously reported transitions in Cu$%
_{x}$Ni$_{1-x}$ junctions were actually the transitions from $"\pi "$ to $%
"0"-$ phase (and not as was assumed, from $"0"$ to $"\pi "-$ phase). It
means also that now it is possible to fabricate the $"\pi "$ $-$ junctions
with a $10^{4}$ times higher critical current. Note, that the first
measurements (Frolov {\it et al.,\ }2004) of the current-phase relation in
S/F/S junction with Cu$_{0.47}$Ni$_{0.53}$ F layer provided no evidence of
the second harmonic in $j\left( \varphi \right) $ relationship at the $"0"$
- $"\pi "$ transition. These measurements were performed using the junction
with F layer thickness around $22$ $nm$, i. e. near the second minimum on
the $I_{c}(2d_{f})$ dependence. The much higher critical current near the
first minimum (at $2d_{f}\thickapprox 11$ $nm$) may occur to be very helpful
for a search of the second harmonic.

The results of Ryazanov {\it et al.\ }(2001a) on the temperature induced
crossover between $0-$and $\pi -$states were recently confirmed in the
experiments of Sellier {\it et al.\ }(2003). Kontos {\it et al.\ }(2002)
observed the damped oscillations of the critical current as a function of F
layer thickness in $Nb/Al/Al_{2}O_{3}/PdNi/Nb$ junctions. The measured
critical current with the theoretical fit (Buzdin and Baladie, 2003) are
presented in Fig. 16. Blum {\it et al.} (2002) reported the strong
oscillations of the critical current with the F layer thickness in $%
Nb/Cu/Ni/Cu/Nb$ junctions.

Bulaevskii {\it et al. }(1977) pointed out that ''$\pi $''-junction
incorporated into a superconducting ring would generate a spontaneous
current and a corresponding magnetic flux would be half a flux quantum $\Phi
_{0}.$ The appearance of the spontaneous current is related to the fact that
the ground state of the ''$\pi $''-junction\ corresponds to the phase
difference $\pi $ and so, this phase difference will generate a supercurrent
in the ring which short circuits the junction. Naturally the spontaneous
current is generated if there are any odd number of ''$\pi $''-junctions in
the ring. This circumstance has been exploited in a elegant way by Ryazanov 
{\it et al. }(2001c) to provide unambiguous proof of the ''$\pi $''-phase
transition. The authors (Ryazanov {\it et al., }2001c) observed the
half-period shift of the external magnetic field dependence of the transport
critical current in triangular S/F/S arrays. The thickness of F layers of
the S/F/S junctions was chosen in such a way that at high temperature the
junctions were the usual ''$0$''-junctions,\ and they transformed into the ''%
$\pi $''-junctions with the decrease of the temperature (Ryazanov {\it et
al., }2001a).

Guichard {\it et al. }(2003) performed similar phase sensitive experiments
using dc SQUID with ''$\pi $''-junction. The total current $I$ flowing
trough the SQUID is the sum of the currents $I_{a}$ and $I_{b}$ flowing
through the two junctions, $I=I_{a}$ +$I_{b}$. If the junctions have the
same critical currents $I_{c}$ and both are ''$0$''-junctions, then $%
I_{a}=I_{c}\sin \varphi _{a}$ and $I_{b}=I_{c}\sin \varphi _{b}$, where $%
\varphi _{a}$ and $\varphi _{b}$ are the phase differences across the
junctions. Neglecting the inductance of the loop of SQUID, the phase
differences satisfy the usual relation (Barone and Paterno, 1982), $\varphi
_{a}-\varphi _{b}=2\pi \Phi /\Phi _{0},$where $\Phi $ is the flux of the
external magnetic field through the loop of the SQUID. The maximum critical
current of the SQUID will be $I_{\max }=2I_{c}\cos (\pi \Phi /\Phi _{0}).$
In the case when one of the junctions (let us say b) is the ''$\pi $%
''-junction with the same critical current, the current flowing through it $%
I_{b}=-I_{c}\sin \varphi _{b}=I_{c}\sin \left( \varphi _{b}+\pi \right) $.
Therefore the maximum critical current of the SQUID in this case will be $%
I_{\max }^{\pi }=2I_{c}\cos (\pi \Phi /\Phi _{0}+\pi /2)$, and the
diffraction pattern will be shifted of half a quantum flux. If both
junctions are the ''$\pi $''-junctions the diffraction pattern will be
identical to the diffraction pattern of the SQUID with two ''$0$%
''-junctions. Namely this was observed on experiment by Guichard {\it et al. 
}(2003) with SQUID containing junctions with PdNi ferromagnetic layers, see
Fig. 17.

Recently Bauer {\it et al. }(2004) measured with the help of micro
Hall-sensor the magnetization of a mesoscopic superconducting loop
containing a PdNi ferromagnetic ''$\pi $''-junction. These measurements also
provided a direct evidence of the spontaneous current induced by the ''$\pi $%
''-junction.

\section{Complex S/F structures}

\subsection{F/S/F spin-valve sandwiches}

The strong proximity effect in superconductor-metallic ferromagnet
structures could lead to the phenomenon of spin-orientation-dependent
superconductivity in F/S/F spin-valve sandwiches. Such type of behavior was
predicted by Buzdin {\it et al.} (1999) and Tagirov (1999) and recently has
been observed on experiment by Gu {\it et al. }(2002). Note that a long time
ago De Gennes (1966b) considered theoretically the system consisting of a
thin S layer in between two ferromagnetic insulators. He argued that the
parallel orientation of the magnetic moments is more harmful for
superconductivity because of the presence of the non-zero averaged exchange
field acting on the surface of the superconductor. This prediction has been
confirmed on experiment by Hauser (1969)\ on In film sandwiched between two
Fe$_{3}$O$_{4}$ films and Deutscher and Meunier (1969), on a In film between
oxidized FeNi and Ni layers, see Fig. 18. Curiously, the experiments of
Deutscher and Meunier (1969) correspond more to the case of the metallic
F/S/F sandwiches as the authors report rather low interface resistance.

To consider the spin-orientation effect in metallic F/S/F sandwiches we use
the notations analogous to that of section 4. More precisely, to have a
direct connection with the corresponding formula of Section 4, we assume
that the thickness of the F layers is $d_{f}$ and the S layer - $2d_{s}$,
see Fig. 19.

Also, to provide a simple theoretical description we consider the case $%
d_{s}\ll \xi _{s}$ with only two orientations of the ferromagnetic moments:
parallel and antiparallel. The case of arbitrary orientations of the
ferromagnetic moments needs the introduction of triplet components of the
anomalous Green's functions. The first attempt of such analysis was made by
Baladi\'{e} {\it et al.} (2001), but on the basis of the incomplete form of
the Usadel equation. The full correct calculations for this case has been
performed by Volkov {\it et al.} (2003), Bergeret {\it et al.} (2003), and
Fominov {\it et al.} (2003a).

In fact, we only need to analyze the case of the antiparallel orientation of
the ferromagnetic moments because the case of the parallel orientation is
completely equivalent to the ''$0"-$phase in S/F multilayered structure
(Section 4) with the F layers two times thinner than in a F/S/F sandwich. In
other words, our choice of notations permits for the parallel orientation
case to use directly the corresponding expressions for the critical
temperature for the ''$0"-$phase from Section 4. To analyze the antiparallel
orientation case, we follow the approach used in Section 4, but we need to
keep the linear over $x$ term in the expansion of the anomalous Green's
function in the S layer Eq. (\ref{exp F}) 
\begin{equation}
F_{s}(x,\omega )=F_{0}\left( 1+\alpha _{\omega }x-\frac{\beta _{\omega }}{2}%
x^{2}\right) .
\end{equation}
With the help of the Usadel equation (\ref{Us eq for Fs}), we readily find
that $F_{0}$ has the form (\ref{F0}) with the pair-breaking parameter $\tau
_{s}^{-1}$ determined by the expression 
\begin{eqnarray}
\frac{4d_{s}\tau _{s}^{-1}}{D_{s}} &=&2d_{s}\beta _{\omega }\simeq \frac{%
F_{s}^{^{\prime }}(-d_{s})}{F_{s}(-d_{s})}-  \nonumber \\
&&-\frac{F_{s}^{^{\prime }}(d_{s})}{F_{s}(d_{s})}-\frac{d_{s}}{2}\left[ 
\frac{F_{s}^{^{\prime }}(d_{s})}{F_{s}(d_{s})}+\frac{F_{s}^{^{\prime
}}(-d_{s})}{F_{s}(-d_{s})}\right] ^{2}.
\end{eqnarray}
Let us suppose that the exchange field is positive ($+h$) in the right F
layer and then for $d_{s}+d_{f}>x>d_{s}$ 
\begin{equation}
F_{f}(x,\omega >0)=A\cosh \left[ \frac{i+1}{\xi _{f}}\left(
x-d_{s}-d_{f}\right) \right] ,
\end{equation}
while for the left F layer, the exchange field is negative and for $%
-d_{s}-d_{f}$ $<x<-d_{s}$ we have 
\begin{equation}
F_{f}(x,\omega >0)=B\cosh \left[ \frac{1-i}{\xi _{f}}\left(
x+d_{s}+d_{f}\right) \right] .
\end{equation}
Taking into account the explicit form of the function $F_{f}(x)$ and the
boundary conditions (\ref{boundary conds}), we may see that for the
antiparallel alignment case $\frac{F_{s}^{^{\prime }}(d_{s})}{F_{s}(d_{s})}%
=-\left( \frac{F_{s}^{^{\prime }}(-d_{s})}{F_{s}(-d_{s})}\right) ^{\ast }$
and the pair-breaking parameter for this case $\tau _{s}^{-1}=\tau
_{s,AP}^{-1}$ may be written as\ 
\begin{equation}
\tau _{s,AP}^{-1}\simeq -\frac{D_{s}}{2d_{s}}%
\mathop{\rm Re}%
\left( \frac{F_{s}^{^{\prime }}(d_{s})}{F_{s}(d_{s})}\right) +\frac{D_{s}}{2}%
\left[ 
\mathop{\rm Im}%
\left( \frac{F_{s}^{^{\prime }}(d_{s})}{F_{s}(d_{s})}\right) \right] ^{2}.
\label{tau AP}
\end{equation}
The second term in the right-hand side of the eq. (\ref{tau AP}) may be
important only in the limit of very small $d_{f}$ and we will omit it
further. The boundary conditions Eqs. (\ref{boundary conds}) permit us to
calculate the parameter $\tau _{s}^{-1}$, provided the anomalous Green
function in the F layer is known. For the parallel alignment of the
ferromagnetic moments it is just $\tau _{s,P}^{-1}=\tau _{s,0}^{-1}$, where $%
\tau _{s,0}^{-1}$ is given by the Eq. (\ref{tau-0}), while for the
antiparallel alignment it is just 
\begin{equation}
\tau _{s,AP}^{-1}=%
\mathop{\rm Re}%
\left( \tau _{s,0}^{-1}\right) =%
\mathop{\rm Re}%
\left( \tau _{sP}^{-1}\right) .
\end{equation}
In result, we obtain the following simple formula for the critical
temperature $T_{c}^{P}$ for the parallel orientation and $T_{c}^{AP}$ for
the antiparallel one 
\begin{equation}
\ln \frac{T_{c}^{P}}{T_{c}}=\Psi \left( \frac{1}{2}\right) -%
\mathop{\rm Re}%
\Psi \left\{ \frac{1}{2}+\frac{1}{2\pi T_{c}^{P}\tau _{s,0}}\right\} ,
\end{equation}
\begin{equation}
\ln \frac{T_{c}^{AP}}{T_{c}}=\Psi \left( \frac{1}{2}\right) -\Psi \left\{ 
\frac{1}{2}+%
\mathop{\rm Re}%
\left( \frac{1}{2\pi T_{c}^{AP}\tau _{s,0}}\right) \right\} .
\end{equation}
The different kinds of $T_{c}(d_{f})$ curves are presented in Fig. 20.

We see that the interface transparency is the important factor, controlling
the spin-valve effect in F/S/F structures. It is interesting that the
optimum condition for the observation of this effect in the case of the
non-negligeable interface transparency is the choice $d_{f}\thicksim \left(
0.1-0.4\right) \xi _{f}$.

In the case when the F layer thickness exceeds $\xi _{f}$, the critical
temperature practically does not depend on $d_{f}$. This case for the
transparent S/F interface ($\gamma _{_{B}}=0$) was considered by Buzdin {\it %
et al.} (1999), and the critical temperatures for the parallel and
antiparallel alignements are presented in Fig. 21. The finite interface
transparency strongly decreases the spin-valve effect, and for the parameter 
$\widetilde{\gamma }_{_{B}}>5$ the dependence of the critical temperatures
on the mutual orientation of ferromagnetic moments is hardly observable.

The thermodynamic characteristics of F/S/F systems were studied
theoretically by Baladi\'{e} and\ Buzdin (2003) and Tollis (2004) in the
framework of Usadel formalism and it was noted that the superconductivity
always remains gapless.

Bagrets {\it et al.} (2003) developed a microscopic theory of F/S/F systems
based on the direct solution of the Gor'kov equations for the normal and
anomalous Green's functions. The main mechanism of the electron scattering
in F layers was supposed to be of the $s-d$ type. The results of this
microscopical analysis were in accordance with the quasiclassical approach
and provided a reasonable quantitative description of the experimental data
of Obi {\it et al.} (1999)\ on $T_{c}(d_{f})$ dependence in Nb/Co
multilayers.

Krunavakarn {\it et al.} (2004) generalized the approach of Fominov {\it et
al.} (2002) to perform exact numerical calculations of the nonmonotonic
critical temperature in F/S/F sandwiches. They demonstrated also that the
Takahashi-Tachiki (1986) theory of the proximity effect is equivalent to the
approach based on the Usadel equations.

Bozovic and Radovic (2002) studied theoretically the coherent transport
current through F/S/F double-barrier junctions. The exchange field and the
interface barrier reduce the Andreev reflection due to the enhancement of
the normal reflection. Interestingly, that the conductance is always higher
for parallel alignment of the ferromagnetic moments. The similar conclusion
was obtained in work of Yamashita {\it et al.} (2003). Such behavior is
related with the larger transmission for the normal tunneling current in
this orientation. The calculations also revealed the periodic vanishing of
Andreev reflection at the energies of geometrical resonance above the
superconducting gap.

The case of insulating F layers (De Gennes, 1966b) corresponds to the
situation when the superconducting electrons feel the exchange field only on
the surface of S layer. We may describe this case taking formally the limit $%
d_{f}\rightarrow 0$ with $\tau _{s0}^{-1}=ih\frac{\widetilde{a}}{d_{s}}$,
where $\widetilde{a}$ is the distance of the order of the interatomic one,
which describes the region near the S/F interface where the exchange
interaction (described by the exchange field $h$ ) with electron spins takes
place. In fact it simply means that, for the parallel orientation case, the
superconductor is under the influence of the averaged exchange field $%
\widetilde{h}=h\frac{\widetilde{a}}{d_{s}}$, while for the antiparallel
orientation this field is absent. Careful theoretical analysis of the system
consisting of the superconducting film sandwiched between two ferromagnetic
semiconducting insulators with differently oriented magnetization was
performed by Kulic and Endres (2000) for both singlet and triplet
superconductivity cases. In the case of a triplet superconductivity, the
critical temperature depends not only on the relative orientation of the
magnetization but also on its absolute orientation.

\subsection{S-F-I-F'-S heterostructures and triplet proximity effect}

A bunch of theoretical works was devoted to the analysis of more complex S/F
systems. Proshin {\it et al. }(2001) (see also Izyumov {\it et al. }2002)
studied the critical temperature of S/F multilayers with alternating
magnetization of adjacent F layers. The same authors (Izyumov {\it et al., }%
2000 and Izyumov {\it et al. }(2002)) also proposed the 3D LOFF state in F/S
contacts. However, this conclusion was based on controversial boundary
conditions, corresponding to the different in plane 2D wave-vectors on the
both sides of the contact - see the comment by Fominov {\it et al. }(2003b)
and the reply of Khusainov and Proshin (2003).

Koshina and Krivoruchko (2001) (see also Golubov {\it et al.}\ 2002a)
studied the Josephson current of two proximity S/F bilayers separated by an
insulating (I) barrier and demonstrated that in such S/F-I-F/S contact the $%
\pi $-phase may appear even at very small F layer thickness (smaller than $%
\xi _{f}$). The mechanism of the $\pi $-phase transition in this case is
related to the rotation on $\pi /2$ of the phase of the anomalous Green's
function $F$ on the S/F boundary in addition to the jump of its modulus. To
demonstrate this we consider the thin F layer of the thickness $d_{f}<<\xi
_{s}$ in contact with a superconductor. If the $x=0$ corresponds to the S/F
interface, and $x=d_{f}$ is the outer surface of the F layer, then the
solution of the linearized Usadel equation in the ferromagnet is

\begin{equation}
F_{f}\left( x,\omega >0\right) =A\cosh \left[ \frac{i+1}{\xi _{f}}\left(
x-d_{f}\right) \right] .
\end{equation}
Using the boundary condition Eq.(\ref{boundary conds}) we may easily obtain

\begin{equation}
F_{f}\left( x,\omega >0\right) \simeq F_{f}\left( 0,\omega >0\right) =\frac{%
F_{s}\left( 0,\omega >0\right) }{1+2i\gamma _{B}\xi _{n}d_{f}/\xi _{f}^{2}}.
\end{equation}
In the case of a rather low interface transparency, $\gamma _{B}\xi
_{n}d_{f}/\xi _{f}^{2}>>1,$ the jump of the phase of the $F$ function at the
interface is practically equal to $-\pi /2$ :

\begin{equation}
F_{f}\left( 0,\omega >0\right) \approx F_{s}\left( 0,\omega >0\right) \exp
(-i\frac{\pi }{2})\frac{\xi _{f}^{2}}{\gamma _{B}\xi _{n}d_{f}}.
\end{equation}
Koshina and Krivoruchko (2001) and Golubov {\it et al.}\ (2002a) argued that
at each S/F interface in the S/F-I-F/S contact the phase jump $-\pi /2$
occurs, and the total phase jump in the equilibrium state would be $\pi $.

Kulic and Kulic (2001) calculated the Josephson current between two
superconductors with a helicoidal magnetic structure. They found that the
critical current depends on the simple manner on the relative orientation $%
\theta $ of the magnetic moments on the\ banks of contact :

\begin{equation}
I_{c}=I_{c0}\left( 1-R_{\pm }\cos \theta \right) ,  \label{helic}
\end{equation}
where $R_{-}(R_{+})$ corresponds to the same (opposite) helicity of the
magnetization in the banks. Depending on the parameters of the helicoidal
ordering, the value of $R_{\pm }$ may be either smaller or larger than $1.$
If $R_{\pm }>1,$ than $I_{c}$ may be negative for some misorientation angles 
$\theta $, which means the realization of the $\pi $ - phase. Interestingly
that tuning the magnetic phase $\theta ,$ it is possible to provoke a switch
between $0$ - and $\pi $ - phase. As it may be seen from Eq. (\ref{helic}),
the critical current of the Josephson junction is maximal for the
antiparallel orientation ($\theta =\pi $) of the magnetizations in the banks.

Bergeret {\it et al. }(2001a) studied the Josephson current between two S/F
bilayers and pointed out the enhancement of the critical current for an
antiparallel alignment of the ferromagnetic moments. They demonstrated that
at low temperatures the critical current in a S/F-I-F/S junction may become
even larger than in the absence of the exchange field (i. e. if the
ferromagnetic layers are replaced by the normal metal layers with $h=0$).
More in details (taking into account different transparency of S/F
interfaces and different orientations of the magnetization in the banks)
these junctions were studied theoretically by Krivoruchko and Koshina
(2001), Golubov {\it et al.}\ (2002a), Chtchelkatchev {\it et al. (}2002)
and Li {\it et al.}\ (2002). Blanter and Hekking (2004) used Eilenberger and
Usadel equations to calculate the current-phase relation of Josephson
junction with the composite F layer, consisting of two ferromagnets with
opposite magnetizations.

Bergeret {\it et al. }(2001b) and Kadigrobov {\it et al. }(2001) analyzed in
the framework of Usadel equations the proximity effect in S/F structures
with local inhomogeneity of the magnetization. They obtained an interesting
conclusion that the varying in space magnetization generates the triplet
component of the anomalous Green's function ($\sim \left\langle \Psi
_{\uparrow }\Psi _{\uparrow }\right\rangle $) which may penetrate in the
ferromagnet at distances much larger than $\xi _{f}$. It is not however the
triplet superconductivity itself because the corresponding triplet order
parameter would be equal to zero, unlike the superfluidity in He$^{3}$, for
example. In general, the triplet components of the anomalous Green's
function always appear at the description of the singlet superconductivity
in the presence of rotating in space exchange field. For example, they were
introduced by Bulaevskii {\it et al. }(1980) in the theory of coexistence of
superconductivity with helicoidal magnetic order. An important finding of
Bergeret {\it et al. }(2001b) and Kadigrobov {\it et al. }(2001) was the
demonstration of the fact that in some sense the triplet component is
insensitive to the pair-breaking by the exchange field. Therefore its
characteristic decaying length is the same as in the normal metal, i. e. $%
\xi _{T,d}=\sqrt{\frac{D_{f}}{2\pi T}}$. The triplet long-range proximity
effect could explain the experiments on S/F mesoscopic structures (Giroud 
{\it et al.}, 1998 and Petrashov {\it et al.}, 1999), where a considerable
increase of the conductunce below the superconducting critical temperature
was observed at distances much larger than $\xi _{f}.$

In their subsequent works Bergeret {\it et al. }(2003) and Volkov {\it et
al. }(2003) studied the unusual manifestation of this triplet component in
S/F multilayered structures. The most striking effect is the peculiar
dependence of the critical current in multilayered S/F structures on the
relative orientation of the ferromagnetic moments. For the collinear
orientation, the triplet component is absent, and provided the thickness of
the ferromagnetic layer $d_{f}$ $>>\xi _{f}$ , the critical current is
exponentially small. On the other hand, if the orientation of the magnetic
moments is noncollinear then the triplet component of the superconducting
condensate appears. Its decaying length $\xi _{T,d}$ is much larger than $%
\xi _{f},$ and namely this triplet component realizes the coupling between
the adjacent superconducting layers. When $\ $the thicknesses of F layers
are in the interval of $\xi _{T,d}>>d_{f}$ $>>\xi _{f},$ then this coupling
occurs to be strong. In result, the critical current is maximal for the
perpendicular orientation of the adjacent ferromagnetic moments, and it may
exceed many times the critical current for their parallel orientation. Due
to the mesoscopic fluctuations (Zyuzin {\it et al. }2003), the decay of the
critical current for collinear orientation of the magnetic moments is not
exponential. Nevertheless, for this orientation it would be very small, and
this circumstance do not change the main conclusion on the existence of the
long range triplet proximity effect. A lot of interesting physics is
expected to emerge in the case of S/F systems with genuine triplet
superconductors. For example, the proximity effect would be strongly
dependent on the mutual orientation of the magnetic moments of the Cooper
pairs and ferromagnets.

The long range triplet proximity effect was predicted to exist in the dirty
limit. An interesting question is how it evolves in the clean limit. In this
regime there is no characteristic decaying length for the anomalous Green's
function in a ferromagnet (see Eqs. (\ref{F clean1}),(\ref{F clean2})), and
the angular behavior of the critical current in S/F multilayers may be quite
different. If, for example, we apply the Eilenberger equations for the
description of clean S/F/F'/S structure with antiparallel ferromagnetic
layers with equal thicknesses, the exchange field completely drops (Blanter
and Hekking, 2004). Therefore, the critical current will be the same as for
the non magnetic interlayers. In this case it is difficult to believe that
for the perpendicular orientation of the magnetic moments the critical
current could be even higher. The microscopical calculations in the
framework of the Bogoliubov-de Gennes equations of the properties of S/F
multilayers with non-collinear orientation of the magnetic moments would be
of substantial interest.

Barash {\it et al. }(2002) studied the Josephson current in S-FIF-S
junctions in clean limit within the quasiclassical theory of
superconductivity, based on the so-called Ricatti parametrization (Schopol
and Maki, 1995). They obtained the striking nonmonotonic dependences of the
critical current on the misorientation angle of the ferromagnetic moments.
However, even for a rather high transparency of I barrier ($D=0.8$), the
maximum of the critical current occurred for the antiparallel orientation of
the magnetic moments.

\section{Atomic thickness S/F multilayers}

\subsection{Layered ferromagnetic superconductors}

In this section, we consider an atomic-scale multilayer F/S system, where
the superconducting (S) and the ferromagnetic (F) layers alternate. When the
electron transfer integral between the S and F layers is small,
superconductivity can coexist with ferromagnetism in the adjacent layers.
Andreev {\it et al.}, (1991) demonstrated that the exchange field in F
layers favors the ''$\pi "-$phase behavior of superconductivity, when the
superconducting order parameter alternates its sign on the adjacent S layers.

Nowdays several type of layered compounds, where superconducting and
magnetic layers alternate, are known. For example in Sm$_{1.85}$Ce$_{0.15}$%
CuO$_{4}$ (Sumarlin {\it et al.}, 1992), which reveals superconductivity at $%
T_{c}=23.5$ K, the superconducting layers are separated by two ferromagnetic
layers with opposite orientations of the magnetic moments and the Neel
temperature is $T_{N}=5.9$ K. Several years ago, a new class of magnetic
superconductors based on the layered perovskite ruthenocuprate compound RuSr$%
_{2}$GdCu$_{2}$O$_{8}$ comprising CuO$_{2}$ bilayers and RuO$_{2}$
monolayers has been syntesized (see for example McLaughlin {\it et al., }%
1999 and references cited there). In RuSr$_{2}$GdCu$_{2}$O$_{8}$, the
magnetic transition occurs at $T_{M}\sim 130-140,$ K and superconductivity
appears at $T_{c}\sim 30-50$ K$.$ Recent measurements of the interlayer
current in small-sized RuSr$_{2}$GdCu$_{8}$ single crystals showed the
intrinsic Josephson effect (Nachtrab {\it et al.}, 2004). Apparently, it is
a week ferromagnetic order which is realized in this compound. Though the
magnetization measurements give evidence of the\ small ferromagnetic
component, the neutron diffraction data on RuSr$_{2}$GdCu$_{2}$O$_{8}$ (Lynn 
{\it et al.,} 1992) revealed the dominant antiferromagnetic ordering in all
three directions. Later, the presence of ferromagnetic in-plane component of
about (0.1-0.3)$\mu _{B\text{ }}$has been confirmed by neutron scattering on
isostructural RuSr$_{2}$YCu$_{2}$O$_{8\text{ }}$(Tokunaga {\it et al.},
2001). In addition, in the external magnetic field the ferromagnetic
component grows rapidly at the expense of the antiferromagnetic one.

Due to the progress of methods of the multilayer preparation, the
fabrication of artificial atomic-scale S/F superlattices becomes possible.
An important example is the YBa$_{2}$Cu$_{3}$O$_{7}/$La$_{2/3}$Ca$_{1/3}$MnO$%
_{3}$ superlattices (Sefrioui {\it et al.}, 2003 and Holden {\it et al.},
2003). The manganite half metallic compound La$_{2/3}$Ca$_{1/3}$MnO$_{3}$
(LCMO) exhibits colossal magnetoresistance and its Curie temperature $\Theta
=240$ $K$. The cuprate high-T$_{c}$ superconductor YBa$_{2}$Cu$_{3}$O$_{7}$
(YBaCuO) with $T_{c}=92$ $K$, have the similar lattice constant as LCMO
which permits to prepare the very high quality YBaCuO/LCMO superlattices
with different ratio of F and S layers thicknesses. The proximity effect in
these \ superlattices occurs to be extremely long-ranged. For a fixed
thickness of the superconducting layer, the critical temperature is
dependent over a thickness of LCMO layer in the $100$ $nm$ range (Sefrioui 
{\it et al.}, 2003 and Pe\~{n}a {\it et al.}, 2004). This is very unusual
behavior because the YBaCuO and LCMO are strongly anisotropic layered
systems with very small coherence length in the direction perpendicular to
the layers ($0.1-0.3$ $nm$). Somewhat similar giant proximity effect has
been recently reported in the non-magnetic trilayer junctions La$_{1.85}$Sr$%
_{0.15}$CuO$_{4}/$La$_{2}$CuO$_{4+d}/$La$_{1.85}$Sr$_{0.15}$CuO$_{4}$
(Bozovic {\it et al.}, 2004) and in the superconductor-antiferromagnet YBa$%
_{2}$Cu$_{3}$O$_{7}/$ La$_{0.45}$Ca$_{0.55}$MnO$_{3}$ superlattices (Pang 
{\it et al.}, 2004). The observed giant proximity effect defies the
conventional explanations. Bozovic {\it et al.} (2004) suggested that it may
be related with resonant tunneling, but at the moment the question about the
nature of this effect is open.

\subsection{Exactly solvable model of the ''$\protect\pi "$-phase}

Let us consider the exactly solvable model (Andreev {\it et al.}, 1991) of
alternating superconducting and ferromagnetic atomic metallic layers. For
simplicity, we assume that the electron's motion inside the F and S layers
is described by the same energy spectrum $\xi \left( {\bf p}\right) $. Three
basic parameters characterize the system : $t$ is the transfer energy
between the F and S layers, $\lambda $ is the Cooper pairing constant which
is assumed to be non zero in S layers only, and $h$ is the constant exchange
field in the F layers. The Hamiltonian of the system can be written as 
\begin{eqnarray*}
H &=&\sum_{\vec{p},n,i,\sigma }\xi ({\bf p})a_{ni\sigma }^{+}({\bf p}%
)a_{ni\sigma }({\bf p})+H_{int1}+H_{int2}+ \\
&&+t\left[ a_{ni\sigma }^{+}({\bf p})a_{n,-i,\sigma }({\bf p}%
)+a_{n+1,-i,\sigma }^{+}({\bf p})a_{ni\sigma }({\bf p})+h.c.\right] ,
\end{eqnarray*}
\begin{equation}
H_{int1}=\;\;\;\;\;\;\;\;\;\;\;\;\;\;\;\;\;\;\;\;\;\;\;\;\;\;\;\;\;\;\;\;\;%
\;\;\;\;\;\;\;\;\;\;\;\;\;\;\;\;\;\;\;\;\;\;\;\;\;\;\;\;\;\;\;\;\;\;\;\;\;\;%
\;\;\;\;  \label{AtMod}
\end{equation}
\[
{\frac{g}{2}}\sum_{\vec{p}_{1},\vec{p}_{2},n,\sigma }a_{n1\sigma }^{+}({\bf p%
}_{1})a_{n,1,-\sigma }^{+}(-{\bf p}_{1})a_{n,1,-\sigma }(-{\bf p}%
_{2})a_{n1\sigma }({\bf p}_{2}), 
\]
\[
H_{int2}=-h\sum_{\vec{p},n,\sigma }\sigma a_{n,-1,\sigma }^{+}({\bf p}%
)a_{n,-1,\sigma }({\bf p}), 
\]
where $a_{ni\sigma }^{+}$ is the creation operator of an electron with spin $%
\sigma $ in the $n^{th}$ elementary cell and a momentum ${\bf p}$ in the
layer $i,$ where $i=1$ for the S layer, and $i=-1$ for the F layer, and $g$
is the pairing constant. The important advantage of this model is that the
quasiparticle Green's functions can be calculated exactly and the complete
analysis of the superconducting characteristic is possible. Assuming that
the order parameter changes from cell to cell in the manner $\Delta
_{n}=\left| \Delta \right| e^{ikn},$ the self-consistency equation for the
order parameter $\left| \Delta \right| $ reads 
\begin{eqnarray}
1 &=&-\lambda T_{c}^{\ast }\lambda \sum_{\omega }\int\limits_{-\infty
}^{\infty }d\xi \\
&&\int\limits_{0}^{2\pi }\frac{dq}{2\pi }\frac{\widetilde{\omega }_{+}%
\widetilde{\omega }_{-}}{\widetilde{\omega }_{+}\widetilde{\omega }%
_{-}\left| \Delta \right| ^{2}-\left( \omega _{-}\widetilde{\omega }%
_{-}-\left| T_{q+k}\right| ^{2}\right) \left( \omega _{+}\widetilde{\omega }%
_{+}-\left| T_{q}\right| ^{2}\right) },  \nonumber
\end{eqnarray}
where $\lambda =gN(0)$ and $\omega _{\pm }=i\omega \pm \xi \left( p\right) ,$
$\widetilde{\omega }_{\pm }=\omega _{\pm }+h.$ The quasimomentum $q$ lies in
the direction perpendicular to the layers, and $T_{q}=2t\cos \left(
q/2\right) e^{iq/2}$. In the limit of a small transfer integral $t<<T_{c},$
where $T_{c}$ is the bare mean-field critical temperature of the S layer in
the absence of coupling ($t=0$), we arrive at the following equation for the
critical temperature $T_{c}^{\ast }$ : 
\begin{eqnarray}
\ln \frac{T_{c}^{\ast }}{T_{c}} &=&-\pi T_{c}^{\ast }t^{2}\sum_{\omega }%
\frac{4}{\left| \omega \right| \left( 4\omega ^{2}+h^{2}\right) }+
\label{eqTc-atomic} \\
&&+\pi T_{c}t^{4}\cos k\sum_{\omega }\frac{12\omega ^{4}-7\omega
^{2}h^{2}-h^{4}}{\left| \omega \right| ^{3}\left( \omega ^{2}+h^{2}\right)
\left( 4\omega ^{2}+h^{2}\right) ^{2}}.  \nonumber
\end{eqnarray}
The critical temperature $T_{c}^{\ast }$ is close to the bare critical
temperature $T_{c}$ and as is seen from Eq. (\ref{eqTc-atomic}), for $h=0$,
the maximal $T_{c}^{\ast }$ corresponds to $k=0$, i.e. the superconducting
order parameter is the same at all layers. It is worth to note that as the
exchange field on the F layers grows, tunneling becomes energetically more
costly, so the leading term second order in $t$ falls as $1/h^{2}$ for large 
$h$ and the critical temperature increases. This is related to the fact
that, due to the decrease of the coupling the effective exchange field
induced on the S layers decreases with the increase of $h.$ For $h>>T_{c},$
the coefficient of the $\cos k$ term has a negative sign and the maximal $%
T_{c}^{\ast }$ corresponds to $k=\pi $, so the transition occurs to the $\pi 
$-phase with an alternating order parameter $\Delta _{n}=\left| \Delta
\right| (-1)^{n}.$ Numerical calculations (Andreev {\it et al.}, 1991) give
for the critical value of the exchange field (at which $k$ changes from $0$
to $\pi $) $h_{c}=3.77T_{c},$ and the complete $(h,T)$ phase diagram is
presented in Fig. 22.

At $T=0$ the transition to the ''$\pi "$-phase occurs at $h_{c0}=0.87T_{c}$\
. The analysis of Proki\'{c} {\it et al.} (1999) and Houzet {\it et al.}
(2001) shows that the perpendicular critical current vanishes at the line of
the transition from the ''$0"$- to the ''$\pi "$-phase and the Josephson
coupled superconducting planes are decoupled. Strictly speaking, the
critical current vanishes only in $\thicksim t^{4\text{ }}$ approximation,
see Eq. (\ref{eqTc-atomic}). The term $\thicksim t^{8}$ gives the
contribution $\thicksim t^{8}\cos 2k,$ and the critical current at the
transition to the ''$\pi "$-phase will drop to the very small value $%
\thicksim I_{c}\left( t/T_{c}\right) ^{8}$. Note that the sign of the second
harmonic in $j(\varphi )$ relation generated by this $\thicksim t^{8}$ term
is positive, and therefore the transition from ''$0"$- to the ''$\pi "$%
-phase is discontinuous.

In result, if the exchange field is in the interval $h_{c0}<h<3.77T_{c},$
the ''$0$-$\pi "$ transition may be easily observed with the lowering of the
temperature due to the nonmonotoneous behavior of the Josephson plasma
frequency and the parallel London penetration (Houzet {\it et al.,} 2001).
However the typical value of the exchange field is rather high and more
probable is the situation $h>>T_{c}$, and so the system will be in the\ ''$%
\pi "$-phase at any temperatures. This is consistent with the recent
experiments of Nachtrab {\it et al.} (2004) on RuSr$_{2}$GdCu$_{2}$O$_{8}$
presenting no evidence of superconducting planes decoupling with
temperature. In RuSr$_{2}$GdCu$_{2}$O$_{8},$ the superconducting pairing is
probably of the d-wave type. This case was analyzed theoretically by
Proki\'{c} and Dobrosavljevi\'{c}-Gruji\'{c} (1999), and the scenario of the
''$\pi "$-phase appearence is very close to the case of the s-wave
superconductivity. Calculations of electronic density of states by
Proki\'{c} and Dobrosavljevi\'{c}-Gruji\'{c} (1999) and Proki\'{c} {\it et
al.} (1999) revealed some changes inherent to the '' $0$-$\pi "$ transition,
but, apparently, the experimental identification of the $\pi $-phase in the
atomic-scale S/F superlattices is an extremely difficult task. In principle,
if the superlattice consists of an even number of superconducting layers,
then the phase of the order parameter at the ends will differ by $\pi $, and
the entire system will function as a Josephson $"\pi "$-junction. The
spontaneous current in a superconducting loop containing such a $"\pi "$%
-junction could be observed at an experiment analogous to the one made by
Bauer {\it et al. }(2004).

The model Eq.(\ref{AtMod}) permits to analyze the transition from the
quasi-2D to 3D system with the increase of the transfer intergral $t$. At $%
t\lesssim T_{c},$ instead of the ''$\pi "$-phase, the LOFF state with
modulation along the superconducting layers appears and the system becomes
analogous to the 3D superconductor in an uniform exchange field (Houzet and
Buzdin, 2002).

Buzdin and Daumens (2003) considered the spin walve effect in the F/S/F
structure consisting of three atomic layers and described by the model Eq. (%
\ref{AtMod}). Analogously to the F/S/F spin-walve sandwiches (see Section
6), the critical temperature is maximal for the antiparallel orientation of
the ferromagnetic moments. However, at low temperature, the situation is
inversed. Namely, the superconducting gap occurs to be larger for the
parallel orientation of the ferromagnetic moments. This counter-intuitive
result of the inversion of the proximity effect may be understood on the
example of the ferromagnetic half-metal. Indeed at $T=0,$ the disappearance
of the Cooper pair in a S layer means that two electrons with opposite spin
must leave it. If the neighbouring F layers of half-metals are parallel,
then, for one spin orientation, they are both insulators and the electron
with this spin orientation can not enter it. It results in the impossibility
of the pair destruction. On the other hand, for the antiparallel orientation
of the F layers, for any electron spin orientation there is an ajacent
normal layer and a Cooper pair can leave the S layer. Such behavior
contrasts with the diffusive model prediction (Baladie and Buzdin, 2003 and
Tollis, 2004) but is in accordance with the $T=0$ results obtained in the
framework of the multiterminal model for S/F hybrid structures (Apinyan and
M\'{e}lin, 2002). Apparently, it is a special property of the clean limit of
the atomic-layer S/F model, and it disappears in the case of several
consequitive S layers per unit cell (M\'{e}lin and Feinberg, 2004).

\section{Superconductivity near the domain wall}

In the previous discussion of the properties of S/F heterostructures, we
have implicitly assumed that the ferromagnet has uniform magnetization, i.
e. there are no domains. It practice, the domains appear in ferromagnets
quite easily and special conditions are usually needed to obtain the
monodomain ferromagnet. In standard situation, the size of the domains is
much larger than the superconducting coherence length, and $\xi _{f}<<\xi
_{s}$, therefore the Cooper pair will sample the uniform exchange field.
However, a special situation with the S/F proximity effect is realized near
the domain wall, where the magnetic moments and the exchange field rotate.
The Cooper pairs feel the exchange field averaged over the superconducting
coherence length. Naturally, such averaged field will be smaller near the
domain wall, which leads to the local decrease of the pair-breaking
parameter. As the result, we may expect that superconductivity would be more
robust near the domain wall. In particular, the critical temperature $T_{cw}$
for the superconductivity localized near the domain wall would be higher
than that of the uniform S/F bilayer $T_{c}^{\ast }$. For bulk ferromagnetic
superconductors, the critical temperature of the superconductivity localized
near the domain wall was calculated by Buzdin {\it et al.,} (1984). The
experimental manifestations of the domain wall superconductivity in Ni$%
_{0.80}$Fe$_{0.20}$/Nb bilayers (with Nb thickness around 20 nm) were
observed by Rusanov {\it et al.} (2004). The N\'{e}el-type domain walls in
Permalloy (Ni$_{0.80}$Fe$_{0.20}$) are responsible for the local increase of
the critical temperature around 10 mK. The width of the domain walls $w$ in
Permalloy films used in (Rusanov {\it et al.,} 2004) is rather large $%
w\thicksim 0.5$ $\mu m$, i. e. much larger than the superconducting
coherence length of niobium. The rotation angle $\alpha $ of the exchange
field at the distance $\xi _{s}$ may be estimated as $\alpha \thicksim \xi
_{s}/w$ , and so the averaged exchange field $h^{av}$ is slightly smaller
than the field$\ h$ far away from the domain wall: $\left( h-h^{av}\right)
/h\thicksim \left( \xi _{s}/w\right) ^{2}$. Therefore, the relative decrease
of the pair-breaking parameter $\tau _{s}^{-1}$ in Eq. (\ref{tau-0}) will be
also of the order $\thicksim \left( \xi _{s}/w\right) ^{2}.$ From Eqs. (\ref
{tau-0}, \ref{temperature critique}) we obtain the following estimate of the
local increase of the critical temperature 
\begin{equation}
\frac{T_{cw}-T_{c}^{\ast }}{T_{c}^{\ast }}\thicksim \left( \xi _{s}/w\right)
^{2},  \label{TcW}
\end{equation}
which is of the same order of magnitude as the effect observed on the Ni$%
_{0.80}$Fe$_{0.20}$/Nb bilayers. Keeping in mind the temperature dependence
of the superconducting coherence length $\xi (T)\thicksim \xi _{s}\sqrt{%
\frac{T_{c}^{\ast }}{\left| T-T_{c}^{\ast }\right| }}$, we see that the
condition of the domain wall superconductivity appearance is simply $\xi
(T_{cw})\thicksim w$.

In the case of a very thin domain wall, the variation of the exchange field
is a step-like and the local suppression of the pair-breaking parameter
occurs at the small distance of the order $\xi _{f}<<\xi _{s}$ near the
domain wall. The situation resembles the enhancement of the superconducting
pairing near the twin planes (Khlyustikov and Buzdin, 1987). The variation
of the pair-breaking occuring over a distance $\xi _{f}$\ induces a
superconducting order parameter over a distance $\xi (T_{cw})$ near the
domain wall and the effective relative decrease of the pair-breaking
parameter will be of the order of $\xi _{f}/\xi (T_{cw}).$ Therefore, if the
shift of the critical temperature of the S/F bilayer is comparable with $%
T_{c}$ itself, i. e. $\left( T_{c}-T_{c}^{\ast }\right) /T_{c}\thicksim 1$,
the critical temperature $T_{cw}$ of the superconductivity, localized near
the domain wall may be estimated from the condition $\frac{%
T_{cw}-T_{c}^{\ast }}{T_{c}^{\ast }}\thicksim \xi _{f}/\xi (T_{cw}).$ In
result we have 
\begin{equation}
\frac{T_{cw}-T_{c}^{\ast }}{T_{c}^{\ast }}\thicksim \left( \xi _{f}/\xi
_{s}\right) ^{2},  \label{TcW-thin}
\end{equation}
which is around (1-5)\% for typical values of $\xi _{f}$\ \ and $\xi _{s}$.
A small width of the domain walls is expected in experiments of Kinsey,
Burnell, and Blamire (2001) on the critical current measurements of Nb/Co
bilayers. The domain walls occured to be responsible for the critical
current enhancement below $T_{c}^{\ast }=(5.24\pm 0.05)$ K. In the presence
of domains walls the non-zero critical current has been observed at $(5.4\pm
0.05)$ K, slightly above $T_{c}^{\ast }$.

It is worth to note that the effect of the increase of the critical
temperature in the vicinity of a domain wall is weak for very large and very
thin domain wall. The optimum thickness, when the effect may be ralatively
strong is $w\thicksim \xi _{s}.$

In the case of a perpendicular easy-axis the branching of the domains may
occur near the surface of magnetic film. If the scale of this branching is
smaller than the superconducting coherence length, the effective exchange
field is averaged, and the pair breaking parameter will be strongly
decreased. This mechanism has been proposed in (Buzdin, 1985) to explain the
presence of traces of superconductivity at low temperature in re-entrant
ferromagnetic superconductors. The similar effect may take place in S/F
bilayers and in such case the superconductivity would be extremely sensitive
to the domain structure. Rather weak magnetic field would suffice to modify
the branching of domains and supress superconductivity.

Up to now we have concentrated on the interplay between superconductivity
and ferromagnetism caused by the proximity effect related to the passing of
electrons across the S/F interface. However, if the magnetic field created
by the ferromagnet penetrates into a superconductor, it switches on the
orbital mechanism of superconductivity and magnetism interaction. The
situation when it is the only one mechanism of superconductivity and
magnetism interaction is naturally realized in the case, when the
ferromagnet is an insulator, or the buffer oxide layer separates the
superconductor and the ferromagnet. The hybrid S/F systems have been
intensively studied in connection with the problem of the controlled flux
pinning. Enhancement of the critical current has been observed
experimentally for superconducting films with arrays of submicron magnetic
dots and antidots(see, for example Van Bael {\it et al.}, 2002a and Van Bael 
{\it et al.}, 2002b, and references cited therein), and for S/F bilayers
with a domain structure in ferromagnetic films (Garc\'{i}a-Santiago {\it et
al.}, 2000). A theory of vortex structures and pinning in S/F systems at
rather low magnetic field has been elaborated by Lyuksyutov and Pokrovsky
(1998), Bulaevskii {\it et al.} (2000), Erdin {\it et al.} (2002) and
Milosevic {\it et al.}, (2002a). This subject is discussed in details in the
recent review by Lyuksyutov and Pokrovsky (2004).

The nucleation of the superconductivity in the presence of domain structure
has been theoretically studied by Buzdin and Melnikov (2003), and Aladyshkin 
{\it et al.}\ (2003) in the case of magnetic film with perpendicular
anisotropy. The conditions of the superconductivity appearance occur to be
more favorable near the domain walls. Recently the manifistation of the
domain wall superconductivity was revealed on experiment by Yang et al. (2004%
$)$. They deposited on the single crystal ferromagnetic BaFe$_{12}$O$_{19}$
substrate a 10 $nm$ Si buffer layer and then a 50 $nm$ Nb film. The strong
magnetic anisotropy of BaFe$_{12}$O$_{19}$ assures that its magnetisation is
perpendicular to the Nb film. The very characteristic $R(T)$ dependences and
pronounced hysteresis effects have been found in the resistance measurements
in the applied field.

A different situation is realized if the magnetization of F layer is lying
in the plane (parallel magnetic anisotropy). Then any type of the domain
walls will be a source of the magnetic field for the adjacent S layer, and
the domain wall locally weakens superconductivity. This idea was proposed by
Sonin (1988) to create in a S layer a superconducting weak link (Josephson
junction) attached to the domain wall.

Lange {\it et al.} (2003) used a nanoengineered lattice of magnetic dots on
the top of the superconducting film for the observation of the field-induced
superconductivity. The applied external magnetic field provided the
compensation of the magnetic field of the dots and increased the critical
temperature. The idea of such compensation effect was proposed a long time
ago by Ginzburg (1956) for the case of the ferromagnetic superconductors.

The analysis of the superconducting states appearing near the magnetic dots
(when the upper critical field depends on the angular momentum of the
superconducting nucleus wave function) was done in the works of Cheng and
Fertig (1999) and Milosevic {\it et al.} (2002b).

\section{Modification of ferromagnetic order by superconductivity}

\subsection{Effective exchange field in thin S/F bilayers}

The influence of ferromagnetism on superconductivity is strong, and it leads
to many experimentally observed consequences. Wherther the inverse is true
also ? In other words, can superconductivity affect or even destroy
ferromagnetism ? To address this question, we start with comparing the
characteristic energy scales for superconducting and magnetic transitions.
The energy gain per atom at the magnetic transition is of the order of the \
Curie temperature $\Theta $. On the other hand the condensation energy per
electron at the superconducting transition (Eq. (\ref{Es})) is much smaller
than $T_{c}$, and it is only about $\thicksim T_{c}\left( T_{c}/E_{F}\right)
<<T_{c}.$ Usually the Curie temperature is higher than $T_{c}$ and
ferromagnetism occurs to be much more robust compared with
superconductivity. Therefore the superconductivity can hardly destroy the
ferromagnetism, but it may nevertheless modify it, if such modification do
not cost too much energy. The example is the bulk ferromagnetic
superconductors ErRh$_{4}$B$_{4}$, HoMo$_{6}$S$_{8}$ and HoMo$_{6}$Se$_{8}$,
where, in superconducting phase, ferromagnetism is transformed into a domain
phase with the domain size smaller than the superconducting coherence length 
$\xi _{s}$ (Maple and Fisher, 1982; \ Bulaevskii {\it et al.}, 1985).
Similar effect has been predicted by Buzdin and Bulaevskii (1988) for a thin
ferromagnetic film on the surface of a superconductor. To illustrate this
effect, we consider the S/F bilayer with S layer thickness $d_{s}$ smaller
than the\ superconducting coherence length $\xi _{s}$ and the F layer
thickness $d_{f}<<\xi _{f}<<d_{s}$, see Fig. 23.

In the case of a transparent S/F interface, the pair-breaking parameter is
given by the Eq. (\ref{tau-0}), and it is 
\begin{equation}
\tau _{s,0}^{-1}(\omega >0)=ih\frac{D_{s}}{D_{f}}\frac{d_{f}}{d_{s}}\frac{%
\sigma _{f}}{\sigma _{s}},
\end{equation}
which simply means that the effective exchange field in the superconductor $%
\widetilde{h}\thickapprox $ $h\frac{d_{f}}{d_{s}}\left( \frac{D_{s}}{D_{f}}%
\frac{\sigma _{f}}{\sigma _{s}}\right) .$ The condition of a transparent
interface implies that the Fermi momenta are equals in both materials and
this permits us to write the effective field as 
\begin{equation}
\widetilde{h}=h\left( d_{f}/d_{s}\right) \left( v_{Fs}/v_{Ff}\right) ,
\label{heff}
\end{equation}
where $v_{Fs}$ and $v_{Ff}$ are the Fermi velocities in S and F layers
respectively. Note however that for strong ferromagnets the condition of
perfect transparency of the interface is different, $v_{F\uparrow
}v_{F\downarrow }=v_{s}^{2}$, where $v_{F\uparrow }$\ and $v_{F\downarrow }$
are the Fermi velocities for two spin polarizations in ferromagnet (Zutic
and Valls, 1999, and Zutic e{\it t al.}, 2004).

In fact, in the considered case of thin F and S layers the situation is
analogous to the magnetic superconductors with an effective exchange field $%
\widetilde{h}$, which may also depend on the coordinates $(y,z)$ in the
plane of bilayer. Let us demonstrate this important point. Keeping in mind
the domain structure, (see Fig. 23), where the exchange field depends only
on the $z\ $coordinate, we may write the Usadel equations in F and S layers 
\begin{eqnarray}
&&\left. -\frac{D_{f}}{2}\left[ G\left( \ F+\frac{\partial ^{2}}{\partial
z^{2}}F\right) -F\left( \frac{\partial ^{2}}{\partial x^{2}}G+\frac{\partial
^{2}}{\partial z^{2}}G\right) \right] \right.  \label{Usadel bilayer} \\
&&\left. +\left( \omega +ih(z)\right) F=0\right.  \nonumber
\end{eqnarray}
\begin{eqnarray}
&&\left. -\frac{D_{s}}{2}\left[ G\left( \frac{\partial ^{2}}{\partial x^{2}}%
F+\frac{\partial ^{2}}{\partial z^{2}}F\right) -F\left( \frac{\partial ^{2}}{%
\partial x^{2}}G+\frac{\partial ^{2}}{\partial z^{2}}G\right) \right] \right.
\\
&&\left. +\omega F=\Delta G.\right.  \nonumber
\end{eqnarray}
Now let us perform the averaging procedure by integrating these equations
over $x$. Due to the small thicknesses of F and S layers, the Green's
functions $G$ and $F$ vary little with $x$ and may be considered as
constants. The integration of the terms with the second derivatives on $x$
will generate $\frac{\partial F}{\partial x}$ and $\frac{\partial G}{%
\partial x}$ terms taken at the interfaces. At the interfaces with vacuum
these derivatives vanish and the boundary conditions Eq.(\ref{boundary conds}%
) permit us to rely on the derivatives of $F$ function on both sides of the
S/F interface (the same relation is true for the $G$ function, due to the
normalization condition Eq. (\ref{Usadel gen})). Excluding the derivatives $%
\left( \frac{\partial F}{\partial x}\right) _{d_{s}}$ and $\left( \frac{%
\partial G}{\partial x}\right) _{d_{s}},$ we obtain the standard Usadel
equation but for the averaged (over the S layer thickness) Green's functions 
$\overline{F}$ and $\overline{G}$%
\begin{equation}
\left( \omega +i\widetilde{h}(z)\right) \overline{F}-\frac{D_{s}}{2}\left[ 
\overline{G}\frac{\partial ^{2}}{\partial z^{2}}\overline{F}-\overline{F}%
\frac{\partial ^{2}}{\partial z^{2}}\overline{G}\right] =\Delta \overline{G},
\label{Usad eff}
\end{equation}
where the effective field $\widetilde{h}(z)=h(z)\frac{d_{f}}{d_{s}}\frac{%
D_{s}}{D_{f}}\frac{\sigma _{f}}{\sigma _{s}}=h\frac{d_{f}}{d_{s}}\frac{v_{Fs}%
}{v_{Ff}}$ and the condition $d_{f}/d_{s}<<1$ is used to neglect the small
renormalization of $D_{s}$ and $\omega $. The possibility to introduce the
effective field $\widetilde{h}(z)$ in the case of a thin bilayer is quite
natural and rather general. The same effective field may be introduced in
the framework of Eilenberger equations.

\subsection{Domain structure}

In the case of the uniform ferromagnetic ordering in the F layer,
superconductivity can exist only if $\widetilde{h}$\ does not exceed the
paramagnetic limit: $\widetilde{h}$ $<1.24T_{c}$. This means that the
thickness of the F layer must be extremely small $d_{f}<\left(
T_{c}/h\right) d_{s}$; even for $d_{s}\thicksim \xi _{s},$ taking $%
T_{c}\thicksim 10$ K and $h\thicksim 5000$ K, the maximum thickness of F
layer only around 1 nm. However, the ferromagnetic superconductors (Maple
and Fisher, 1982; \ Bulaevskii {\it et al.}, 1985) give us the example of
domain coexistence phases with the exchange field larger than the
paramagnetic limit.

We may apply the theory of magnetic superconductors (Bulaevskii {\it et al.}%
, 1985) to the description of the domain structure with wave vector $Q>>\xi
_{s}^{-1}$ in the S/F bilayer, Fig. 23. The pair-breaking parameter
associated with\ the domain structure is $\tau _{s}^{-1}\thicksim \frac{%
\widetilde{h}^{2}}{vQ}$ (Bulaevskii {\it et al.}, 1985), where $v=v_{Fs}$ is
the Fermi velocity in S layer. Let us write the domain wall energy per unit
area as $\sigma /\pi a^{2}$, where $a$ is the interatomic distance. The
domain wall energy in the F film per unit length of the wall will be $%
d_{f}\left( \sigma /\pi a^{2}\right) $. Note that we consider the case of
relatively small domain wall thickness $w<<Q^{-1}$%
\mbox{$<$}%
\mbox{$<$}%
$\xi _{s}$ and the constant $\sigma $, describing the domain wall energy is
of the order of Curie temperature $\Theta $ for the atomic thickness domain
wall but may be smaller for the thick domain wall. The change of the density
of the superconducting condensation energy due to the pair-breaking effect
of domain structure is of the order of $N(0)\Delta ^{2}/\left( \Delta \tau
_{s}\right) $. Therefore the density (per unit area) of the energy $E_{DS}$
related to the domain structure reads 
\begin{equation}
E_{DS}\thicksim N(0)d_{s}\Delta \frac{\widetilde{h}^{2}}{vQ}\ +d_{f}\frac{%
\sigma Q}{a^{2}}.  \label{Energy DS}
\end{equation}
Its minimum is reached at 
\begin{equation}
Q^{2}=\frac{d_{s}}{d_{f}}\frac{N(0)\Delta a^{2}\widetilde{h}^{2}}{\sigma v}%
\thicksim \frac{1}{a\xi _{0}}\frac{d_{s}}{d_{f}}\frac{\widetilde{h}^{2}}{%
\sigma E_{F}},  \label{Q^2}
\end{equation}
where $\xi _{0}=\hslash v/\left( \pi \Delta \right) $. The factor which
favors the existence of the domain structure is the superconducting
condensation energy $E_{s}\thicksim -N(0)d_{s}\Delta ^{2}$ per unit area.
The domain structure decreases the total energy of the system if $%
E_{DS}+E_{s}<0,$ and we obtain the following condition of its existence 
\begin{equation}
T_{c}\gtrsim \left( \widetilde{h}^{2}\sigma d_{f}/d_{s}\right) ^{1/3}=%
\widetilde{h}\left( \sigma /h\right) ^{1/3}.  \label{Tc}
\end{equation}
Due to the small factor $\left( \sigma /h\right) <<1$ this condition is less
restrictive than the paramagnetic limit ($T_{c}>0.66\widetilde{h}$).
Nevertheless the conditions of the formation of the domain structure remain
rather stringent. To minimize the $d_{f}/d_{s}$ ratio (and so the effective
exchange field) it is better to choose the largest possible $d_{s}$
thickness. However, the maximum thickness of the region, where
superconductivity will be affected by the presence of F layer is of the
order of $\xi _{s}.$ Then, even in the case of the bulk superconductor $%
d_{s}^{\max }\thicksim \xi _{s}$ and the condition of the domain phase
formation in such a case reads 
\begin{equation}
T_{c}\gtrsim h\frac{d_{f}}{\xi _{s}}\left( \sigma /h\right) ^{1/3}.
\label{Tc DW bulk}
\end{equation}
We may conclude that for the domain phase observation it is better to choose
a superconductor with a large coherence length $\xi _{s}$ and the
ferromagnet with low Curie temperature and small energy of the domain walls.

The transition into the domain state is a first order one, and as all
transitions related with the domain walls, it would be highly hysteretic.
This circumstance may strongly complicate its experimental observation. To
overcome this difficulty, it may be helpful to fabricate the S/F bilayer
with a feromagnet with a low Curie temperature $\Theta <T_{c}.$ In such
case, from the very beginning we may expect the appearence of the
non-uniform magnetic structure below $\Theta $. This system in many senses
would be analogous to the ferromagnetic superconductors ErRh$_{4}$B$_{4}$,
HoMo$_{6}$S$_{8}$ and HoMo$_{6}$Se$_{8}.$

Bergeret et al. (2000) argued that the appearance of a nonhomogeneous
magnetic order in a F film deposited on the bulk superconductor occurs via
the second order transition and the period of the structure goes to infinity
at the critical point. They considered the helicoidal magnetic structure
with a wave vector $Q$ and the magnetic moment lying in the plane of the
film. The increase of the magnetic energy due to the rotation of the moments
was taken to be proportional to $Q^{2}$. However, the considered magnetic
structure is known to generate the magnetic field at distance $\thicksim
Q^{-1}$ from the film. The contribution coming from this field makes the
magnetic energy to be proportional\ to $Q$ and not to $Q^{2}$ at a small
wave-vector regime. This circumstance qualitatively change the conclusions
of Bergeret et al. (2000) and makes the transition into a nonhomogeneous
magnetic state a first-order one.

The experiments of M\"{u}hge {\it et al.}, (1998) on the ferromagnetic
resonance measurements in the Fe/Nb bilayers revealed some decrease of the
effective magnetisation below $T_{c}$ for the bilayers with $d_{f}<1\ $nm.
This thickness is compatible with the estimate Eq. (\ref{Tc DW bulk}), but
the analysis of these experimental data by Garifullin (2002) reveals the
possibility of the formation of islands at a small thickness of Fe layer,
which may strongly complicate the interpretation of experimental results.

\subsection{Negative domain wall energy}

In the previous analysis, the energy of the domain walls was considered to
be constant independent of the presence of the superconducting layer. It is
a good approximation for a thin domain wall $w<<\xi _{s}$. However, the
phenomenon of superconductivity localized near the domain walls is the
manifestation of the local enhancement of the superconducting condensation
energy, which may give a negative contribution to the domain wall energy. We
estimate this effect for a thick $w>>\xi _{s}$ domain wall. The effect is
maximum for the S/F bilayer with the relative variation of the critical
temperature $\left( T_{c}-T_{c}^{\ast }\right) T_{c}\thicksim 1$ at $%
d_{s}\backsim \xi _{s}.$ We will suppose these conditions to be satisfied.
Following the same reasoning as in the case of the domain wall
superconductivity, we may estimate the relative local decrease of the
pair-breaking parameter as $\delta \left( \tau _{s}^{-1}\right) /\tau
_{s}^{-1}\thicksim \left( \xi _{s}/w\right) ^{2}$. Therefore the local
negative contribution to the domain wall energy (per its unit length) coming
from the superconductivity reads 
\begin{equation}
\delta E_{s}\thicksim -N(0)\Delta ^{2}\left( \xi _{s}/w\right) ^{2}wd_{s}.
\end{equation}
The proper magnetic energy of the domain wall is $E_{DW}\thicksim
d_{f}\left( \sigma /\pi a^{2}\right) ,$ and for a large domain wall $\sigma
\thicksim \Theta \left( a/w\right) $. The condition of the vanishing of the
total energy of the domain wall $\delta E_{s}+E_{DW}=0$ gives 
\begin{equation}
\frac{T_{c}^{2}}{E_{F}}\frac{\xi _{s}^{3}}{wa}\thicksim d_{f}\sigma
\thicksim \Theta \frac{a}{w}d_{f},  \label{zeroDWE}
\end{equation}
where the estimate $d_{s}\backsim \xi _{s}$ is used. Finally, we may
conclude that the energy of the domain wall may be negative for the system
with 
\begin{equation}
T_{c}\gtrsim \Theta \frac{a}{l}\frac{d_{f}}{\xi _{s}},  \label{TcDWE=0}
\end{equation}
where $l$ is the electron mean free path. We have taken into account that $%
\xi _{s}\thicksim \sqrt{\xi _{0}l}$ and $a/\xi _{0}\thicksim T_{c}/E_{F}.$
If the condition Eq. (\ref{TcDWE=0}) is fulfilled, the following scenario
emerges. The decrease of the temperature below $T_{c}^{\ast }$ will decrease
the energy of the domain walls, which are practically always present in a
ferromagnet. The concentration of the domain walls will increase and
finally, when the domain wall energy will change its sign, the relatively
dense domain structure will appear. The average distance between the domains
walls in such a structure would be of the order of the domain wall thickness
itself. Note that in the case of the small thickness of the domain wall the
superconducting contribution to its energy is negligeable and instead of Eq.
(\ref{TcDWE=0}) we obtain the non-realistic condition $T_{c}\gtrsim \Theta
\left( d_{f}/\xi _{f}\right) \left( \xi _{s}/l\right) .$ We have taken into
account only the exchange mechanism of the interaction between magnetism and
superconductivity. The orbital effect gives an opposite contribution to the
domain wall energy, related with the out of plane magnetic field near the
domain wall, which generates the screening currents in the superconducting
layer.

At the present time, there are no clear experimental evidences for the
domain structure formation in S/F bilayers. The experiments of M\"{u}hge 
{\it et al.}, (1998) on the ferromagnetic resonance measurements in the
Fe/Nb bilayers revealed some decrease of the effective magnetization below $%
T_{c}^{\ast }$ for the bilayers with $d_{f}<1\ $nm. This thickness is
compatible with the estimate Eq. (\ref{Tc DW bulk}), but the magnetic moment
decreases continuously below $T_{c}^{\ast }$. In addition the analysis of
these experimental data by Garifullin (2002) reveals the possibility of the
formation of islands at small thickness of iron layer thus reducing its
magnetic stiffness. The condition Eq. (\ref{TcDWE=0}) is apparently
fulfilled in the experiments of M\"{u}hge {\it et al.}, (1998). Therefore
the decrease of the domain wall energy may be at the origin of the observed
effect.

\subsection{Ferromagnetic film on a superconducting substrate}

Bulaevskii and Chudnovsky (2000) and Bulaevskii {\it et al.} (2002)
demonstrated that the pure orbital effect could decrease the equilibrium
domain width in the ferromagnetic film on the superconducting substrate. The
ferromagnet with a perpendicular magnetic anisotropy is either an insulator,
or it is separated from the superconductor by a thin insulating (e. g.
oxide) layer, see Fig.24.

In such case the ferromagnetic film and the superconductor are coupled only
by the magnetic field. It is well-known\ (Landau and Lifshitz, 1982) that
the positive energy of the magnetic field favors small domains, so that the
stray field does not spread at large distance. On the other hand, the
positive domains wall energy favors a large domain size. The balance of
these two contributions gives the equilibrium domain width $l_{N}\thicksim 
\sqrt{wd_{f}}$. In the presence of a superconductor, the screening currents
modify the distribution of the magnetic field near the S/F interface and
give an additional positive contribution to the energy of the magnetic
field. This results in the shrinkage of the domain width. The energy $E_{D}$
of the domain structure on the superconducting substrate reads (Bulaevskii
and Chudnovsky, 2000 and Bulaevskii {\it et al.}, 2002) 
\begin{eqnarray}
E_{D} &\thicksim &3\overline{l}+\frac{2\overline{l}_{N}^{2}}{\overline{l}}-
\\
&&-\frac{16\overline{l}}{7\zeta (3)}\sum\limits_{k\geq 0}\frac{1}{\left(
2k+1\right) ^{2}\left( 2k+1+\sqrt{\left( 2k+1\right) ^{2}+16\overline{l}^{2}}%
\right) }.  \nonumber
\end{eqnarray}
Here $\overline{l}=l/(4\pi \lambda )$ and $\overline{l}_{N}=l_{N}/(4\pi
\lambda )$ are the reduced widths of domains on a superconducting and normal
substrate respectively, and $\lambda $ is the London penetration depth. The
minimization of $E_{D}$ over $\overline{l}$ gives the equilibrium width of
domains. In the limit $\lambda \rightarrow \infty $ the influence of
superconductivity vanishes and $l=l_{N}.$ The limit $\lambda \rightarrow 0$,
when the magnetic field does not penetrate inside the superconductor was
considered by Sonin (2002). In this limit the shrinkage of the width of the
domains is maximum and $l=\sqrt{2/3}l_{N}$. Then we may conclude that the
influence of superconductivity on the domain structure is not very large and
it is even less pronounced in S/F bilayer when the thickness of the S layer
becomes smaller than the London penetration depth (Daumens and Ezzahri,
2003).

Helseth {\it et al.} (2002) studied the change of the Bloch domain wall
structure in a ferromagnetic thin film on the superconducting substrate with
the in-plane magnetization of the domains. It occurs that the wall
experiences a small shrinkage, which corresponds to the increase of the
energy of the domain wall.

Recently, Dubonos {\it et al.} (2002) demonstrated experimentally the
influence of the superconducting transition on the distribution of the
magnetic domains in mesoscopic ferromagnet-superconductor structures. This
finding makes quite plausible the observation of the effect predicted by
Bulaevskii and Chudnovsky (2000) and Bulaevskii {\it et al.} (2002).
Rearrangement of the domains normally results in the resistance change in
metallic ferromagnets. In this context Dubonos {\it et al.} (2002) noted
that domain walls' displacement due to the superconducting transition could
be the actual mechanism of the long-range resistive proximity effects
previously observed in mesoscopic Ni/Al structures (Petrashov {\it et al.},
1999) and Co/Al structures (Giroud {\it et al.}, 1998). Note also that
Aumentado and Chandrasekhar (2001) studied the electron transport in
submicron ferromagnet (Ni) in contact with a mesoscopic superconductor (Al)
and demonstrated that the interface resistance is very sensitive to the
magnetic state of the ferromagnetic particle.

\section{Conclusions}

The most striking peculiarity of the proximity effect between superconductor
and ferromagnet is the damped oscillatory behavior of the Cooper pair wave
function in ferromagnet. It results in the non-monotonous dependence of the
critical temperature of S/F bilayers and multilayers on the F layer
thickness, as well as in the formation of ''$\pi "-$ junctions in S/F/S
systems. The minimum energy of the ''$\pi "-$ junction is realized for the
phase difference $\pm \pi $, and a spontaneous supercurrent may appear in a
circuit containing the ''$\pi "-$ junction. Two possible directions of the
supercurrent reflect the double-degenerate ground state. In contrast to the
usual junction such a state is achieved without external applied field. The
qubit (or quantum bit) is the analog of a bit for quantum computation,
describing by state in a two level quantum system (Nielsen and Chuang,
2000). The S/F systems open a way to create an environmentally decoupled (so
called ''quiet'') qubit (Ioffe {\it et al.,} 1999) on the basis of the S/F/S
junction.

The ''$\pi "-$ junctions allow for a realization of the concept of the
complimentary logic. In the metal-oxide superconductor logic family the
combination of the semiconducting n-p-n junctions with the complimentary
p-n-p ones permits to significantly simplify the circuitery. The similar is
possible for the Josephson junctions devices and circuits when the ''$\pi "-$
junctions are used (Terzioglu and Beasley, 1998).The logic cells with the ''$%
\pi "-$ junctions play a role of the complimentary devices to the usual
Josephson logic cells.

Recently, Ustinov and Kaplunenko (2003) proposed to use the ''$\pi "-$
junction as a phase shifter in the rapid single-flux quantum circuits. The
relatively large geometrical inductance, which is required by the
single-flux quantum storage, may be replaced by the much smaller ''$\pi "-$
junction. The advantage of the implementation of the ''$\pi "-$ junctions is
the possibility to scale the dimension of superconducting logic circuits
down to the submicron size. In addition, the use of the ''$\pi "-$ junction
as a phase shifter substantially increases the parameter margins of the
circuits.

As it has been discussed in Section III.D the exchange interaction strongly
affects the Andreev reflection at the F/S interface presenting a powerful
tool to probe ferromagnets and measure their spin polarization.

The structures consisting of ''$0"$ and ''$\pi "-$ Josephson junctions can
exhibit quite unusual properties. Bulaevskii {\it et al. }(1978)
demonstrated that the spontaneous Josephson vortex carrying the flux $\Phi
_{0}$/$2$ appears at the boundary between ''$0"$ and ''$\pi "-$junctions. A
periodic structure consisting of small (comparing with Josephson length)
alternating ''$0"$ and ''$\pi "-$Josephson junctions may have any value of
an equilibrium averaged phase difference $\varphi _{0}$ in the interval $%
-\pi <\varphi _{0}<\pi $, depending on the ratio of lengths of ''$0"$ and ''$%
\pi "-$ junctions (Mints, 1998; Buzdin and Koshelev, 2003). The S/F
heterostructures provide the possibility of the realization of such ''$%
\varphi "-$ junction with very special two maxima current-phase relation and
Josephson vortices carrying partial fluxes $\Phi _{0}\left( \varphi _{0}/\pi
\right) $ and $\Phi _{0}\left( 1-\varphi _{0}/\pi \right) $.

The possibility to combine in a controlled manner paramagnetic and orbital
mechanisms of the interaction between superconductivity and magnetism makes
the physics of S/F heterostructures quite rich and promising for potential
applications. Let us mention in this context the recent observation of
strong vortex pinning in S/F hybrid structures, the spin valve effect in
F/S/F systems and the domain wall superconductivity, which open a large
perspective to the creation of new electronics devices. The progress of
controllable fabrication of high-quality heterostructures and especially the
high-quality interfaces was crucial for the recent breakthrough in this
domain. Further development of the microfabrication technology permits to
expect another interesting findings in the near future.

\section{Acknowledgments}

It is my pleasure to thank J. Aarts, A. Abrikosov, M. Aprili, I.
Baladi\'{e}, S. Bader, J.-P. Brison, L. Bulaevskii, H. Courtois, M. Daumens,
M. Faur\'{e}, D. Feinberg, J. Flouquet, E. Goldobin, A. Golubov, M. Houzet,
A. Koshelev, M. Kulic, M. Kuprianov, J. Lynn, B. Maple, A. Melnikov, B.
Pannetier, A. Ustinov, Z. Radovic, V. Ryazanov, N. Ryzhanova, L. Tagirov, A.
Volkov, and A. Vedyaev for useful discussions and helpful comments. I would
like to acknowledge the help of C. Meyers and M. Faur\'{e} for performing
the numerical calculations. I am grateful to M. Aprili, C. Chien, Ya.
Fominov, J. Jiang and V. Ryazanov for providing illustrative figures from
their works.

The special thanks are due to T. Chameeva and J. Leandri for their help in
preparing the manuscript.

This work was supported in part by the ESF ''Pi-shift'' program.

\section{APPENDIX:}

\subsection{Bogoliubov-de Gennes equations}

As the characteristic length of the induced superconductivity variation in a
ferromagnet is small compared with a superconducting length, it implies the
use of the microscopic theory of superconductivity to describe the proximity
effect in S/F structures. The very convenient microscopical approach to
study the superconducting properties in the ballistic regime (the clean
limit) in the presence of spatially varying field is the use of the
Bogoliubov-de Gennes equations (de Gennes, 1966a). The equations for
electron and hole wave functions $u_{\uparrow }\left( {\bf r}\right) $ and $%
v_{\downarrow }\left( {\bf r}\right) $ are

\begin{eqnarray}
\left( H_{0}-h\left( {\bf r}\right) \right) u_{\uparrow }\left( {\bf r}%
\right) +\Delta \left( {\bf r}\right) v_{\downarrow }\left( {\bf r}\right)
&=&E_{\uparrow }u_{\uparrow }\left( {\bf r}\right)  \label{Bog.eq} \\
\Delta ^{\ast }\left( {\bf r}\right) u_{\uparrow }\left( {\bf r}\right)
-\left( H_{0}+h\left( {\bf r}\right) \right) v_{\downarrow }\left( {\bf r}%
\right) &=&E_{\uparrow }v_{\downarrow }\left( {\bf r}\right) ,  \nonumber
\end{eqnarray}

where $E_{\uparrow }$ is the quasiparticle excitation energy, $H_{0}=-\hbar
^{2}\frac{\nabla ^{2}}{2m}-E_{F}$ is the single particle Hamiltonian, $%
h\left( {\bf r}\right) $ is the exchange field in the ferromagnet, and the
spin quantization axis is chosen along its direction. The equations for the
wave functions with opposite spin orientation $u_{\downarrow }\left( {\bf r}%
\right) $ and $v_{\uparrow }\left( {\bf r}\right) $ and the excitation
energy $E_{\downarrow }$ are obtained from Eq. (\ref{Bog.eq}) by\ the
substitution $h\rightarrow -h$. Note that the solution $\left( u_{\downarrow
}{\bf ,}v_{\uparrow }\right) $ with energy $E_{\downarrow }$ may be
immediately obtained from the solution of Eq. (\ref{Bog.eq}), if we choose $%
u_{\downarrow }=v_{\uparrow ,}$ $v_{\uparrow }=-u_{\downarrow }$ and $%
E_{\downarrow }=-E_{\uparrow }$.\ The pair potential in the superconductor
is determined by the self-consistent equation

\begin{equation}
\Delta ({\bf r})=\lambda \sum_{E_{\uparrow }>0}u_{\uparrow }\left( {\bf r}%
\right) v_{\downarrow }^{\ast }\left( {\bf r}\right) \left( 1-2f\left(
E_{\uparrow }\right) \right) ,  \label{sef-Bog}
\end{equation}

where $f\left( E\right) $ is the Fermi distribution function $f\left(
E\right) =1/\left( 1+\exp \left( E/T\right) \right) ,$ and $\lambda $ is the
BCS coupling constant.

Assuming that the Cooper pairing is absent in the ferromagnet, we have $%
\Delta ({\bf r})=0$ there. The situations when it is possible to obtain the
analytical solutions of the Bogoliubov-de Gennes equations with spatially
varying pair potential are very rare. However, these equations provide a
good basis for the numerical calculations to treat different aspects of S/N
and S/F proximity effects.

\subsection{Eilenberger and Usadel equations for ferromagnets}

Another microscopical approach in the theory of superconductivity uses the
electronic Green's functions. The Green's functions technique for
superconductors has been proposed by Gor'kov who introduced in addition to
the normal Green's function $G({\bf r}_{_{1}},{\bf r}_{_{2}})$ the anomalous
(Gor'kov) function $F({\bf r}_{_{1}},{\bf r}_{_{2}})$ (see, for example,
Abrikosov {\it et al.}, 1975). This technique is a very powerful tool, but
the corresponding Green's functions in a general case occur to be rather
complicated and oscillate as a function of the relative coordinate ${\bf r}%
_{_{1}}-{\bf r}_{_{2}}$ on the scale of the interatomic distance. On the
other hand, the characteristic length scales for superconductivity in S/F
systems are of the order of the layers thicknesses or damping dacay length
for the induced superconductivity and, then, they are much larger than the
atomic length. This smooth variation is described by the center of mass
coordinate ${\bf r}=\left( {\bf r}_{_{1}}+{\bf r}_{_{2}}\right) /2$ in the
Green's functions. The very convenient quasiclassical equations for the
Green's functions averaged over the rapid oscillations on the relative
coordinate has been proposed by Eilenberger (1968) (and also by Larkin and
Ovchinnikov (1968)).

Eilenberger equations are transport-like equations for the energy-integrated
Green's functions $f({\bf r,}\omega ,{\bf n})$ and $g({\bf r,}\omega ,{\bf n}%
)$, depending on the center of mass coordinate ${\bf r}$, Matsubara
frequencies $\omega =\pi T\left( 2n+1\right) $ and the direction of the unit
vector ${\bf n}$ normal to the Fermi surface. For the case of S/F
multilayers we may restrict ourselves to the situations when all quantities
only depend on one coordinate $x${\bf ,} chosen perpendicular to the layers.
Introducing the angle $\theta $ between the {\bf x} axis and the direction
of the vector ${\bf n}$ (direction of the Fermi velocity), we may write the
Eilenberger equations in the presence of the exchange field $h(x)$ in the
form (see, for example Bulaevskii {\it et al.} (1985) and a recent review on
the physics of Josephson junctions by Golubov {\it et al.} (2004)) 
\begin{gather}
\left( \omega +ih(x)+\frac{1}{2\tau }G(x,\omega )\right) f\left( x,\theta
,\omega \right) +\frac{1}{2}v_{F}\cos \theta \frac{\partial f\left( x,\theta
,\omega \right) }{\partial x}  \nonumber \\
=\left( \Delta (x)+\frac{1}{2\tau }F(x,\omega )\right) g\left( x,\theta
,\omega \right) ,  \label{Eilenberger} \\
G(x,\omega )=\int \frac{d\Omega }{4\pi }g\left( x,\theta ,\omega \right)
,\;F(x,\omega )=\int \frac{d\Omega }{4\pi }f\left( x,\theta ,\omega \right) ,
\nonumber \\
f\left( x,\theta ,\omega \right) f^{+}\left( x,\theta ,\omega \right)
+g^{2}\left( x,\theta ,\omega \right) =1,  \nonumber
\end{gather}
where the function $f^{+}(x,{\bf n,}\omega )$ satisfies the same equation as 
$f(x,-{\bf n,}\omega )$ with $\Delta \rightarrow \Delta ^{\ast }$ and the
presence of impurities is descibed by the elastic scattering time $\tau
=l/v_{f}.$ The functions $G(x,\omega )$ and $F(x,\omega )$ are the Green's
functions averaged over the Fermi surface. The Eilenberger equations are
completed by the self-consistency equation for the pair potential $\Delta
(x) $ in a superconducting layer : 
\begin{equation}
\Delta (x)=\pi T\lambda \sum_{\omega }F(x,\omega ).
\label{Auto-cohérence générale}
\end{equation}
The BCS coupling constant $\lambda $ is spatially independent in a
superconducting layer, while in a ferromagnetic layer it is equal to zero.
In a superconducting layer, the self-consistency equation may also be
written in the following convenient form 
\begin{equation}
\Delta (x)\ln \frac{T}{T_{c}}+\pi T\sum_{\omega }\left( \frac{\Delta (x)}{%
\left| \omega \right| }-F(x,\omega )\right) =0,
\end{equation}
where $T_{c0}$ is the bare transition temperature of the superconducting
layer in the absence of proximity effect.

Note that the presented form of the Eilenberger equations implies the
natural choice of the spin quantization axis along the direction of the
exchange field, and the only difference with the standard form of these
equations is the substitution of the Matsubara frequency $\omega $ by $%
\omega +ih(x)$.

Usually, the electron scattering mean free path in S/F/S systems is rather
small. In such a dirty limit, the angular dependence of the Green's
functions is weak, and the Eilenberger equations can be replaced by the much
simpler Usadel (1970) equations. In fact, the conditions of the
applicability of the Usadel equations are $T_{c}\tau \ll 1$ and $h\tau \ll
1. $ The second condition is much more restrictive due to a large value of
the exchange field ($h\gg T_{c}$). The Usadel equations only deal with the
Green's functions $G(x,\omega )$ and $F(x,\omega )$ averaged over the Fermi
surface $:$%
\begin{eqnarray}
&&\left. -\frac{D}{2}\left[ G(x,\omega ,h)\frac{\partial ^{2}}{\partial x^{2}%
}F\left( x,\omega ,h\right) -F(x,\omega ,h)\frac{\partial ^{2}}{\partial
x^{2}}G\left( x,\omega ,h\right) \right] \right.  \nonumber \\
&&\left. +\left( \omega +ih(x)\right) F(x,\omega ,h)=\Delta (x)G\left(
x,\omega ,h\right) ,\right.  \label{Usadel gen}
\end{eqnarray}
\[
G^{2}\left( x,\omega ,h\right) +F(x,\omega ,h)F^{+}(x,h,\omega )=1, 
\]
$D=\frac{1}{3}v_{F}l$ is the diffusion coefficient which is different in S
and F regions and the equation for the function $F^{+}(x,h,\omega )$ is the
same as for $F(x,\omega ,h)$ with the substitution $\Delta \rightarrow
\Delta ^{\ast }$. Here also the only difference with the standard form of
the Usadel equations is the substitution $\omega $ by $\omega +ih(x).$

The equations for the Green's functions in F and S regions must be completed
by the corresponding boundaries conditions at the interfaces. For the
Eilenberger equations they were derived by Zaitsev (1984) and for the Usadel
equations by Kupriyanov and Lukichev (1988). These boundary conditions take
into account the finite transparency (resistance) of the interfaces - see
Eq. (\ref{boundary conds}).

The most important pair-breaking mechanism in the ferromagnet is the
exchange field $h$. However a disorder in the lattice of magnetic atoms
creates centers of magnetic scattering. In ferromagnetic alloys, used as the
F layer in S/F/S Josephson junctions, the role of the magnetic scattering
may be quite important. Note that even in the case of a perfect ordering of
the magnetic atoms, the spin-waves will generate magnetic scattering. The
natural choice of the spin-quantization axis used implicitly above is along
the direction of the exchange field. The magnetic scattering and spin-orbit
scattering mix up the up and down spin states. Therefore to describe this
situation it is needed to introduce two normal Green's functions $G_{1}\sim
\left\langle \psi _{\uparrow }\psi _{\uparrow }^{+}\right\rangle $, $%
G_{2}\sim \left\langle \psi _{\downarrow }\psi _{\downarrow
}^{+}\right\rangle $ and two anomalous ones $F_{1}\sim \left\langle \psi
_{\uparrow }\psi _{\downarrow }\right\rangle $, $F_{2}\sim \left\langle \psi
_{\uparrow }\psi _{\uparrow }\right\rangle .$ The microscopical Green's
function theory of superconductors with magnetic impurities and spin-orbit
scattering was proposed by Abrikosov and Gorkov (1960, 1962). The
generalization of the Usadel equations (\ref{Usadel gen}) to this case gives 
\begin{eqnarray*}
&&\left. -\frac{D}{2}\left[ G_{1}\frac{\partial ^{2}}{\partial x^{2}}%
F_{1}-F_{1}\frac{\partial ^{2}}{\partial x^{2}}G_{1}\right] +\left( \omega
+ih+\left( \frac{1}{\tau _{z}}+\frac{2}{\tau _{x}}\right) G_{1}\right)
F_{1}+\right. \\
&&\left. G_{1}\left( F_{2}-F_{1}\right) \left( \frac{1}{\tau _{x}}-\frac{1}{%
\tau _{so}}\right) +F_{1}\left( G_{2}-G_{1}\right) \left( \frac{1}{\tau _{x}}%
+\frac{1}{\tau _{so}}\right) =\Delta (x)G_{1},\right.
\end{eqnarray*}
\[
G_{1}^{2}\left( x,\omega ,h\right) +F_{1}(x,\omega ,h)F_{1}^{+}(x,h,\omega
)=1, 
\]
and the similar equation for $F_{2}$ with the indices substitution $%
1\leftrightarrow 2$. Here $\tau _{so}^{-1}$ is the spin-orbit scattering
rate, while the magnetic scattering rates are $\tau _{z}^{-1}=\tau
_{2}^{-1}\left\langle S_{z}^{2}\right\rangle /S^{2}$ and $\tau
_{x}^{-1}=\tau _{2}^{-1}\left\langle S_{x}^{2}\right\rangle /S^{2}$. The
rate $\tau _{2}^{-1}$ describes the intensity of the magnetic scattering via
exchange interaction and we follow the notation of the paper of Fulde and
Maki (1966). In the spatially uniform case the equations (\ref{Usadel
magscatt}) are equivalent to those of the Abrikosov-Gorkov theory (1960,
1962) (see also Fulde and Maki, 1966). Demler {\it et al.} (1997) analyzed
the influence of the spin-orbit scattering on the critical temperature of
the S/F multilayers. The equations (Demler {\it et al.}, 1997) corresponds
to the limit $\Delta \rightarrow 0$, $G_{1,2}=1$ in (\ref{Usadel magscatt}).

The ferromagnets used as F layers in S/F heterostructures reveal strong
uniaxial anisotropy. Then the magnetic scattering in the plane ($xy$)
perpendicular to the anisotropy axis is negligeable. Moreover due to the
relatively small atomic numbers of the F layers atoms the spin-orbit
scattering is expected to be weak. In such case there is no spin mixing
scattering anymore and the Usadel equations retrieve the initial form (\ref
{Usadel gen}) with the substitution of the Matsubara frequencies by $\omega
\rightarrow \omega +G/\tau _{s}$, where $\tau _{s}^{-1}=\tau _{z}^{-1}=\tau
_{2}^{-1}\left\langle S_{z}^{2}\right\rangle /S^{2}$ may be considered as a
phenomenological parameter describing the intensity of the magnetic
scattering (Buzdin, 1985).

\bigskip The linearized Usadel equation in the ferromagnet reads 
\begin{equation}
\left( \left| \omega \right| +ihsgn\left( \omega \right) +\frac{1}{\tau _{s}}%
\right) F_{f}-\frac{D_{f}}{2}\frac{\partial ^{2}F_{f}}{\partial x^{2}}=0.
\label{Us + magn scatt}
\end{equation}
If $\tau _{s}T_{c}<<1,$ we may neglect $\left| \omega \right| $ in Eq. (\ref
{Us + magn scatt}) and the exponentially decaying solution has the form 
\begin{equation}
F_{f}\left( x,\omega >0\right) =A\exp \left( -x(k_{1}+ik_{2})\right) ,
\end{equation}
with $k_{1}=\frac{1}{\xi _{f}}\sqrt{\sqrt{1+\alpha ^{2}}+\alpha }$ and $%
k_{2}=\frac{1}{\xi _{f}}\sqrt{\sqrt{1+\alpha ^{2}}-\alpha }$,where $\alpha
=1/(\tau _{s}h)$. In the absence of magnetic scattering, the decaying and
oscillating wave vectors are the same $k_{1}=k_{2}$. The magnetic scattering
decreases the characteristic decaying length and increases the period of
oscillations. In practice, it means that the decrease of the critical
current of S/F/S junction with the increase of $d_{f\text{ }}$ will be more
strong. Note that the spin-orbit scattering (in contrast to the magnetic
scattering) decreases the pair-breaking effect of the exchange field (Demler 
{\it et al.}, 1997) and both scattering mechanisms decrease the amplitude of
the oscillations of the Cooper pair wave function. In some sense the
spin-orbit scattering is more harmful for these oscillations because they
completely disappear at $\tau _{so}^{-1}>h$. The observation on experiment
of the oscillatory behavior of $T_{c}$ in S/F multilayers is an indirect
proof of the weakness of the spin-orbit scattering.

The expression for $I_{c}(2d_{f})$ dependence (\ref{Ic(2df) low temp}) may
be generalized to take into account the magnetic scattering

\begin{equation}
I_{c}R_{n}=64\frac{\pi T}{e}%
\mathop{\rm Re}%
\left[ \stackrel{\infty }{%
\mathrel{\mathop{\sum }\limits_{\omega >0}}%
}\frac{2q_{\omega }y\exp (-2q_{\omega }y)\Phi _{\omega }}{\left[ \sqrt{%
(1-\eta _{\omega }^{2})\Phi _{\omega }+1}+1\right] ^{2}}\right] ,
\label{Ic general}
\end{equation}
where the functions 
\begin{equation}
\Phi _{\omega }=\frac{\Delta ^{2}}{\left( \Omega +\omega \right) ^{2}},\text{
}q_{\omega }=\sqrt{2i+2\alpha +2\omega /h},\text{ }\eta _{\omega }^{2}=\frac{%
\alpha }{\alpha +i+\omega /h}.
\end{equation}

Near $T_{c}$ and in the limit $h>>T_{c}$ and $2d_{f}k_{2}>>1$ it possible to
obtain the following simple analytical expression for the critical current

\begin{equation}
I_{c}=\frac{\pi S\sigma _{f}\Delta ^{2}k_{1}}{2eT_{c}}\left[ \cos \left(
2d_{f}k_{2}\right) +\frac{k_{2}}{k_{1}}\sin \left( 2d_{f}k_{2}\right) \right]
\exp (-2d_{f}k_{1}).  \label{Ic near Tc with spin flip}
\end{equation}
We see that due to the magnetic scattering the decaying length of the
critical current $\xi _{f1}=1/k_{1}$ may be substantially smaller than the
oscillating length $\xi _{f2}=1/k_{2}$.

As it has been noted above, the condition of the applicability of the Usadel
equations, $h\tau \ll 1,$ is rather restrictive in ferromagnets due to the
large value of the exchange field. Therefore, it is of interest to retain in
the Usadel equations the first correction in the parameter $h\tau $. The
first attempts to calculate this correction were made by Tagirov (1998) and
Proshin and Khusainov (1998) and resulted in the renormalization of the
diffusion constant of the F layer $D_{f}\rightarrow D_{f}(1-2ih\tau
sign\left( \omega \right) ).$ Later on, the similar renormalization has been
proposed by Bergeret {\it et al.} (2001c) and Baladi\'{e} and Buzdin (2001).
The critical analysis of this renormalization by Fominov {\it et al.} (2002)
(see also Fominov {\it et al.} 2003b and Khusainov and Proshin, 2003)
revealed the inaccuracy of this renormalization, but did not provided the
answer. The careful derivation of the Usadel equation for an F layer
retaining the linear correction over\ the parameter $h\tau $ was made by
Buzdin and Baladi\'{e} (2003) and simply resulted in a somewhat different
renormalization of the diffusion constant $D_{f}\rightarrow
D_{f}(1-0.4ih\tau sign\left( \omega \right) ).$ The coefficient in the
parameter $h\tau $ occurs to be rather small which provides more confidence
in the description of F layers in the framework of the Usadel equations.
Note that this renormalization of the diffusion constant\ increases the
decaying characteristic length and decreases the period of oscillations,
which is opposite to the influence of the magnetic scattering.

The Usadel equations give the description of Green's functions only on
average. Zyuzin {\it et al.} (2003) pointed out that due to the mesoscopic
fluctuations, the decay of the anomalous Green's function $F_{f}$ at
distances much larger than $\xi _{f}$ is not exponential. In result, the
Josephson effect in S/F/S systems may be observed even with a thick
ferromagnetic layer.

The Eilenberger and Usadel equations adequately describe the weak
ferromagnets, where $h<<E_{F}$ and the spin-up $v_{F\uparrow ,}$and
spin-down $v_{F\downarrow }$ Fermi velocities are the same. When the
parameters of the electrons spectra of the spin-up and spin-down bands are
very different, the quasiclassical approach fails. However, if the
characteristics of the spin bands are similar, the Eilenberger and Usadel
equations are still applicable. Performing the derivation of the Eilenberger
equation in such case, it may be demonstrated that the Fermi velocity $v_{F}$
in Eq. (\ref{Eilenberger}) must be substituted by $\left( v_{F\uparrow
}+v_{F\downarrow }\right) /2$ and the scattering rate $1/\tau $ by $\left(
1/\tau _{\uparrow }+1/\tau _{\downarrow }\right) /2$. In consequence, the
diffusion coefficient $D_{f}$ in the Usadel equation becomes $\left(
1/6\right) \left( v_{F\uparrow }+v_{F\downarrow }\right) ^{2}/\left( 1/\tau
_{\uparrow }+1/\tau _{\downarrow }\right) .$ Let us stress that such
renormalization is justified only for close values of $v_{F\uparrow ,}$and $%
v_{F\downarrow }$ (as well as $\tau _{\uparrow }$ and $\tau _{\downarrow }$%
). Otherwise the Bogoliubov-de Gennes equations must be used for the
description of the proximity effect in the strong ferromagnets.

\section{References}

Aarts, J., J. M. E. Geers, E. Br\"{u}ck, A. A. Golubov, and R. Coehoorn,
1997, ''Interface transparency of superconductor/ferromagnetic
multilayers,'' Phys. Rev. B {\bf 56}, 2779-2787.

Abrikosov, A. A., and L. P. Gor'kov, ''Contribution to the theory of
superconducting alloys with paramagnetic impurities,'' 1960, Zh. Eksp.
Theor. Fiz. {\bf 39}, 1781-1796 [Sov. Phys. JETP {\bf 12}, 1243-1253 (1961)].

Abrikosov, A. A., and L. P. Gor'kov, ''Spin-orbit interaction and the Knight
shift in superconductors,'' 1962, Zh. Eksp. Theor. Fiz. {\bf 42}, 1088-1096
[Sov. Phys. JETP {\bf 15}, 752-757 (1962)].

Abrikosov, A. A., L. P. Gor'kov, and I. E. Dzyaloshinski, 1975,{\it \
Methods of Quantum Field Theory in Statistical Physics} (Dover, New York).

Aladyshkin, A. Yu., A. I. Buzdin, A. A. Fraerman, A. S. Mel'nikov, D. A.
Ryzhov, and A. V. Sokolov, 2003, ''Domain-wall superconductivity in hybrid
superconductor-ferromagnet structures,'' Phys. Rev. B {\bf 68}, 184508.

Aumentado, J., and V. Chandrasekhar, 2001, ''Mesoscopic
ferromagnet-superconductor junctions and the proximity effect,'' Phys. Rev.
B {\bf 64}, 054505.

Anderson, P. W., and H. Suhl, 1959, ''Spin alignment in the superconducting
state,'' Phys. Rev. {\bf 116}, 898-900.

Andersson, M., J. C. Cuevas, and M. Fogelst\"{o}m, 2002, ''Transport through
superconductor/magnetic dot/superconductor structures,'' Physica C {\bf 367}%
, 117-122.

Andreev, A. F., 1964 ''The thermal conductivity of the intermediate state in
superconductors,'' Zh. Eksp. Theor. Fiz. {\bf 46}, 1823-1825 [Sov. Phys.
JETP {\bf 19}, 1228-1231 (1964)].

Andreev, A. V., A. I. Buzdin, and R. M. Osgood III, 1991, ''$\pi $ phase in
magnetic-layered superconductors,'' Phys. Rev. B {\bf 43}, 10124-10131.

Aoki, D., A. Huxley, E. Ressouche, D. Braithwaite, J. Flouquet, J.-P.
Brison, E. Lhotel, and C. Paulsen, 2001, ''Coexistence of superconductivity
and ferromagnetism in URhGe,'' Nature {\bf 413}, 613-616.

Apinyan, V., and R. R. M\'{e}lin, 2002, ''Microscopic theory of non local
pair correlations in metallic F/S/F trilayers,'' Eur. Phys. J. B, {\bf 25},
373-389.

Aslamazov, L. G., 1968, ''Influence of impurities on the existence of an
inhomogeneous state in a ferromagnetic superconductor,'' Zh. Eksp. Teor. Fiz.%
{\it \ {\bf 55}, }1477-1482 [Sov. Phys. JETP\ {\bf 28}, 773-775 (1969)].

Bagrets, A., C. Lacroix, and A.\ Vedyayev, 2003, ''Theory of proximity
effect in superconductor/ferromagnet heterostructures,'' Phys. Rev. B {\bf 68%
}, 054532.

Baladi\'{e}, I., A. Buzdin, N. Ryzhanova, and A. Vedyayev, 2001, ''Interplay
of superconductivity and magnetism in superconductor/ferromagnet
structures,'' Phys. Rev. B {\bf 63}, 54518.

Baladi\'{e}, I., and\ A. Buzdin, 2001, ''Local quasiparticle density of
states in ferromagnet/superconductor nanostructures,'' Phys. Rev. B {\bf 64}%
, 224514.

Baladi\'{e}, I., and\ A. Buzdin, 2003, ''Thermodynamic properties of
ferromagnet/superconductor/ferromagnet nanostructures,'' Phys. Rev. B {\bf 67%
}, 014523.

Balicas, L., J. S.\ Brooks, K. Storr, S. Uji, M. Tokumoto, H. Tanaka, H.
Kobayashi, A. Kobayashi, V. Barzykin, and L.\ P. Gorkov, 2001,
''Superconductivity in an organic insulator at very high magnetic fields,''
Phys. Rev. Lett. {\bf 87}, 067002.

Barash, Yu. S., I. V. Bobkova, and T. Kopp, 2002,\ ''Josephson current in
S-FIF-S junctions: nonmonotonic dependence on misorientation angle,'' Phys.
Rev. B {\bf 66}, 140503(R).

Barone, A., and G. Paterno, 1982, {\it Physics and Applications of the
Josephson Effect} (Wiley, New York).

Bauer, A., J. Bentner, M. Aprili, and M. L. Della Rocca, 2004, ''Spontaneous
supercurrent induced by ferromagnetic $\pi $ junctions,'' Phys. Rev. Lett. 
{\bf 92}, 217001{\it .}

Beenakker, C. W. J., 1997, ''Random-matrix theory of quantum transport,''
Rev. Mod. Phys. {\bf 69}, 731-808.

Bergeret, F. S., K. B. Efetov, and A. I. Larkin, 2000, ''Nonhomogeneous
magnetic order in superconductor-ferromagnet multilayers,'' Phys. Rev. B 
{\bf 62}, 11872-11878.

Bergeret, F., A. F. Volkov, and K. B. Efetov, 2001a, ''Enhancement of the
Josephson current by an exchange field in superconductor-ferromagnet
structures,'' Phys. Rev. Lett. {\bf 86}, 3140-3143.

Bergeret, F., A. F. Volkov, and K. B. Efetov, 2001b, ''Long-range proximity
effects in superconductor-ferromagnet structures,'' Phys. Rev. Lett. {\bf 86}%
, 4096-4099.

Bergeret, F., A. F. Volkov, and K. B. Efetov, 2001c, ''Josephson current in
superconductor-ferromagnet structures with nonhomogeneous magnetization,''
Phys. Rev. B {\bf 64}, 134506.

Bergeret, F. S., A. F. Volkov, and K. B. Efetov, 2002, ''Local density of
states in superconductor-strong ferromagnet structures,'' Phys. Rev. B {\bf %
65}, 134505.

Bergeret, F. S., A. F. Volkov, and K. B. Efetov, 2003, ''Manifestation of
the triplet superconductivity in superconductor-ferromagnet structures,''
Phys. Rev. B {\bf 68}, 064513.

Bergeret, F. S., A. F. Volkov, and K. B. Efetov, 2004a, ''Induced
ferromagnetism due to superconductivity in superconductor-ferromagnet
structures,'' Phys. Rev. B {\bf 69}, 174504.

Bergeret, F. S., A. F. Volkov, and K. B. Efetov, 2004b, ''Spin screening of
magnetic moments in superconductors,'' Europhys. Lett. {\bf 66}, 111-117.

Blanter, Ya. M., and F.\ W.\ J.\ Hekking, 2004, ''Supercurrent in long SFFS
junctions with antiparallel domain configuration,'' Phys. Rev. B {\bf 69},
024525.

Blonder, G. E. , M. Tinkham, and T. M. Klapwijk, 1982, ''Transition from
metallic to tunneling regimes in superconducting microconstrictions: Excess
current, charge imbalance, and supercurrent conversion,'' Phys. Rev. B {\bf %
25}, 4515-4532.

Blum, Y., A. Tsukernik, M. Karpovski, and A. Palevski, 2002, ''Oscillations
of the superconducting critical current in Nb-Cu-Ni-Cu-Nb junctions,'' Phys.
Rev. Lett. {\bf 89}, 187004.

Bourgeois, O, and R. C. Dynes, 2002, ''Strong coupled superconductor in
proximity with a quench-condensed ferromagnetic Ni film: a search for
oscillating T$_{c}$,'' Phys. Rev. B {\bf 65}, 144503.

Bozovic, M., and Z. Radovic, 2002, ''Coherent effects in double-barrier
ferromagnet/superconductor/ferromagnet junctions,'' Phys. Rev. B {\bf 66},
134524.

Bozovic, I., G. Logvenov, M. A. Verhoeven, P. Caputo, E. Goldobin, and M. R.
Beasley, 2004, ''Giant proximity effect in cuprate superconductors,'' Phys.
Rev. Lett. {\bf 93}, 157002.

Bulaevskii, L. N., 1973, ''Magnetic properties of layered superconductors
with weak interaction between the layers,'' Zh. Eksp. Teor. Fiz.{\it \ {\bf %
64}, }2241-2247 [Sov. Phys. JETP\ {\bf 37}, {\it 1133-1136} (1973)]{\it .}

Bulaevskii, L. N., V. V. Kuzii, and A. A. Sobyanin, 1977, ''Superconducting
system with weak coupling with a current in the ground state,'' Pis'ma Zh.
Eksp. Teor. Phys. {\bf 25}, 314-318 [JETP Lett., {\bf 25}, 290-294 (1977)].

Bulaevskii, L. N., V. V. Kuzii, and A. A. Sobyanin, 1978, ''On possibility
of the spontaneous magnetic flux in a Josephson junction containing magnetic
impurities,'' Solid St. Comm., {\bf 25}, 1053-1057.

Bulaevskii, L. N. , A. I. Rusinov, and M. L. Kulic, 1980, ''Helical ordering
of spins in a supercoductor,'' J. Low Temp. Phys. {\bf 39}, 255-272.

Bulaevskii, L. N., A. I. Buzdin, M. L. Kuli\'{c}, and S. V. Panjukov, 1985,
''Coexistence of superconductivity and magnetism. Theoretical predictions
and experimental results,'' Advances in Physics, {\bf 34}, 175-261.

Bulaevskii, L. N., and E. M. Chudnovsky, 2000, ''Ferromagnetic film on a
superconducting substrate,'' Phys. Rev. B {\bf 63}, 012502.

Bulaevskii, L. N., E. M. Chudnovsky, and M.\ P.\ Maley, 2000, ''Magnetic
pinning in superconductor-ferromagnet multilayers,'' Appl. Phys. Lett. {\bf %
76}, 2594-2596.

Bulaevskii, L. N., E. M. Chudnovsky, and M. Daumens, ''Reply to ''Comment on
`Ferromagnetic film on a superconducting substrate', '' 2002, Phys. Rev. B 
{\bf 66}, 136502.

Buzdin, A. I., 1985, ''Surface superconductivity in ferromagnets,'' Pis'ma
Zh. Eksp. Teor. Phys. {\bf 42}, 283-285\ [JETP Lett. {\bf 42, }350-352
(1985)].

Buzdin, A., 2000, ''Density of states oscillations in a ferromagnetic metal
in contact with superconductor,'' Phys. Rev. B {\bf 62}, 11377-11379.

Buzdin, A., 2003, ''$\pi -$ junction realization due to tunneling through a
thin ferromagnetic layer,'' Pis'ma Zh. Eksp. Teor. Phys. {\bf 78, }1073-1076
[JETP Lett. {\bf 78, }583-586 (2003)].

Buzdin, A. I., L. N. Bulaevskii, and S. V. Panyukov, 1982,
''Critical-current oscillations as a function of the exchange field and
thickness of the ferromagnetic metal (F) in a S-F-S Josephson junction,''
Pis'ma Zh. Eksp. Teor. Phys. {\bf 35}, 147-148 [JETP Lett. {\bf 35}, 178-180
(1982)].

Buzdin, A. I., L. N. Bulaevskii, and S. V. Panyukov, 1984, ''Existence of
superconducting domain walls in ferromagnets,'' Zh. Eksp. Teor. Phys. {\bf 87%
}, 299-309 [Sov. Phys. JETP {\bf 60}, 174-179 (1984)].

Buzdin, A. I., and Polonskii S. V., 1987, ''Nonuniform state in quasi-1D
superconductors,'' Zh. Eksp. Teor. Phys. {\bf 93}, 747-761 [Sov. Phys. JETP%
{\it \ {\bf 66}, }422-429 (1987)].

Buzdin, A. I., and L. N. Bulaevskii, 1988, ''Ferromagnetic film on the
surface of a superconductor: possible onset of inhomogeneous magnetic
ordering,'' Zh. Eksp. Teor. Phys. {\bf 94}, 256-261 [Sov. Phys. JETP {\bf 67}%
, 576-578 (1988)].

Buzdin, A. I., and M. Y. Kuprianov, 1990, ''Transition temperature of a
superconductor-ferromagnet superlattice,'' Pis'ma Zh. Eksp. Teor. Phys. {\bf %
52}, 1089-1091\ [JETP Lett. {\bf 52, }487-491 (1990)].

Buzdin, A. I., and M. Y. Kuprianov, 1991, ''Josephson junction with a
ferromagnetic layer,'' Pis'ma Zh. Eksp. Teor. Phys. {\bf 53, }308-312 [JETP
Lett. {\bf 53, }321-326 (1991)].

Buzdin, A. I., B. Vujicic', and M. Y. Kuprianov, 1992,
''Superconductor-ferromagnet structures,'' Zh. Eksp. Teor. Phys. {\bf 101},
231-240 [Sov. Phys. JETP {\bf 74}, 124-128 (1992)].

Buzdin, A. I. and H. Kachkachi, 1997, ''Generalized Ginzburg-Landau theory
for nonuniform FFLO superconductors,'' Physics Letters, {\bf A 225}, 341-348.

Buzdin, A. I., A. V. Vedyayev, and N. V. Ryzhanova, 1999,
''Spin-orientation-dependent superconductivity in S/F/S structures,''
Europhys. Lett. {\bf 48}, 686-691.

Buzdin, A. and I. Baladie, 2003, ''Theoretical description of ferromagnetic $%
\pi -$junctions near the critical temperature,'' Phys. Rev. B {\bf 67},
184519.

Buzdin, A. I., and M. Daumens, 2003, ''Inversion of the proximity effect in
hybrid ferromagnet-superconductor-ferromagnet structures,'' Europhys. Lett. 
{\bf 64}, 510-516 .

Buzdin, A. I., and A.\ S. Melnikov, 2003, ''Domain wall superconductivity in
ferromagnetic superconductors,'' Phys. Rev. B {\bf 67}, R020503.

Buzdin, A. I., and A. Koshelev, 2003, ''Periodic alternating 0- and $\pi $%
-junction structures as realization of $\varphi $-Josephson junctions,''
Phys. Rev. B {\bf 67}, R220504.

Casalbuoni, R., and G. Nardulli, 2004, ''Inhomogeneous superconductivity in
condensed matter and QCD,'' Rev. Mod. Phys. {\bf 76}, 263-321.

Cayssol, J., and G. Montambaux, 2004, ''Incomplete Andreev reflection in a
clean Superconductor/Ferromagnet/Superconductor junction'', Cond-mat/0404215.

Chandrasekhar, B. S., 1962, ''Maximum critical field of high-field
superconductors,'' Appl. Phys. Lett., {\bf 1}, 7-8.

Cheng, S.-L., and H.\ A. Fertig, 1999, ''Upper critical field H$_{c3}$ for a
thin-film superconductor with a ferromagnetic dot,'' Phys. Rev. B {\bf 60},
13107-13111.

Chien, C. L., and Reich D. H., 1999, ''Proximity effects in
superconducting/magnetic multilayers,'' J. Magn. Magn. Mater. {\bf 200},
83-94.

Chtchelkatchev, N. M., W Belzig, Yu. V. Nazarov and C. Bruder, 2001, ''$\pi
-0$ transition in superconductor-ferromagnet-superconductor junctions,''
Pis'ma Zh. Eksp. Teor. Phys. {\bf 74, }357-361 [JETP Lett. {\bf 74, }323-327
(2001)].

Chtchelkatchev, N. M., W Belzig, and C. Bruder, 2002, ''Josephson effect in S%
$_{F}$XS$_{F}$ junctions,'' Pis'ma Zh. Eksp. Teor. Phys. {\bf 75, }772-776
[JETP Lett. {\bf 75, }646-650 (2002)].

Clogston, A. M., 1962, ''Upper limit for the critical field in hard
superconductors,'' Phys. Rev. Lett. {\bf 9}, 266-267.

de Gennes, P. G., 1966a, {\it Superconductivity of Metals and Alloys (}New
York : Benjamin).

de Gennes, P. G., 1966b, ''Coupling between ferromagnets through a
superconducting layer,''Phys. Lett. {\bf 23}, 10-11.

de Jong, M.\ J.\ M., and C.\ W.\ J. Beenakker, 1995, ''Andreev reflection in
ferromagnet-superconductor junctions,'' Phys. Rev. Lett. {\bf 74}, 1657-1660.

Daumens, M., and Y. Ezzahri, 2003, ''Equilibrium domain structure in a
ferromagnetic film coated by a superconducting film,'' Phys. Lett. A {\bf 306%
}, 344-347.

Demler, E. A., G.\ B. Arnold, and M. R. Beasley, 1997, ''Superconducting
proximity effects in magnetic metals,'' Phys. Rev. B {\bf 55}, 15174-15182.

Deutscher, G. and P. G. De Gennes, 1969, ''Proximity effects,'' in {\it %
Superconductivity}, edited by R. D. Parks (Marcel Dekker, New York),
p.1005-1034.

Deutscher, G., and F. Meunier, 1969, ''Coupling between ferromagnetic layers
through a superconductor,'' Phys. Rev. Lett. {\bf 22}, 395-396.

Deutscher, G., and D. Feinberg, 2000, ''Coupling
superconducting-ferromagnetic point contacts by Andreev reflections,'' Appl.
Phys. Lett. {\bf 76}, 487-489.

Deutscher, G., 2004, ''Andreev-Saint James reflections: a probe of cuprate
superconductors,'' Rev. Mod. Phys.

Dobrosavljevic-Grujic, L., R. Zikic, and Z. Radovic, 2000, ''Quasiparticle
energy spectrum in ferromagnetic Josephson weak links,'' Physica C, {\bf 331}%
, 254-262.

Dubonos, S. V., A. K. Geim, K. S. Novoselov, and I. V. Grigorieva, 2002,
''Spontaneous magnetization changes and nonlocal effects in mesoscopic
ferromagnet-superconductor structures,'' Phys. Rev. B {\bf 65}, 220513(R).

Eilenberger G., 1968, ''Transformation of Gorkov's equation for type II
superconductors into transport-like equations,'' Z. Phys. {\bf 214}, 195-213.

Erdin, S., I.\ F. Lyuksyutov, V.\ L. Pokrovsky, and V.\ M.\ Vinokur, 2002,
''Topological textures in a ferromagnet-superconductor bilayer,'' Phys. Rev.
Lett. {\bf 88}, 017001.

Fazio, R., and C. Lucheroni, 1999, ''Local density of states in
superconductor-ferromagnetic hybrid systems,'' Europhysics Lett. {\bf 45},
707-713.

Fogelst\"{o}m, M., 2000, ''Josephson currents through spin-active
interfaces,'' Phys. Rev. B {\bf 62}, 11812-11819.

Fominov, Ya. V., N. M. Chtchelkatchev, and A. A. Golubov, 2002,
''Nonmonotonic critical temperature in superconductor/ferromagnet
bilayers,'' Phys. Rev. B {\bf 66,} 14507.

Fominov, Ya., A. A. Golubov, and M. Yu. Kupriyanov, 2003a, ''Triplet
proximity effect in FSF trilayers,'' Pis'ma Zh. Eksp. Teor. Phys. {\bf 77, }%
609 [JETP Lett. {\bf 77}, 510-515 (2003)].

Fominov, Ya. V., M. Yu. Kupriyanov, and M. V. Feigelman, 2003b, ''A comment
on the paper ''Competition of superconductivity and magnetism in
ferromagnet-superconductor heterostructures'' by Yu A Izyumov, Yu N Proshin,
and M G Khusainov,'' Uspekhi Fiz. Nauk {\bf 173}, 113-115 [Sov. Phys. Usp., 
{\bf 46}, 105-107 (2003)].

Frolov, S. M., D. J. Van Harlingen, V. A. Oboznov, V. V. Bolginov, and V. V.
Ryazanov, 2004, ''Measurement of the current-phase relation of
superconductor/ferromagnet/superconductor Josephson junctions,'' Phys. Rev.
B {\bf 70}, 144505.

Fulde, P., and R. A. Ferrell, 1964, ''Superconductivity in a strong
spin-exchange field,'' Phys. Rev. {\it {\bf 135},} A550-A563{\it .}

Fulde, P., and K. Maki, 1966, ''Theory of superconductors containing
magnetic impurities,'' Phys. Rev. {\it {\bf 141},} 275-280{\it .}

Garifullin, I. A. , D. A. Tikhonov, N. N. Garif'yanov, L. Lazar, Yu. V.
Goryunov, S. Ya. Khlebnikov, L. R. Tagirov, K. Westerholt, and H. Zabel,
2002, ''Re-entrant superconductivity in the superconductor/ferromagnet V/Fe
layered system,'' Phys. Rev. B {\bf 66}, R020505.

Garc\'{i}a-Santiago, A., F. S\'{a}nchez , M. Varela, and J. Tejada, 2000,
''Enhanced pinning in a magnetic-superconducting bilayer,'' Appl. Phys.
Lett. {\bf 77}, 2900-20902.

Garifullin, I. A., 2002, ''Proximity effects in ferromagnet/superconductor
heterostructures,'' J. Magn. Magn. Mater. {\bf 240}, 571-576.

Ginzburg, V. L., 1956, ''Ferromagnetic superconductors,'' Zh. Eksp. Teor.
Fiz. {\bf 31}, 202-214 [Sov. Phys. JETP {\bf 4}, 153-161 (1957)].

Giroud, M., H. Courtois, K. Hasselbach, D. Mailly, and B. Pannetier, 1998,
''Superconducting proximity effect in a mesoscopic ferromagnetic wire,''
Phys. Rev. B {\bf 58}, R11872-R11875.

Goldobin, E., D. Koelle, and R. Kleiner, 2002, ''Semifluxons in long
Josephson $0-\pi $ junctions,'' Phys. Rev. B {\bf 66}, 100508.

Golubov, A. A., M. Yu. Kupriyanov, and Ya. V. Fominov, 2002a, ''Critical
current in SFIFS junctions,'' Pis'ma Zh. Eksp. Teor. Phys. {\bf 75, }223-227
[JETP Lett. {\bf 75}, 190-194 (2002)].

Golubov, A. A., M. Yu. Kupriyanov, and Ya. V. Fominov, 2002b,
''Nonsinusoidal current-phase relation in SFS Josephson junction,'' Pis'ma
Zh. Eksp. Teor. Phys. {\bf 75, }709-713 [JETP Lett. {\bf 75}, 588-592
(2002)].

Golubov, A. A., M. Yu. Kupriyanov, and E. Il'ichev, 2004, ''The
current-phase relation in Josephson junctions,'' Rev. Mod. Phys. {\bf 76},
411-469.

Gu, J. Y., C.-Y. You, J. S. Jiang, J. Pearson, Ya. B. Bazaliy, and S. D.
Bader, 2002, ''Magnetization-orientation dependence of the superconducting
transition temperature in the ferromagnet-superconductor-ferromagnet system:
CuNi/Nb/CuNi,'' Phys. Rev. Lett. {\bf 89}, 267001.

Guichard, W., M. Aprili, O. Bourgeois, T. Kontos, J. Lesueur, and P. Gandit,
2003, ''Phase sensitive experiments in ferromagnetic-based Josephson
junctions,'' Phys. Rev. Lett. {\bf 90}, 167001.

Halterman, K., and O. T. Valls, 2001, ''Proximity effects at
ferromagnet-superconductor interfaces,'' Phys. Rev. B {\bf 65}, 014509.

Halterman, K., and O. T. Valls, 2002, ''Proximity effects and characteristic
lengths in ferromagnet-superconductor structures,'' Phys. Rev. B {\bf 66},
224516.

Halterman, K., and O. T. Valls, 2003, ''Energy gap of
ferromagnet-superconductor bilayers,'' Physica C {\bf 397}, 151-158.

Halterman, K., and O. T. Valls, 2004a, ''Layered ferromagnet-superconductor
structures: The $\pi $ state and proximity effects,'' Phys. Rev. B {\bf 69},
014517.

Halterman, K., and O. T. Valls, 2004b, ''Stability of $\pi -$junction
configurations in ferromagnet-superconductor heterostructures,'' Phys. Rev.
B {\bf 70}, 104516.

Hauser, J. J., 1969, ''Coupling between ferromagnetic insulators through a
superconducting layer,'' Phys. Rev. Lett. {\bf 23}, 374-377.

Helseth, L.\ E., P. E.\ Goa, H.\ Hauglin, M. Baziljevich, and T.\ H.\
Johansen, 2002, ''Interaction between a magnetic domain wall and a
superconductor,'' Phys. Rev. B {\bf 65}, 132514.

Holden, T., H.-U. Habermeier, G. Cristiani, A. Golnik, A. Boris, A. Pimenov,
J. Huml\'{i}ek, O. I. Lebedev, G. Van Tendeloo, B. Keimer, and C. Bernhard,
2004, ''Proximity induced metal-insulator transition in YBa$_{2}$Cu$_{3}$O$%
_{7}$/La$_{2/3}$Ca$_{1/3}$MnO$_{3}$ superlattices,'' Phys. Rev. B {\bf 69},
064505.

Houzet, M., Y. Meurdesoif, O. Coste and, A. Buzdin, 1999, ''Structure of the
non-uniform Fulde-Ferrell-Larkin-Ovchinnikov state in 3D superconductors,''
Physica C {\bf 316}, 89-96.

Houzet, M., and A. Buzdin, 2000, ''Influence of the paramagnetic effect on
the vortex lattice in 2D superconductors,'' Europhysics Lett. {\bf 50},
375-381.

Houzet, M., and A. Buzdin, 2001, ''Structure of the vortex lattice in the
Fulde-Ferrell-Larkin-Ovchinnikov state,'' Phys. Rev. B {\bf 63}, 184521.

Houzet, M., A. Buzdin, and M. Kuli\'{c}, 2001, ''Decoupling of
superconducting layers in the magnetic superconductor RuSr$_{2}$GdCu$_{2}$O$%
_{8}$,'' Phys. Rev. B {\bf 64}, 184501.

Houzet, M., and A. Buzdin, 2002, ''Nonuniform superconducting phases in a
layered ferromagnetic superconductor,'' Europhysics Lett. {\bf 58}, 596-602.

Houzet, M., A. I. Buzdin, L. N. Bulaevskii, and M. Maley, 2002, ''New
superconducting phases in field-induced organic superconductor lambda-(BETS)$%
_{2}$FeCl$_{4}$,'' Phys. Rev. Lett. {\bf 88}, 227001.

Ioffe, L. B., V. B. Geshkenbein, M. V. Feigel'man, A. L. Fauch\`{e}re, G.
Blatter, 1999, ''Environmetally decoupled sds-wave Josephson junctions for
quantum computing,'' Nature {\bf 398}, 679-681.

Izyumov, Yu. A., Yu. N. Proshin, and M. G. Khusainov, 2000, ''Multicritical
Behavior of the Phase Diagrams of Ferromagnet/Superconductor Layered
Structures,'' Pis'ma Zh. Eksp. Teor. Phys. {\bf 71, }202-209 [JETP Lett. 
{\bf 71}, 138-143 (2000)].

Izyumov, Yu. A., Yu. N. Proshin, and M. G. Khusainov, 2002, ''Competition
between superconductivity and magnetism in ferromagnet/superconductor
heterostructures,'' Uspekhi Fiz. Nauk {\bf 172}, 113-154 [Sov. Phys. Usp., 
{\bf 45}, 109-148 (2002)].

Jaccarino, V. and M. Peter, 1962, ''Ultra-high-field superconductivity,''
Phys. Rev. Lett. {\bf 9}, 290-292.

Jiang, J. S., D. Davidovi\'{c}, D. H. Reich, and C. L. Chien, 1995,
''Oscillatory superconducting transition temperature in Nb/Gd multilayers,''
Phys. Rev. Lett. {\bf 74}, 314-317.

Kadigrobov, A., R.\ I. Shekhter, and M. Jonson, 2001, ''Quantum spin
fluctuations as a source of long-range proximity effects in diffusive
ferromagnet-super conductor structures,'' Europhysics Lett. {\bf 54},
394-400.

Khlyustikov, I. N., and A. I. Buzdin, ''Twinning-plane superconductivity,''
1987, Adv. in Physics, {\bf 36}, 271-330.

Khusainov, M. G., and Yu. N. Proshin, 2003, ''Inhomogeneous superconducting
states in ferromagnetic metal/superconductor structures (Reply to the
comment by Ya V Fominov, M Yu Kupriyanov, and M V Feigel'man on the review
''Competition between superconductivity and magnetism in
ferromagnet/superconductor heterostructures'' by Yu A Izyumov, Yu N Proshin,
and M G Khusainov),'' Uspekhi Fiz. Nauk {\bf 173}, 1385-1386 [Sov. Phys.
Usp., {\bf 46}, 1311-1311 (2003)].

Kinsey, R. J., G. Burnell, and M. G. Blamire, 2001, ''Active supercurrent
control in superconductor/ferromagnet heterostructures,'' IEEE Trans. Appl.
Superc. {\bf 11}, 904-907.

Koelle, R. Kleiner, C. Bernhard, and C. T. Lin, 2004, ''Intrinsic Josephson
Effects in the Magnetic Superconductor RuSr$_{2}$GdCu$_{2}$O$_{8}$,'' Phys.
Rev. Lett. {\bf 92}, 117001.

Kontos, T., M. Aprili, J. Lesueur, and X. Grison, 2001, ''Inhomogeneous
superconductivity induced in a ferromagnet by proximity effect,'' Phys. Rev.
Lett. {\bf 86}, 304307.

Kontos, T., M. Aprili, J. Lesueur, F. Genet, B. Stephanidis, and R.
Boursier, 2002, ''Josephson junction through a thin ferromagnetic layer:
negative coupling,'' Phys. Rev. Lett. {\bf 89}, 137007.

Koorevaar, P., Y. Suzuki, R. Coehoorn, and J. Aarts J., 1994, ''Decoupling
of superconducting V by ultrathin Fe layers in V/Fe multilayers,'' Phys.
Rev. B, {\bf 49}, 441-449.

Koshina, E. A., and V. N. Krivoruchko, 2001, ''Spin polarization and $\pi $%
-phase state of the Josephson contact: Critical current of mesoscopic SFIFS
and SFIS junctions,'' Phys. Rev. B, {\bf 63}, 224515.

Krivoruchko, V. N., and E. Koshina, 2001, ''From inversion to enhancement of
the dc Josephson current in S/F-I-F/S tunnel structures,'' Phys. Rev. B, 
{\bf 64}, 172511.

Krivoruchko, V. N., and E. A. Koshina, 2002, ''Inhomogeneous magnetism
induced in a superconductor at a superconductor-ferromagnet interface,''
Phys. Rev. B, {\bf 66}, 014521.

Krivoruchko, V. N, and R. V. Petryuk, 2002, ''Spin-orbit scattering effect
on critical current in SFIFS tunnel structures,'' Phys. Rev. B, {\bf 66},
134520.

Krunavakarn, B., W. Sritrakool, and S. Yoksan{\it , }2004, ''Nonmonotonic
critical temperature in ferromagnet/superconductor/ferromagnet trilayers,''
Physica C, {\bf 406}, 46-52.

Kuli\'{c}, M. L., and M. Endres, 2000,
''Ferromagnetic-semiconductor-singlet-(or triplet)
superconductor-ferromagnetic-semiconductor systems as possible logic
circuits and switches,'' Phys. Rev. B {\bf 62}, 11846-11853.

Kuli\'{c}, M. L., and I. M. Kuli\'{c}, 2001, ''Possibility of a $\pi $
Josephson junction and swith in superconductors with spiral magnetic
order,'' Phys. Rev. B{\bf \ 63}, 104503.

Kulik, I. O., 1965,\ ''Magnitude of the critical Josephson tunnel current,''
Zh. Eksp. Teor. Fiz. {\bf 49}, 1211-1214 [Sov. Phys. JETP {\bf 22}, 841-843
(1966)].

Kuprianov, M. Y., and V. F. Lukichev, 1988, ''Influence of boundary
transparency on the critical current of ''durty'' SS'S structures,'' Zh.
Eksp. Teor. Fiz. {\bf 94}, 139-149 [Sov. Phys. JETP {\bf 67, }1163-1168
(1988)].

Landau, L.\ D., and Lifshitz, 1982, {\it Electrodynamics of Continuous Media
(}Moscow: Nauka).

Lange, M., M. J. Van Bael, Y. Bruynseraede, and V. V. Moshchalkov, 2003,
''Nanoengineered Magnetic-Field-Induced Superconductivity,'' Phys. Rev.
Lett. {\bf 90}, 197006.

Larkin, A. I., and Y. N. Ovchinnikov, 1964, ''Inhomogeneous state of
superconductors,'' Zh. Eksp. Teor. Fiz.{\it \ {\bf 47}, }1136-1146 [Sov.
Phys. JETP\ 20, 762-769 (1965)].

Larkin, A. I., and Y. N. Ovchinnikov, 1968, ''Quasiclassical method in the
theory of superconductivity,'' Zh. Eksp. Teor. Fiz.{\it \ {\bf 55}, }%
2262-2272{\it \ }[Sov. Phys. JETP\ 28, 1200-1205 (1965)].

Lazar, L., K. Westerholt, H. Zabel, L. R. Tagirov, Yu. V. Goryunov, N. N.
Garifyanov, and I. A. Garifullin, 2000, ''Superconductor/ferromagnet
proximity effect in Fe/Pb/Fe trilayers,'' Phys. Rev. B, {\bf 61}, 3711-3722.

Li, X., Z. Zheng, D. Y. Xing, G. Sun, and Z. Dong, 2002, ''Josephson current
in ferromagnet-superconductor tunnel junctions,'' Phys. Rev. B {\bf 65},
134507.

Lyuksyutov, I.\ F., and V.\ L. Pokrovsky, 1998, ''Magnetization controlled
superconductivity in a film with magnetic dots,'' Phys. Rev. Lett. {\bf 81},
2344-2347.

Lyuksyutov, I.\ F., and V.\ L. Pokrovsky, 2004, ''Ferromagnet-superconductor
hybrids,'' Cond-mat., 0409137.

Lynn, J. W, B. Keimer, C. Ulrich, C. Bernhard, and J. L. Tallon, 2000,
''Antiferromagnetic ordering of Ru and Gd in superconducting RuSr$_{2}$GdCu$%
_{2}$O$_{8}$,'' Phys. Rev B {\bf 61}, R14964-R14967.

McLaughlin, A. C., W. Zhou, J. P. Attfield, A. N. Fitch, and J. L.\ Tallon,
1999, ''Structure and microstructure of the ferromagnetic superconductor RuSr%
$_{2}$GdCu$_{2}$O$_{8}$,'' Phys. Rev. B {\bf 60}, 7512-7516.

Maple, M. B., and O. Fisher, 1982, in {\it Superconductivity in Ternary
Compounds II, Topics in Current Physics, }edited by M. B. Maple and \O .
Fischer, (Springer-Verlag, Berlin).

Mazin, I., 1999, ''How to define and calculate the degree of spin
polarization in ferromagnets,'' Phys. Rev. Lett. {\bf 83}, 1427-1430.

M\'{e}lin, R., and D. Feinberg, 2004, ''What is the value of the
superconducting gap of a F/S/F trilayer? ,'' Europhysics Lett. {\bf 65},
96-102.

Mercaldo, V., C. Affanasio, C. Coccorese, L. Maritato, S. L. Prischepa, and
M. Salvato, 1996, ''Superconducting-critical-temperature oscillations in
Nb/CuMn multilayers,'' Phys. Rev. B, {\bf 53}, 14040-14042.

Meservey, P., and P.\ M.\ Tedrow, 1994, ''Spin-polarized electron
tunneling,'' Physics Reports {\bf 238}, 173-243.

Milosevic, M.\ V., S.\ V.\ Yampolskii, and F.\ M.\ Peters, 2002a, ''Magnetic
pinning of vortices in a superconducting film: the (anti)vortex-magnetic
dipole interaction energy in the London approximation,'' Phys. Rev B {\bf 66}%
, 174519.

Milosevic, M.\ V., S.\ V.\ Yampolskii, and F.\ M.\ Peters, 2002b, ''Vortex
structure of thin mesoscopic disks in the presence of an inhomogeneous
magnetic field,'' Phys. Rev B {\bf 66}, 024515.

Mints, R. G., 1998, ''Self-generated flux in Josephson junctions with
alternating critical current density,'' Phys. Rev B {\bf 57}, R3221-R3224.

Moussy, N., H. Courtois, and B. Pannetier 2001, ''Local spectroscopy of a
proximity superconductor at very low temperature,'' Europhys. Lett. {\bf 55}%
, 861-867.

M\"{u}ller, K.-H., and V. N. Narozhnyi, 2001, ''Interaction of
superconductivity and magnetism in borocarbide superconductors,'' Rep. Prog.
Phys. {\bf 64}, 943-1008.

M\"{u}hge, Th., N. N. Garif'yanov, Yu. V. Goryunov, G. G. Khaliullin, L. R.
Tagirov, K. Westerholt, I. A. Garifullin, and H. Zabel, 1996, ''Possible
origin for oscillatory superconducting transition temperature in
superconductor/ferromagnet multilayers,'' Phys. Rev. Lett. {\bf 77},
1857-1860.

M\"{u}hge, Th., N. N. Garif'yanov, Yu. V. Goryunov, K. Theis-Br\"{o}hl, K.
Westerholt, I. A. Garifullin, and H. Zabel, 1998, ''Influence of
superconductivity on magnetic properties of superconductor/ferromagnet
epitaxial bilayers,'' Physica C {\bf 296}, 325-336.

Nachtrab, T. , D. Koelle, R. Kleiner, C. Bernhard, and C. T. Lin, 2004,
''Intrinsic Josephson effects in the magnetic superconductor RuSr$_{2}$GdCu$%
_{2}$O$_{8}$,'' Phys. Rev. Lett. {\bf 92}, 117001.

Nielsen, M., and Chuang, I., 2000 {\it Quantum computation and quantum
information} (Cambridge University Press, Cambridge, New York).

Obi, Y., M. Ikebe, T. Kubo, and H. Fujimori, 1999, ''Oscillation phenomenon
of transition temperatures in Nb/Co and V/Co superconductor/ferromagnet
multilayers,'' Physica C, {\bf 317-318}, 149-153.

Pang, B. S. H., R. I. Tomov, C. Bell, Z., and \ M.\ G. Blamire, 2004,
''Effect of ferromagnetism on superconductivity in manganite/cuprate
heterostructures,'' Physica C {\bf 415}, 118-124.

Pannetier, B., and H. Courtois, 2000, ''Andreev reflection and proximity
effect,'' Journal of Low Temperature Physics {\bf 118}, 599-615.

Pe\~{n}a, V., Sefrioui, Z., D. Arias, C. Le\'{o}n, J. Santamaria, M. Varela,
S. J. Prnnycook, and \ J. L. Martinez, 2004, ''Coupling of superconductors
through a half-metallic ferromagnet: evidence for a long-range proximity
effect,'' Phys. Rev. B {\bf 69}, 011422.

Petrashov, V. T., I. A. Sosnin, I. Cox, A. Parsons, and C. Troadec, 1999,
''Giant mutual proximity effects in ferromagnetic/superconducting
nanostructures,'' Phys. Rev. Lett. {\bf 83}, 3281-3284.

Proki\'{c}, V., and Lj. Dobrosavljevi\'{c}-Gruji\'{c}, 1999, ''Zero-energy
bound states in superconductor/ferromagnet superlattices,'' Physica C, {\bf %
320}, 259-266.

Proki\'{c}, V., A. I. Buzdin, and Lj. Dobrosavljevi\'{c}-Gruji\'{c}, 1999,
''Theory of the $\pi $ junctions formed in atomic-scale
superconductor/ferromagnet superlattices,'' Phys. Rev. B {\bf 59}, 587-595.

Proshin, Yu. N., and M. G. Khusainov, 1997, ''Manifestation of the
Larkin-Ovchinnikov-Fulde-Ferrell state in bimetal ferromagnet-superconductor
structures,'' Pis'ma Zh. Eksp. Teor. Phys. {\bf 66, }527-532 [JETP Lett. 
{\bf 66, }562-568 (1997)].

Proshin, Yu. N., and M. G. Khusainov, 1998, ''Nonmonotonic behavior of the
superconducting transition temperature in bimetallic
ferromagnetsuperconductor structures,'' Zh. Eksp. Teor. Fiz. {\bf 113},
1708-1730 [Sov. Phys. JETP {\bf 86, }930-942 (1988)].

Proshin, Yu. N., Yu. A. Izyumov, and M. G. Khusainov, 2001, ''$\pi $
magnetic states of ferromagnet/superconductor superlattices,'' Phys. Rev. B 
{\bf 64}, 064522.

Radovic, Z., L. Dobrosaljevic-Grujic, A. I. Buzdin, and J. R. Clem, 1988,
''Upper critical field of superconductor-ferromagnet multilayers,'' Phys.
Rev. B {\bf 38}, 2388-2393.

Radovic, Z., M. Ledvij, L. Dobrosaljevic-Grujic, A. I. Buzdin, and J. R.
Clem, 1991, ''Transition temperature of superconductor-ferromagnet
superlattices,'' Phys. Rev. B {\bf 44}, 759-764.

Radovic, Z., L. Dobrosaljevic-Grujic, and B. Vujicic, 1999, ''Spontaneous
currents in Josephson devices,'' Phys. Rev. B {\bf 60}, 6844-6849.

Radovic, Z., L. Dobrosaljevic-Grujic, and B. Vujicic, 2001, ''Coexistence of
stable and metastable 0 and $\pi $ states in Josephson junctions,'' Phys.
Rev. B {\bf 63}, 214512.

Radovic, Z., N. Lazarides, and N. Flytzanis, 2003, ''Josephson effect in
double-barrier superconductor-ferromagnet junctions,'' Phys. Rev. B {\bf 68}%
, 014501.

Rusanov, A. Yu., M. Hesselberth, J. Aarts, and A. I. Buzdin, 2004,
''Enhancement of the superconducting transition temperature in Nb/Permalloy
bilayers by controlling the domain state of the ferromagnet,'' Phys. Rev.
Lett. {\bf 93}, 057002.

Ryazanov, V. V., V. A. Oboznov, A. Yu. Rusanov, A. V. Veretennikov, A. A.
Golubov, and J. Aarts, 2001a, ''Coupling of two superconductors through a
ferromagnet: evidence for a $\pi $-junction,'' Phys. Rev. Lett. {\bf 86},
2427-2430.

Ryazanov, V. V., V. A. Oboznov, A. V. Veretennikov, A. Yu. Rusanov, A. A.
Golubov, and J. Aarts, 2001b, ''Coupling of two superconductors through a
ferromagnet. SFS $\pi $-junctions and intrinsically-frustrated
superconducting networks,'' Usp. Fiz. Nauk (Suppl.), {\bf 171}, 81-86.

Ryazanov, V. V., V. A. Oboznov, A. V. Veretennikov, and A. Yu. Rusanov,
2001c, ''Intrinsically frustrated superconducting array of
superconductor-ferromagnet-superconductor $\pi $ junctions,'' Phys. Rev. B 
{\bf 65}, R020501.

Ryazanov, V. V., V. A. Oboznov, A. S. Prokof'ev, and S. V. Dubonos, 2003,
''Proximity effect and spontaneous vortex phase in planar SF structures,''
Pis'ma Zh. Eksp. Teor. Phys. {\bf 77, }43-47 [JETP Lett. {\bf 77, }39-43
(2003)].

Ryazanov, V. V., V. A. Oboznov, A. S. Prokof'ev, V. V. Bolginov, and A. K.
Feofanov, 2004, ''Superconductor-ferromagnet-superconductor $\pi $
junctions,'' J. Low Temp. Phys. {\bf 136, }385-400.

Ryazanov, V. V., V. A. Oboznov, V. V. Bolginov, A. K. Feofanov and A.
Buzdin, 2005, to be published.

Saint-James, D., D. Sarma , and E. J. Thomas, 1969, {\it Type II
Superconductivity }(Pergamon, New York ).

Saxena, S. S., P. Agarwal, K. Ahilan, F. M. Grosche, R. K. W. Haselwimmer,
M. J. Steiner, E. Pugh, I. R. Walker, S. R. Julian, P. Monthoux, G. G.
Lonzarich, A. Huxley, I. Sheikin, D. Braithwaite, J. Flouquet, 2000,
''Superconductivity on the border of itinerant-electron ferromagnetism in UGe%
$_{2}$,'' Nature {\bf 406}, 587-592.

Schopohl, N., and K. Maki, 1995, ''Quasiparticle spectrum around a vortex
line in a d-wave superconductor,'' Phys. Rev. B {\bf 52}, 490-493.

Sefrioui, Z., D. Arias, V. Pe\~{n}a, J. E. Villegas, M. Varela, P. Prieto,
C. Le\'{o}n, J. L. Martinez, and J. Santamaria, 2003,
''Ferromagnetic/superconducting proximity effect in La$_{0.7}$Ca$_{0.3}$MnO$%
_{3}$/YBa$_{2}$Cu$_{3}$O$_{7}$ superlattices,'' Phys. Rev. B {\bf 67},
214511.

Sellier, H., Baraduc C., Lefloch F., and Calemczuk R., 2003,
''Temperature-induced crossover between 0 and $\pi $ states in S/F/S
junctions,'' Phys. Rev. B {\bf 68}, 05453.

Sidorenko, A. S. , V. I. Zdravkov, A. A. Prepelitsa, C. Helbig, Y. Luo, S.
Gsell, M. Schreck, S. Klimm, S. Horn, L. R. Tagirov and R. Tidecks, 2003,
''Oscillations of the critical temperature in superconducting Nb/Ni
bilayers,'' Ann. Phys. {\bf 12}, 37-50.

Sonin, E. B., 1988, ''Suppression of superconductivity (weak link) by a
domain wall in a two-layer superconducto-ferromagnet film,'' Pis'ma \ Zh.
Tekh. Phys. {\bf 14}, 1640-1644 [Sov. Tech. Phys. Lett. {\bf 14}, 714-716
(1988)].

Sonin, E. B., 2002, ''Comment on ''Ferromagnetic film on a superconducting
substrate'','' Phys. Rev. B {\bf 66}, 136501.

Soulen Jr, R.\ J., M. Byers, M.\ S. Osofsky, B.\ Nadgorny, T.\ Ambrose, S.\
F.\ Cheng, P.\ R. Broussard, C.\ T.\ Tanaka, J.\ Nowak, J.\ S.\ Moodera, A.
Bary, and J.\ M.\ D.\ Coey, 1998, ''Measuring the spin polarization of a
metal with a superconducting point contact,'' Science, {\bf 282}, 85-88.

Strunk, C., C. S\"{u}rgers, U. Paschen, and H. v. L\"{o}hneysen, 1994,
''Superconductivity in layered Nb/Gd films,'' Phys. Rev. B {\bf 49},
4053-4063.

Sumarlin, I. W., S. Skanthakumar, J. W. Lynn, J. L. Peng, Z. Y. Li, W.
Jiang, and R. L. Greene, 1992, ''Magnetic ordering of Sm in Sm$_{2}$CuO$_{4}$%
,'' Phys. Rev. Lett. {\bf 68}, 2228-2231.

Sun, G., D. Y. Xing, J. Dong, and M. Liu, 2002, ''Gapless superconductivity
in ferromagnet/superconductor junctions,'' Phys. Rev. B {\bf 65}, 174508.

Tagirov, L. R., 1998, ''Proximity effect and superconducting transition
temperature in superconductor/ferromagnet sandwiches,'' Physica C {\bf 307},
145-163.

Tagirov, L. R., 1999,\ ''Low-field superconducting spin switch based on a
superconductor/ferromagnet multilayer,'' Phys. Rev. Lett. {\bf 83},
2058-2061.

Takahashi, S., and M. Tachiki, 1986, ''Theory of the upper critical field of
superconducting superlattices,'' Phys. Rev. B {\bf 33}, 4620-4631.

Tanaka, Y., and S. Kashiwaya, 1997, ''Theory of Josephson effect in
superconductor-ferromagnetic-insulator-superconductor junction,'' Physica C 
{\bf 274}, 357-363.

Terzioglu, E., and M.\ R.\ Beasley, 1998, ''Complementary Josephson junction
devices and circuits: a possible new approach to superconducting
electronics,'' IEEE Trans. Appl. Supercond. {\bf 8}, 48-53.

Tokunaga, Y., H. Kotegawa, K. Ishida, Y. Kitaoka, H. Takagiwa, and J.
Akimitsu, 2001, ''NMR evidence for coexistence of superconductivity and
ferromagnetic component in magnetic superconductor RuSr$_{2}$YCu$_{2}^{{}}$O$%
_{8}$: $^{99,101}$Ru and $^{63}$Cu NMR,'' Phys. Rev. Lett. {\bf 86},
5767-5770.

Tollis, S., 2004, ''First-order phase transitions in
ferromagnetic/superconducting/ferromagnetic trilayers,'' Phys. Rev. B 69,
104532.

Uji, S., H. Shinagawa, T. Terashima, T. Yakabe, Y. Terai, M. Tokumoto, A.\
Kobayashi, H. Tanaka, and H. Kobayashi, 2001, ''Magnetic-field-induced
superconductivity in a two-dimensional organic conductor,'' Nature {\bf 410}%
, 908-910.

Usadel, L., 1970, ''Generalized diffusion equation for superconducting
alloys,'' Phys. Rev. Lett. {\bf 25,} 507-509.

Ustinov, A.\ V., and V.\ K.\ Kaplunenko, 2003, ''Rapid single-flux quantum
logic using $\pi $-shifters,'' J. App. Phys. {\bf 94}, 5405-5407.

Upadhyay, S.\ K., A. Palanisami, R.\ N.\ Louie, and R.\ A.\ Buhrman, {\it %
1998, }''Probing ferromagnets with{\it \ }Andreev reflection,''{\it \ }Phys.
Rev. Lett. {\bf 81}, 3247-3250.

Van Harlingen, D.\ J., 1995, ''Phase-sensitive tests of the symmetry of the
pairing state in the high-temperature superconductors: evidence for d$%
_{x2-y2}$ symmetry,'' Rev. Mod. Phys. {\bf 67}, 515-535.

Van Bael, M. J., S. Raedts, K. Temst, J. Swerts, V. V. Moshchalkov, and Y.
Bruynseraede, 2002, ''Magnetic domains and flux pinning properties of a
nanostructured ferromagnet/superconductor bilayer,'' J. Appl. Phys. {\bf 92}%
, 4531-4537.

Van Bael, M. J., L. Van Look, M. Lange, J. Bekaert, S. J. Bending, A. N.
Grigorenko, K. Temst, V. V. Moshchalkov, and Y. Bruynseraede, 2002,
''Ferromagnetic pinning arrays,'' Physica C {\bf 369}, 97-105.

V\'{e}lez, M., M. C. Cyrille, S. Kim, J. L. Vicent, and I. K. Schuller,
1999, ''Enhancement of superconductivity by decreasing magnetic spin-flip
scattering: nonmonotonic T$_{c}$ dependence with enchanced magnetic
ordering,'' Phys. Rev. B {\bf 59}, 14659-14662.

Vodopyanov, B. P., and L. R. Tagirov, 2003a, ''Andreev conductance of a
ferromagnet-superconductor point contact,'' Pis'ma Zh. Eksp. Teor. Phys. 
{\bf 77, }126-131 [JETP Lett. {\bf 77, }153-158 (2003)].

Vodopyanov, B. P., and L. R. Tagirov, 2003b, ''Oscillations of
superconducting transition temperature in strong ferromagnet-superconductor
bilayers,'' Pis'ma Zh. Eksp. Teor. Phys. {\bf 78, }1043-1047 [JETP Lett. 
{\bf 78, }555-559 (2003)].

Volkov, A. F., F. S. Bergeret, and K. B. Efetov, 2003, ''Odd triplet
superconductivity in superconductor-ferromagnet multilayered structures,''
Phys. Rev. Lett. {\bf 90}, 117006.

White, R. M , and T. H. Geballe, 1979, {\it Long Range Order in Solids}
(Academic Press, New York).

Wong, H. K., B.\ Y. Jin, H.\ Q. Yang, J.\ B.\ Ketterson, and J.\ E.\
Hilliard, 1986, ''Superconducting properties of V/Fe superlattices,'' J. Low
Temp. Phys. {\bf 63}, 307-315.

Xu, J. H., J. H. Miller, Jr., and C. S. Ting, 1995, ''$\pi $ -vortex state
in a long $0-$ $\pi $ Josephson junction,'' Phys. Rev. B {\bf 51},
11958-11961.

Yamashita, T., S. Takahashi, and S. Maekawa, 2003, '' Crossed Andreev
reflection in structures consisting of a superconductor with ferromagnetic
leads,'' Phys. Rev. B {\bf 68}, 174504.

Yamashita, T., H. Imamura, S. Takahashi, and S. Maekawa, 2003, ''Andreev
reflection in ferromagnet/superconductor/ferromagnet double junction
systems,'' Phys. Rev. B {\bf 67}, 094515.

Yang, Z., M. Lange, A. Volodin, R. Szymczak, and V. Moshchalkov, 2004,
''Domain-wall superconductivity in

superconductor-ferromagnet hybrids,'' Nature Materials {\bf 3}, 793-798.

Zaitsev, A. V., 1984, ''Quasiclassical equations of the theory of
superconductivity for contiguous metals and the properties of constricted
microcontacts,'' Zh. Eksp. Teor. Fiz.{\it \ }{\bf 86}{\it , }1742-1758 [Sov.
Phys. JETP{\it \ }{\bf 59}, 1015-1024 (1984)].

Zareyan, M., W. Belzig, and Yu. V. Nazarov, 2001, ''Oscillations of Andreev
states in clean ferromagnetic films,'' Phys. Rev. Lett. {\bf 86}, 308-311.

Zareyan, M., W. Belzig , and Yu. V. Nazarov, 2002, ''Superconducting
proximity effect in clean ferromagnetic layers,'' Phys. Rev. B {\bf 65},
184505.

Zikic, R., L. Dobrosavljevic-Grujic, and Z. Radovic, 1999, ''Phase-dependent
nergy spectrum in Josephson weak links'' Josephson weak links,'' Phys. Rev.
B {\bf 59}, 14644-14652.

Zutic I., and S. Das Sarma, 1999, ''Spin-polarized transport and Andreev
reflection in semiconductor/superconductor hybrid structures,'' Phys. Rev. B 
{\bf 60}, R16322.

Zutic I., and O.\ T.\ Valls, 1999, ''Spin-polarized tunneling in
ferromagnet/unconventional superconductor junctions,'' Phys. Rev. B 60,
6320-6323.

Zutic I., and O.\ T.\ Valls, 2000, ''Tunneling spectroscopy for
ferromagnet/superconductor junctions,'' Phys. Rev. B {\bf 61}, 1555-1561.

Zutic I., J. Fabian, and S. Das Sarma, 2004, ''Spintronics: Fundamentals and
applications,'' Rev. Mod. Phys. {\bf 76}, 323-410.

Zyuzin, A. Yu., and B. Spivak, 2000, ''Theory of $\pi /2$ superconducting
Josephson junctions,'' Phys. Rev. B {\bf 61}, 5902-5904.

Zyuzin, A. Yu., B. Spivak, and M. Hruska, 2003, ''Mesoscopic effects in
superconductor-ferromagnet-super- conductor junctions,'' Europhys. Lett. 
{\bf 62}, 97-102.

\qquad {\LARGE TABLE I. Characteristic length scales of S/F proximity effect.%
}

\begin{tabular}{|c|c|}
\hline
Thermal diffusion length $L_{T}$ & $\sqrt{\frac{D}{2\pi T}}$ \\ \hline
Superconducting coherence length $\xi _{s}$ & 
\begin{tabular}{l}
$\frac{v_{Fs}}{2\pi T_{c}}$ in pure limit \\ 
$\sqrt{\frac{D_{s}}{2\pi T_{c}}}$ in dirty limit
\end{tabular}
\\ \hline
Superconducting correlations decaying length $\xi _{1f}$ in a ferromagnet & 
\begin{tabular}{l}
$\frac{v_{Ff}}{2\pi T}$ in pure limit \\ 
$\xi _{f}=\sqrt{\frac{D_{f}}{h}}$ in dirty limit
\end{tabular}
\\ \hline
Superconducting correlations oscillating length $\xi _{2f}$ in a ferromagnet
& 
\begin{tabular}{l}
$\frac{v_{Ff}}{2h}$ in pure limit \\ 
$\xi _{f}=\sqrt{\frac{D_{f}}{h}}$ in dirty limit
\end{tabular}
\\ \hline
\end{tabular}

\begin{center}
{\LARGE Figure captions}

\bigskip
\end{center}

FIG. 1. The $(T,H)$ phase diagram for 3D superconductor. At temperature
below $T^{\ast }=0.56T_{c}$ the second order transition occurs from the
normal to the non-uniform superconducting FFLO phase. The dashed line
corresponds to the first order transition into the uniform superconducting
state, and the dotted line presents the second order transition into the
uniform superconducting state.

FIG. 2. Energy band of 1D superconductor near the Fermi energy. Due to the
Zeeman splitting the energy of the electrons with spin orientation along the
magnetic field ($\uparrow $) decreases - dotted line, while the energy of
the electrons with the opposite spin orientation ($\downarrow $) increases -
dotted line. The splitting of the Fermi momenta is $\pm \delta k_{F}$, where 
$\delta k_{F}=\mu _{B}H/v_{F\text{ }}$. The Cooper pair comprises one
electron with the spin ($\uparrow $) and momentum $k_{F}+\delta k_{F}$, and
another electron with the spin ($\downarrow $) and momentum $-k_{F}+\delta
k_{F}$. The resulting momentum of the Cooper pair is non-zero : $%
k_{F}+\delta k_{F}+\left( -k_{F}+\delta k_{F}\right) =2\delta k_{F}\neq 0.$%
\medskip

FIG. 3. Schematic behavior of the superconducting order parameter near the
interface (a) superconductor-normal metal, and (b)
superconductor-ferromagnet. The continuity of the order parameter at the
interface implies the absence of the potential barrier. In general case at
the interface the jump of the superconducting order parameter occurs.

FIG. 4. Measurements of the differential conductance by Kontos {\it et al.}
(2001) for two Al/Al$_{2}$O$_{3}$/PdNi/Nb junctions with two different
thicknesses (50 \AA\ and 75 \AA ) of the ferromagnetic PdNi layer.\medskip A
1500-\AA -thick aluminium layer was evaporated on SiO and then quickly
oxidized to produce a Al$_{2}$O$_{3}$ tunnel barrier. Tunnel junction areas
were defined by evaporating 500 \AA\ of SiO \ through masks. A PdNi thin
layer was deposited and then backed by a Nb layer.

FIG. 5. Experimental data of Jiang {\it et al.} (1995) on the oscillation of
the critical temperature of Nb/Gd multilayers vs thickness of Gd layer $%
d_{G} $ for two different thicknesses of Nb layers : (a) $d_{Nb}=$600 \AA\
and (b) $d_{Nb}=$500 \AA . Dashed line in (a) is a fit by the theory of
Radovic {\it et al.} (1991).\medskip

FIG. 6. S/F multilayer. The axe $x$ is chosen perpendicular to the planes of
S and F layers with the thicknesses $2d_{s}$ and $2d_{f}$ respectively. (a)
The curve $\Psi (x)$ represents schematically the behavior of the Cooper
pair wave function in ''$0$''- phase. Due to the symmetry reasons the
derivative of $\Psi $ (and $F$) is zero at the centers of S and F layers.
The case of the ''$0$''- phase is equivalent to the S/F bilayer with S and F
layers thicknesses $d_{s}$ and $d_{f}$ respectively. (b) The Cooper pair
wave function in ''$\pi $''- phase vanishes at the centers of F layers and $%
\Psi (x)$ is antisymmetric toward the center of F layer.\medskip

FIG. 7. The dependence of the critical temperature on the thickness of F
layer for ''$0$''-phase (solid line) and ''$\pi $''- phase (dotted line) in
the case of the transparent S/F interface. Note that the highest transition
temperature $T_{c}^{\ast }$ corresponds to the lowest point. The
dimensionless thickness of F layer $2y=2d_{f}/\xi _{f}$ and the first
transition from ''$0$''- to ''$\pi $''- phase occurs at $2d_{f}=2.36\xi
_{f}. $\medskip The parameter $\tau _{0}=\frac{2d_{s}\xi _{f}}{D_{s}}\frac{%
\sigma _{s}}{\sigma _{f}}.$

FIG. 8. The critical temperature of ''$0$''- phase (solid line) and ''$\pi $%
''- phase (dashed line) as a function of the dimensionless thickness of F
layer $2y=2d_{f}/\xi _{f}$ for different S/F interface barriers $\tilde{%
\gamma}=\gamma _{B}\left( \xi _{n}/\xi _{f}\right) .$

(a) The dimensionless pair-breaking parameter $\tilde{\tau}_{0}=4\pi T_{c}%
\frac{2d_{s}\xi _{f}}{D_{s}}\frac{\sigma _{s}}{\sigma _{f}}=21.$

(b) The dimensionless pair-breaking parameter $\tilde{\tau}_{0}=20.05.$%
\medskip

FIG. 9. Variation of the critical temperature of\ the Nb/Cu$_{0.43}$Ni$%
_{0.57}$ bilayer with the F layer thickness (Ryazanov {\it et al. }2003).
Theoretical fit (Fominov {\it et al.}, 2002) gives the exchange field value $%
h\thicksim 130$ K and the interface transparency parameter $\gamma
_{B}\thicksim 0.3.$\medskip

FIG. 10. Geometry of the S/F/S junction. The thickness of the ferromagnetic
layers is $2d_{f}$ and the both S/F interfaces have the same transparencies,
characterized by the coefficient $\gamma _{B}$.\medskip

FIG. 11. Critical current of the S/F/S Josephson junction near $T_{c}$ as a
function of the dimensionless thickness of F layer $2y=2d_{f}/\xi _{f}$.
There are no barriers at the S/F interfaces $\left( \gamma _{B}=0\right) $, $%
R_{n}$ is the resistance of the junction and $V_{0}=\frac{\pi \Delta ^{2}}{%
2eT_{c}}.$\medskip

FIG. 12. Temperature dependences of the critical thickness $2d_{f\text{ }%
}^{c}$ of F layer, corresponding to the crossover from ''$0$''- to ''$\pi $%
''- phase in the limit of very small boundary transparency for different
values of the exchange field.\medskip

FIG. 13. Non-monotonous temperature dependences of the normalized critical
current for low transparency limit: curve 1: $h/T_{c}$ $=10$ and $2d_{f}/\xi
_{f}=0.84$; curve 2: $h/T_{c}$ $=40$ and $2d_{f}/\xi _{f}=0.5$; curve 3: $%
h/T_{c}$ $=100$ and $2d_{f}/\xi _{f}=0.43.$\medskip

FIG. 14. Critical current $I_{c}$ as a function of temperature for Cu$%
_{0.48} $Ni$_{0.52}$ junctions with different F layers thicknesses $2d_{F}.$
At the thickness of F layer of $27$ nm the temperature mediated transition
between ''$0$''- and ''$\pi $''- phases\ occurs. Adapted from (Ryazanov {\it %
et al.,\ }2001a).\medskip

FIG. 15. Critical current $I_{c}$ at $T=4.2$ $K$ of Cu$_{0.47}$Ni$_{0.53}$
junctions as a function of F layer thickness (Ryazanov {\it et al.,\ }%
2005).Two $"0"-"\pi "$ transitions are revealed. The theoretical fit
corresponds to the Eq. (\ref{Ic general}) in Appendix B, taking into account
the presence of the magnetic scattering with parameters $\alpha =1/(\tau
_{s}h)=1.33$ and $\xi _{f}=2.4$ $nm$. The inset shows the temperature
mediated $"0"-"\pi "$ transition for the F layer thickness $11$ $nm$.

FIG. 16. The experimental points correspond to the measurements of the
critical current, done by Kontos {\it et al.\ }(2002) vs the $PdNi$ layer
thickness. The theoretical curve is the fit of Buzdin and Baladie (2003).
The fitting parameters are\ $\xi _{f}\thicksim $30 \AA\ and $\frac{\pi
\Delta ^{2}}{eT_{c}}\thicksim 110$ $\mu V.$\medskip

FIG. 17. Experiments of Guichard {\it et al. }(2003) on the diffraction
pattern of SQUID with ''$0$''- and ''$\pi $''-junctions. There is no shift
of the pattern between a ''$0-0$'' and ''$\pi -\pi $'' SQUIDs. The $\Phi
_{0}/2$ shift is observed between a ''$0-\pi $'' and ''$0-0$'' or ''$\pi
-\pi $'' SQUIDs. The ''$0$''- and ''$\pi $''-junctions were obtained by
varying the PdNi layer thickness.\medskip

FIG. 18. Earlier observation by Deutscher and Meunier (1969) of the
spin-walve effect on In film between oxidized FeNi and Ni layers. The figure
presents the resistive measurements of the critical temperature in zero
field: \ dashed line, after application of 1 T field parallel to the
ferromagnetic layers; solid line, after application of the -1 T field\ and
subsequently +0.03 T field to return the magnetization of FeNi layer.\medskip

FIG. 19. Geometry of the F/S/F sandwich. The thickness of S layer is $2d_{s}$
and two F layers have identical thicknesses $d_{f}.$\medskip

FIG. 20. Influence of the S/F interface transparency (parameter $\tilde{%
\gamma}=\gamma _{B}\left( \xi _{n}/\xi _{f}\right) $)\ on the $T_{c}^{\ast }$
vs $d_{f}$ dependence (Baladi\'{e} and\ Buzdin, 2003). The thickness of F
layer is normalized to the $\xi _{f}.$ The dimensionless pair-breaking
parameter $\tilde{\tau}_{0}=4\pi T_{c}\frac{2d_{s}\xi _{f}}{D_{s}}\frac{%
\sigma _{s}}{\sigma _{f}}$ is chosen constant and equal to 4. The full line
corresponds to the antiparallel case, and the dashed line to the parallel
case. One can distinguish four characteristic types behavior: (a) weakly
non-monotonous decay to a finite value of $T_{c}^{\ast }$ , (b) reentrant
behavior for the parallel orientation, and (c) and (d) monotonous decay to $%
T_{c}^{\ast }=0$ with (d) or without (c) switching to a first-order
transition in the parallel case. In (d), the dotted line presents
schematically the first order transition line.\medskip

FIG. 21. The calculate dependence of the superconducting transition
temperature vs inverse reduced half-thickness $d^{\ast }/d_{s}$ of the
superconducting layer for parallel and antiparallel alignments for the
transparent interface ($\gamma _{B}=0$) and thick ferromagnetic layer ($%
d_{f}>>\xi _{f}$). The effective length $d^{\ast }=\left( \sigma _{f}/\sigma
_{s}\right) \left( D_{s}/4\pi T_{c}\right) \left( h/D_{f}\right) ^{1/2}$%
.\medskip

FIG. 22. The ($T,h$) - phase diagram of the atomic S/F multilayer in the
limit of the small transfer integral $t<<T_{c}.$\medskip

FIG. 23. S/F bilayer with domain structure in the ferromagnetic layer. The
period $\ D$ of the domain structure ($D=2\pi /Q$) is smaller than the
superconducting coherence length $\xi _{s}.$\medskip

FIG. 24. The ferromagnetic film with perpendicular anisotropy on a
superconducting substrate.

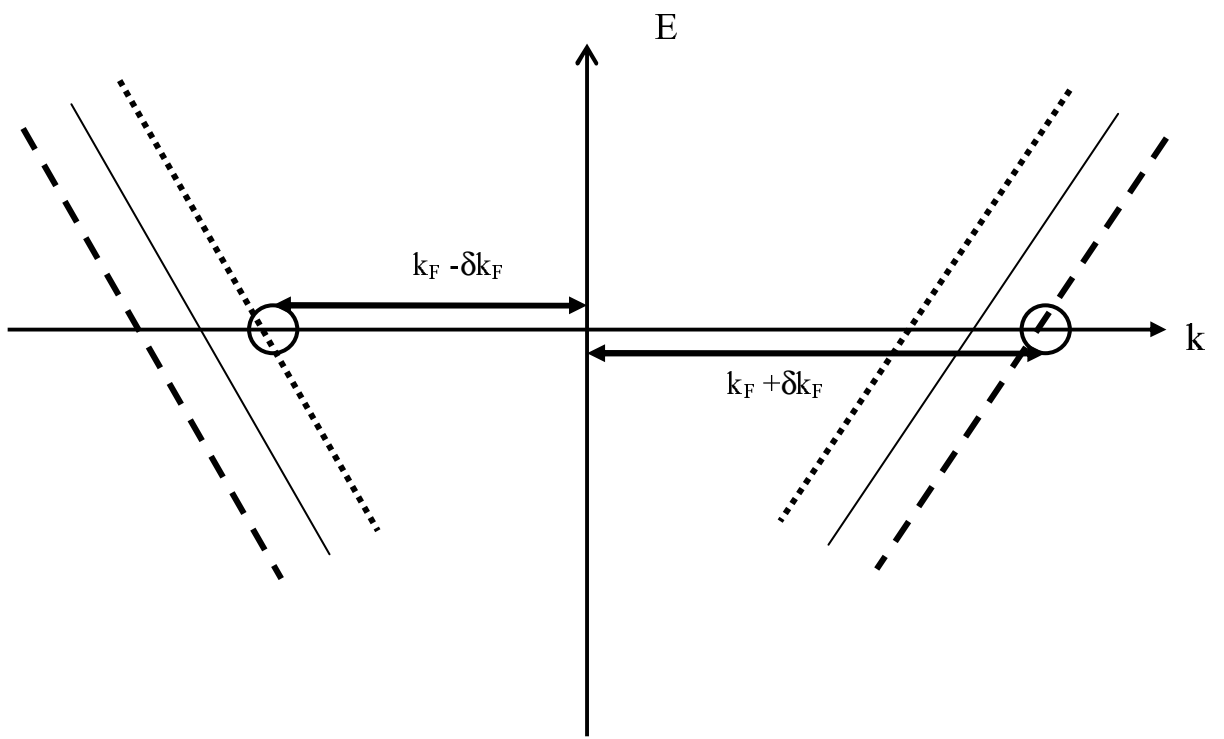
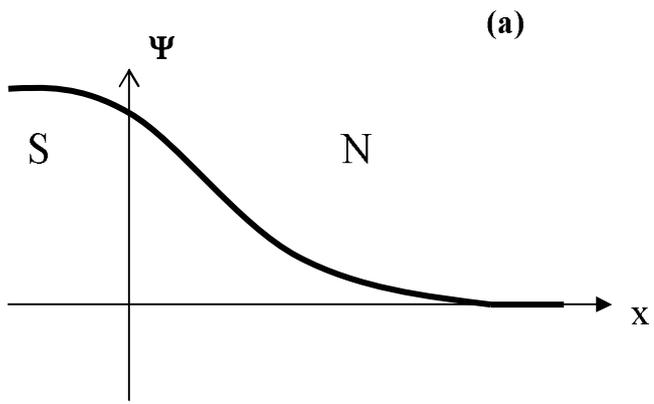
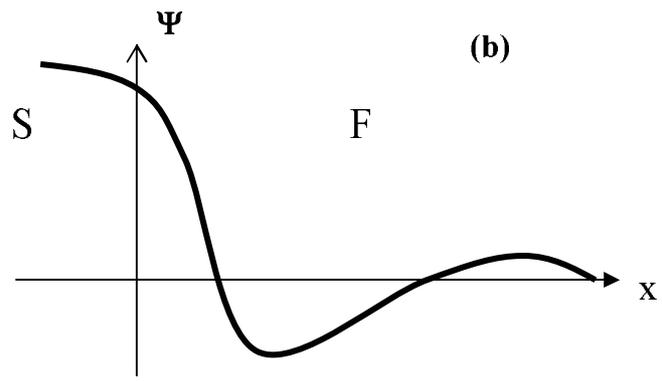
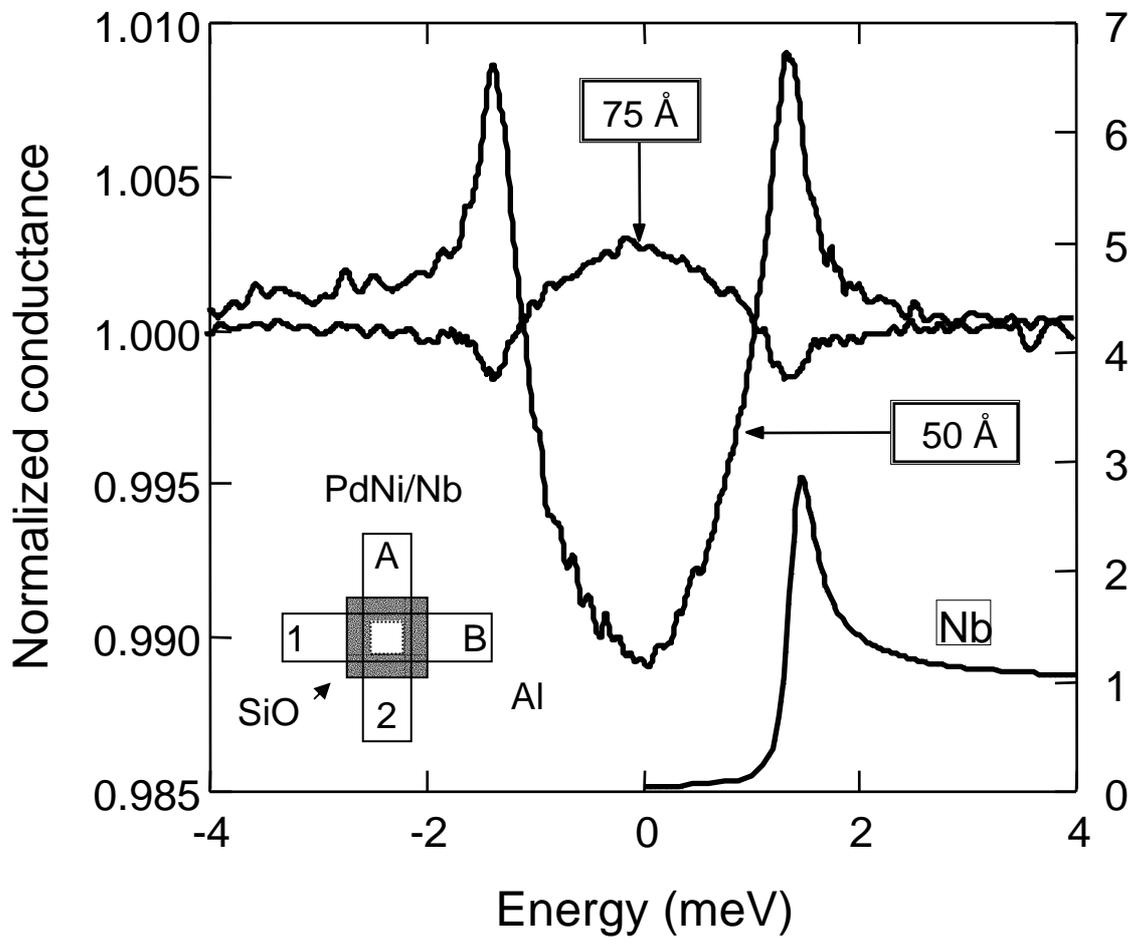
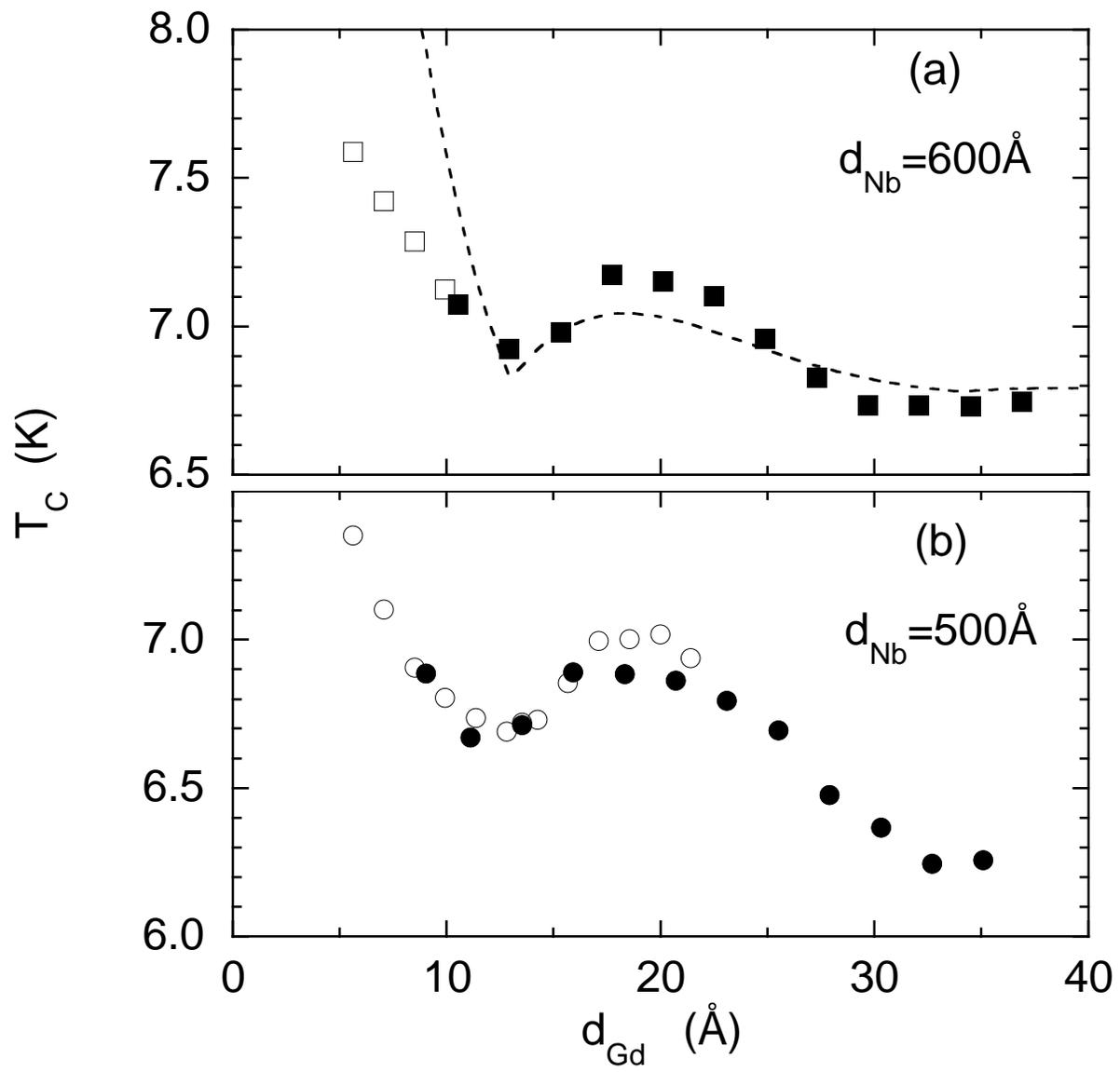
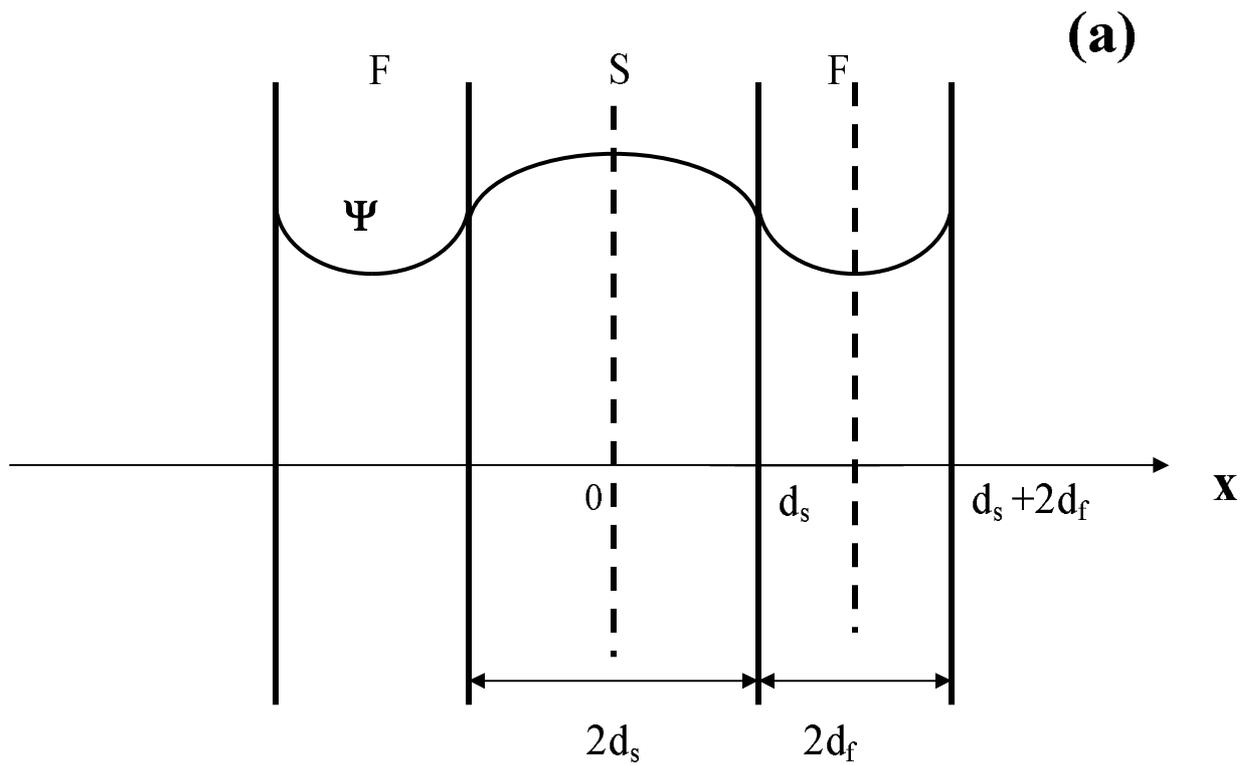
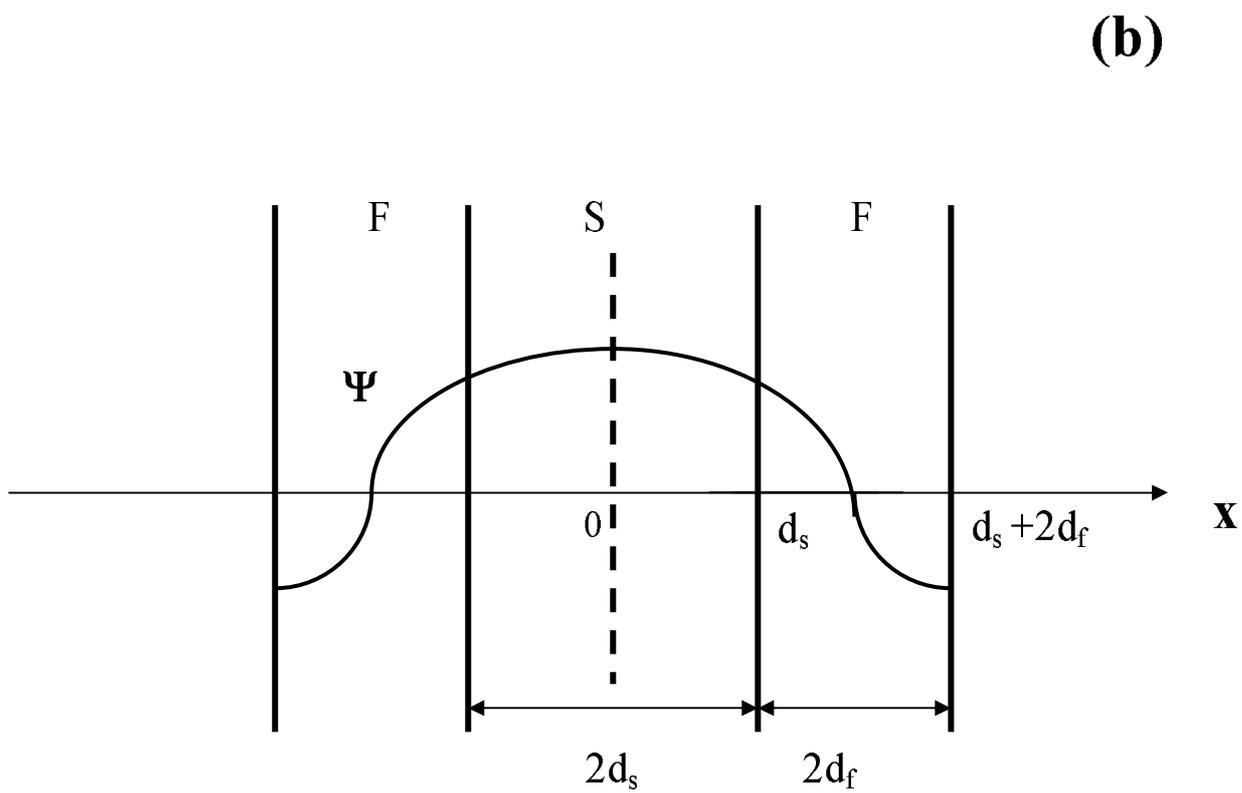
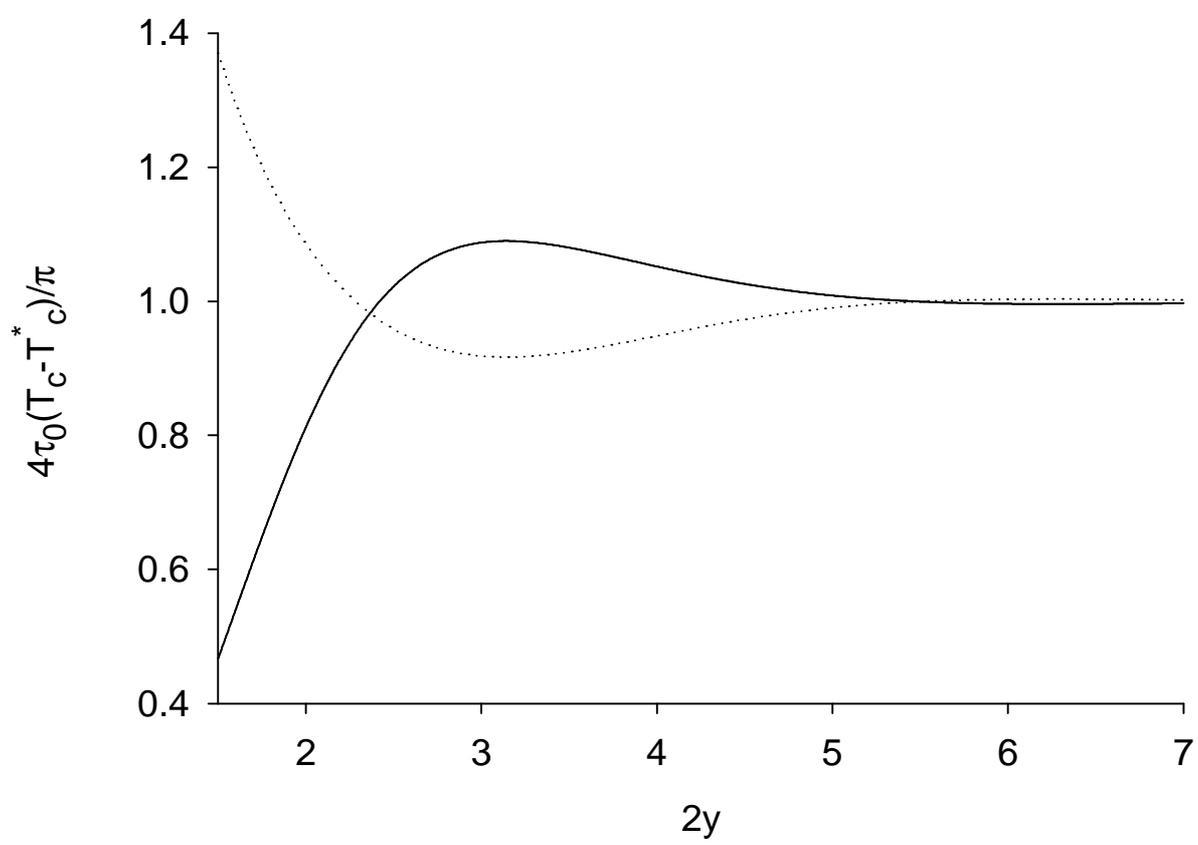
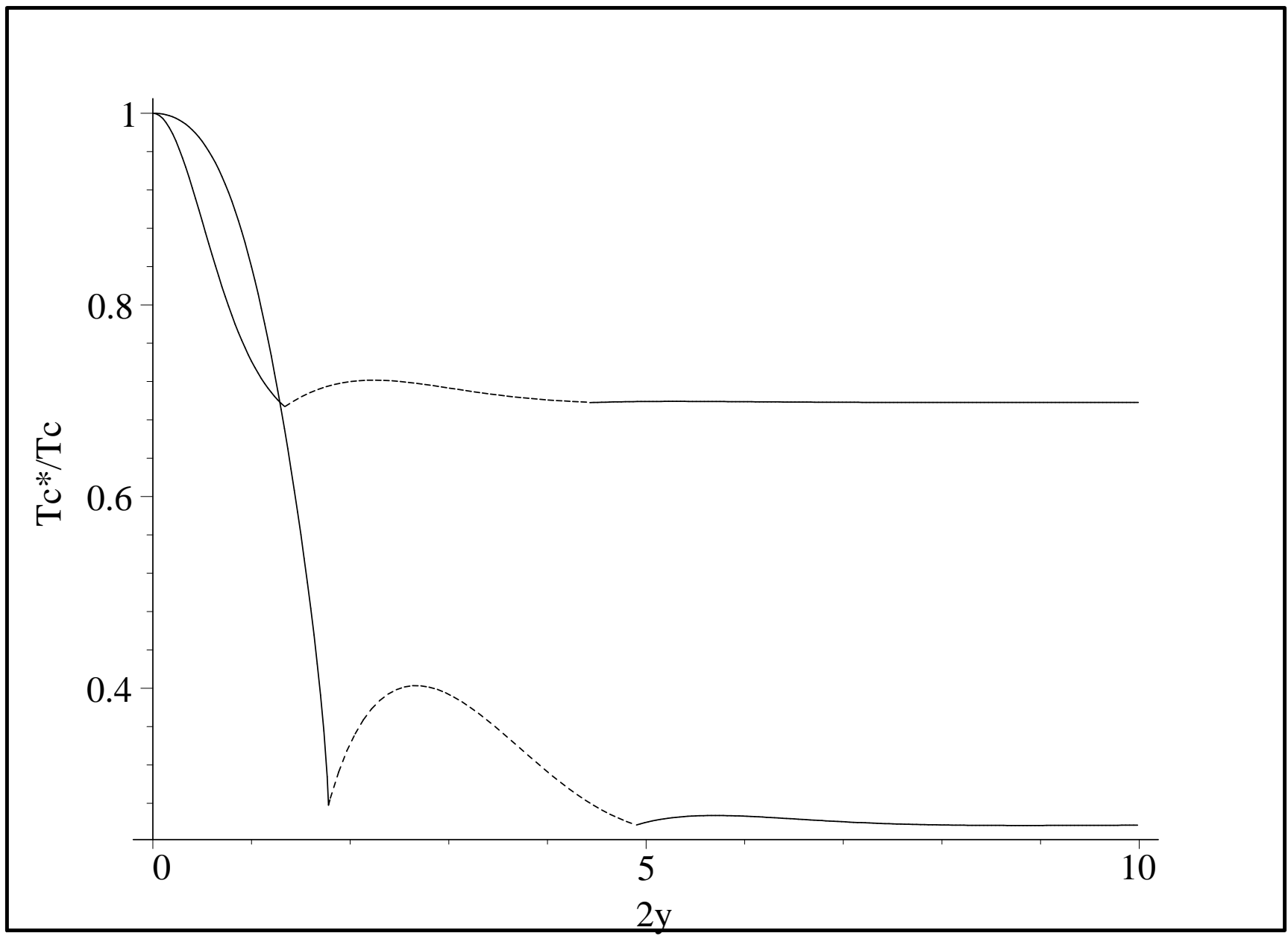
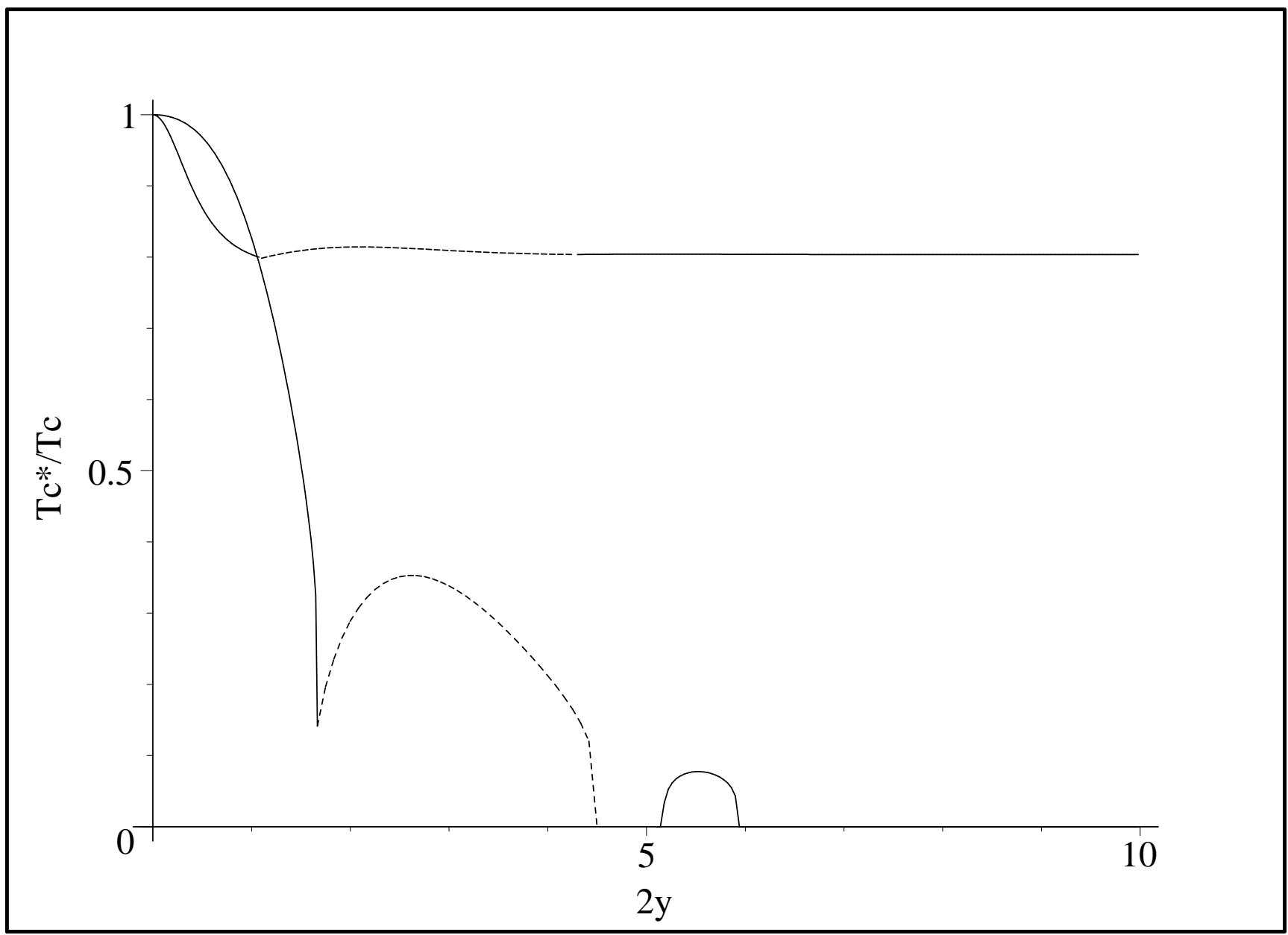
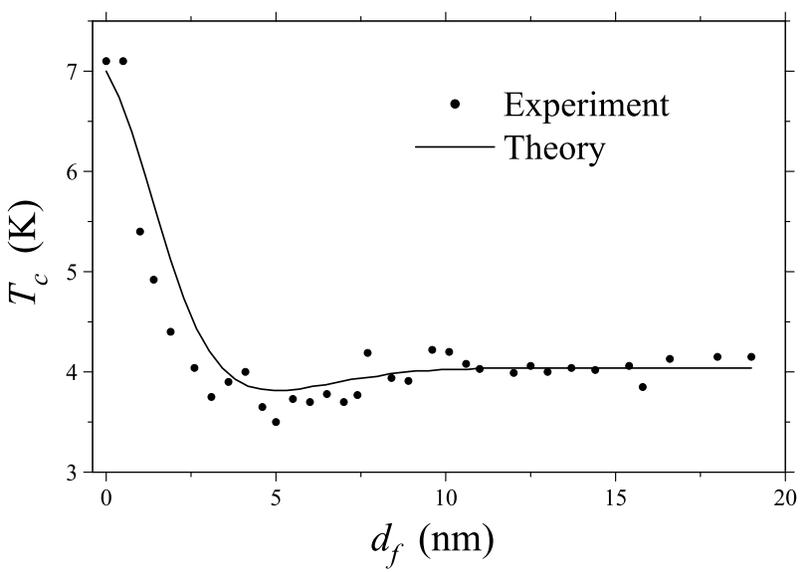
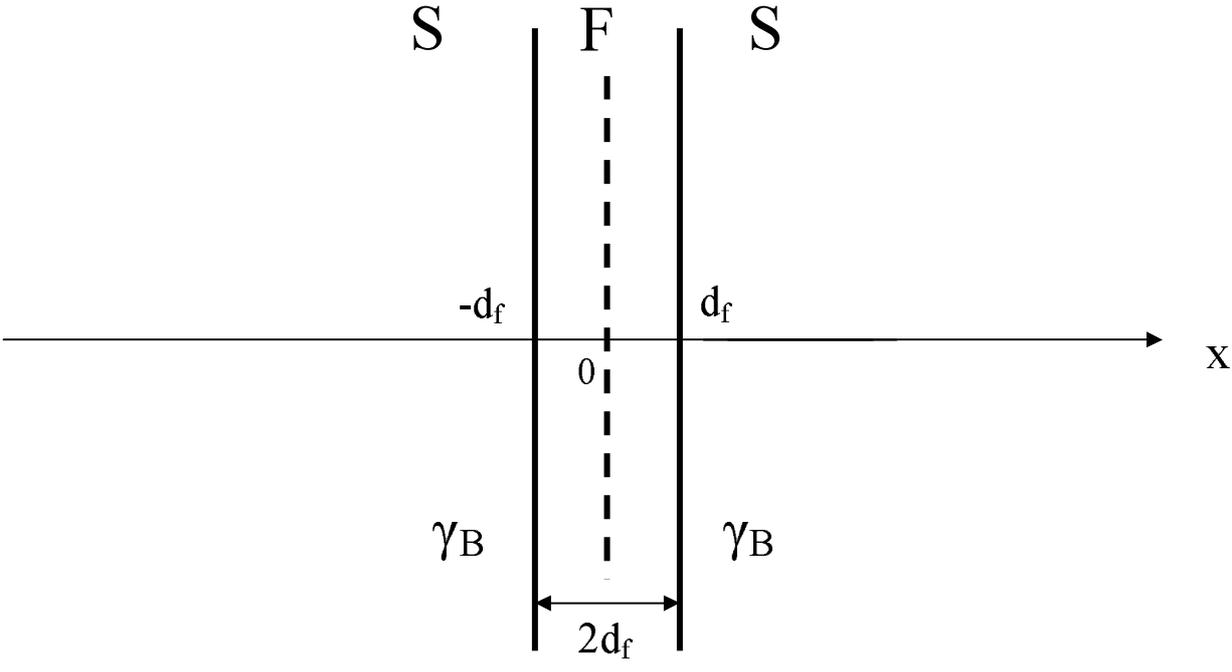
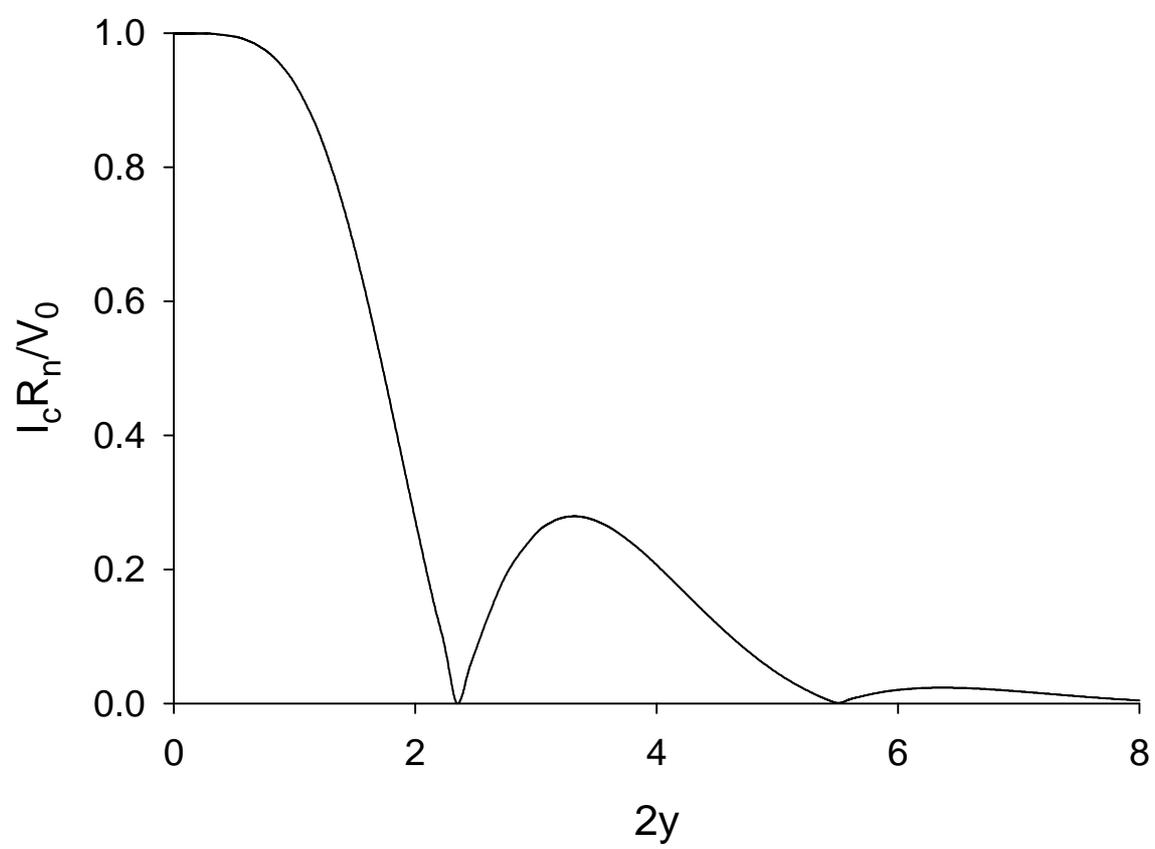
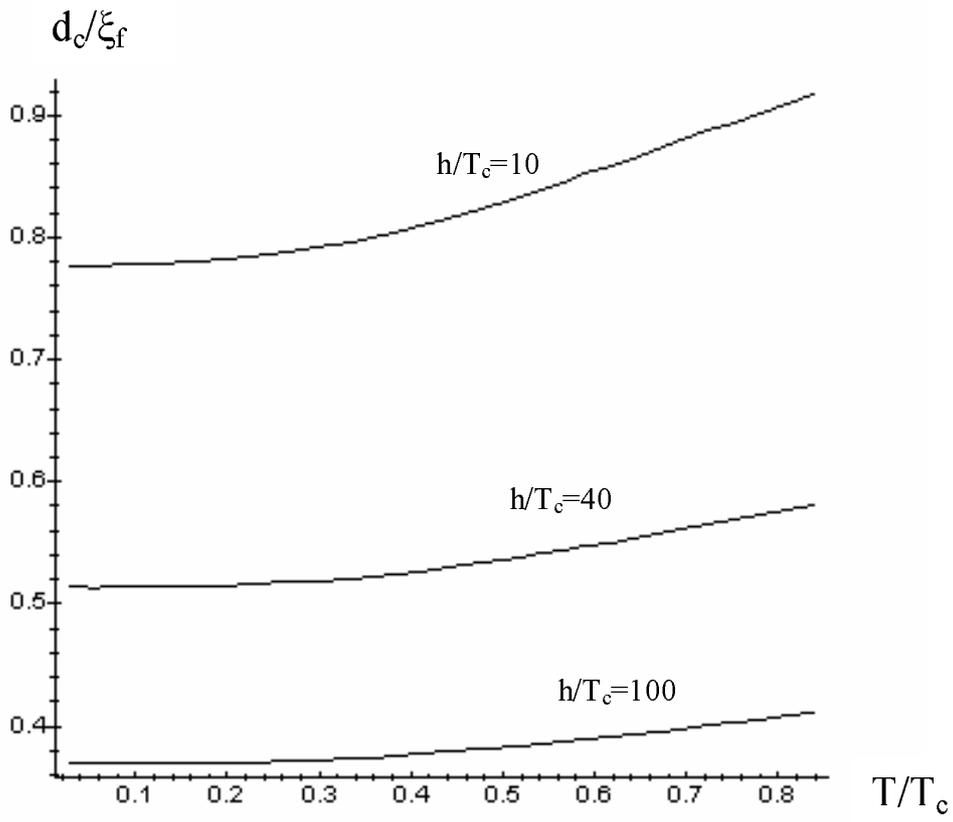
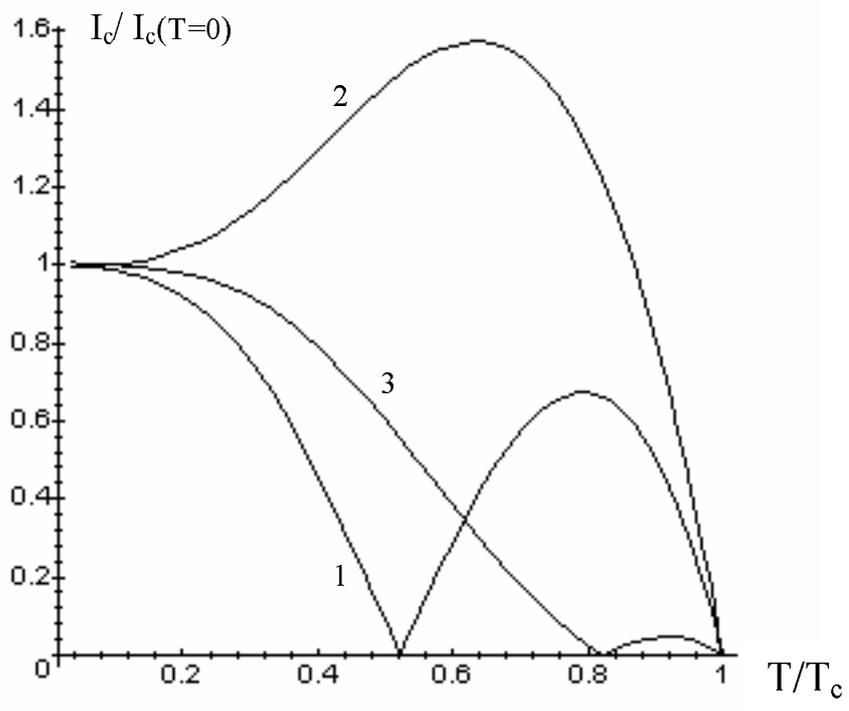
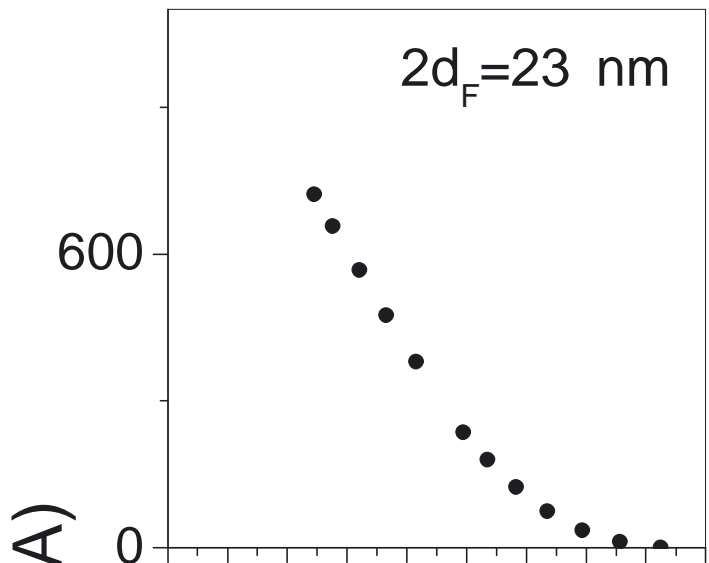
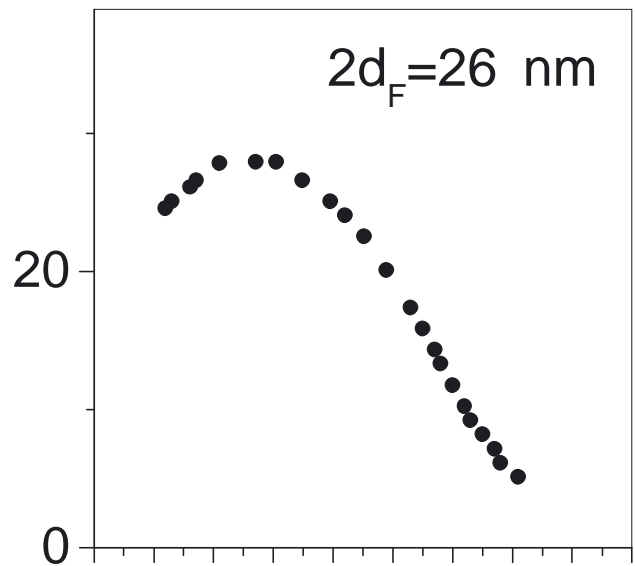
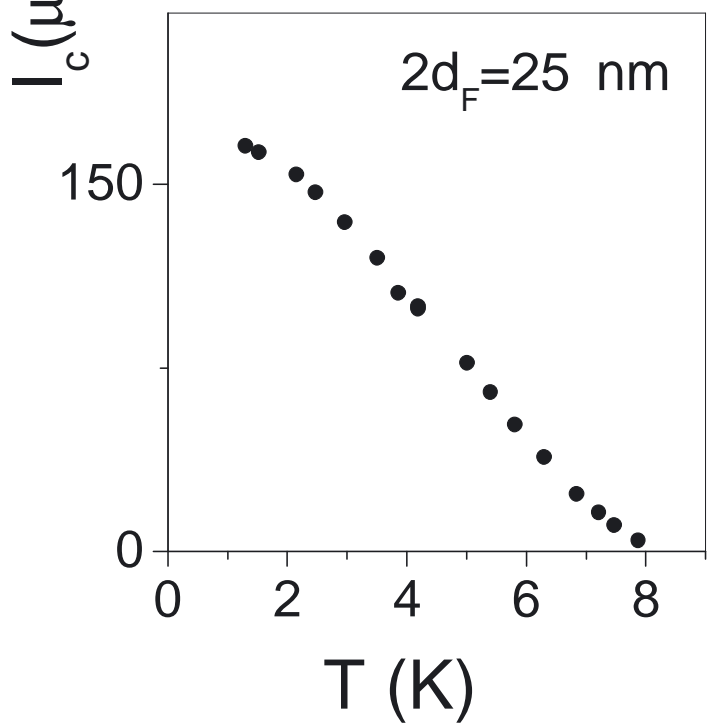
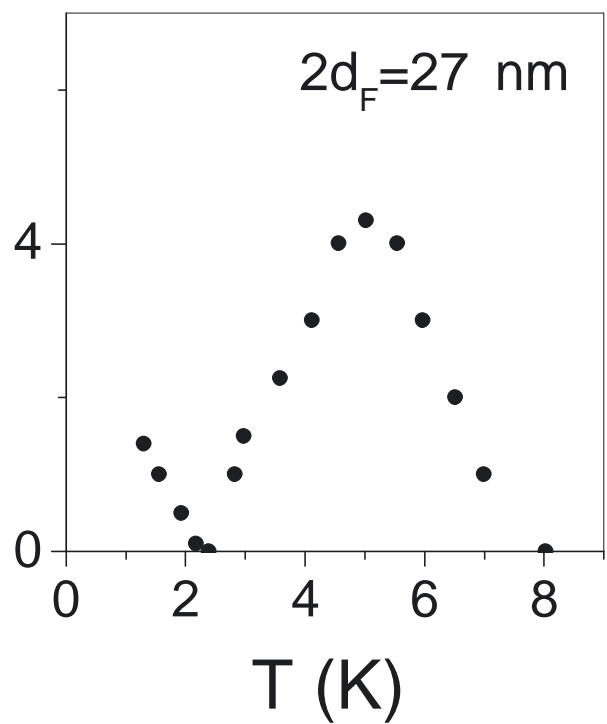
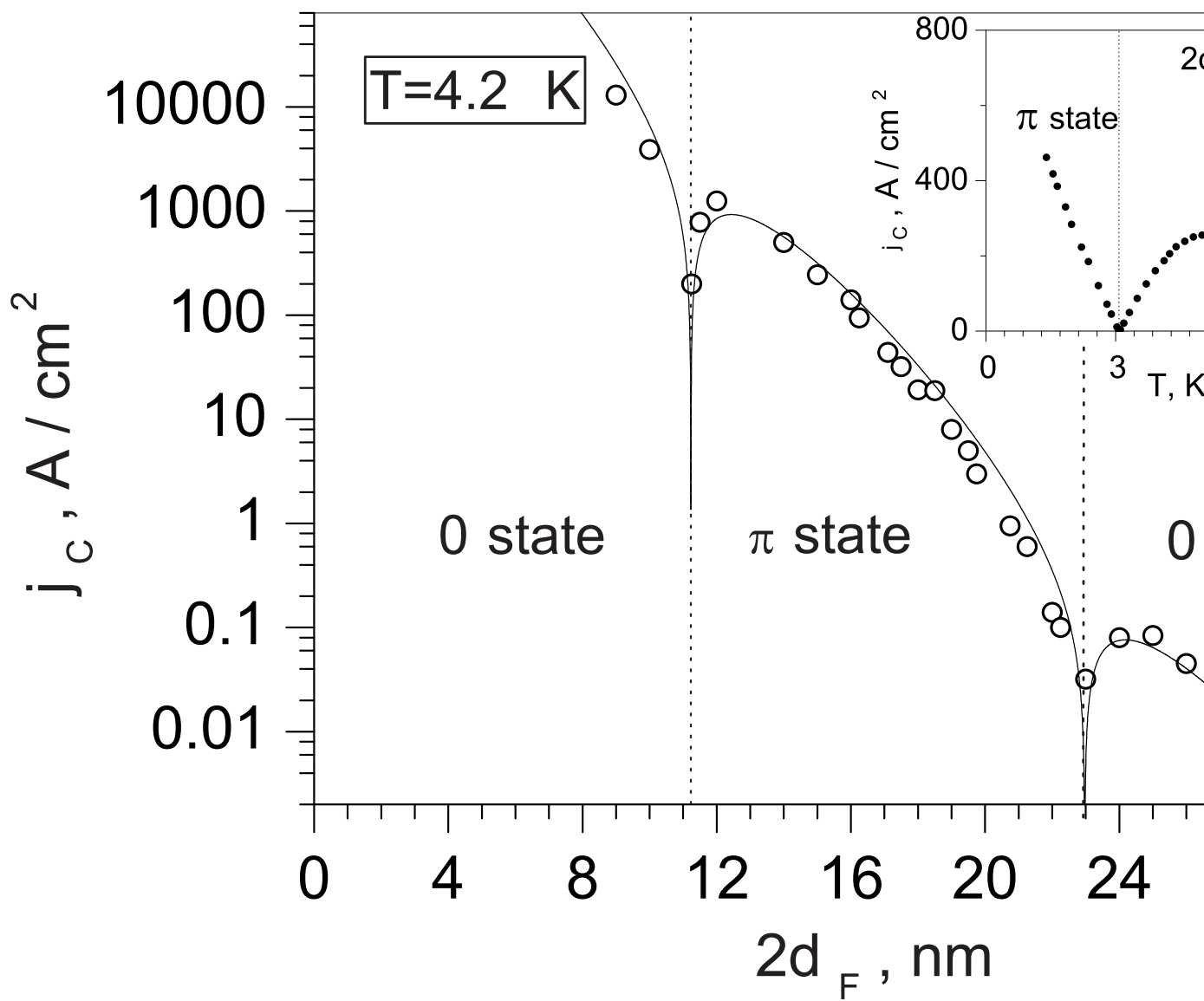
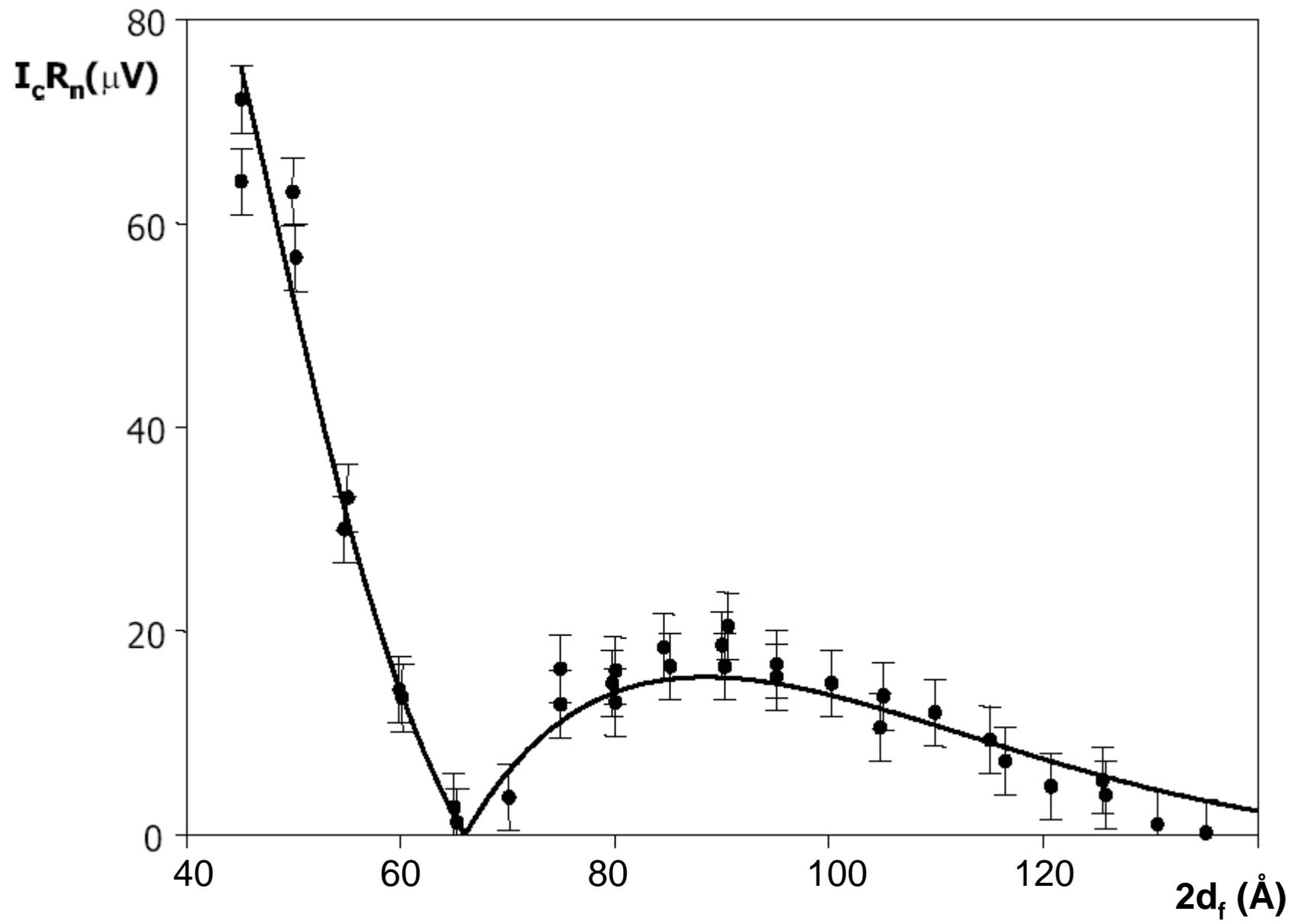
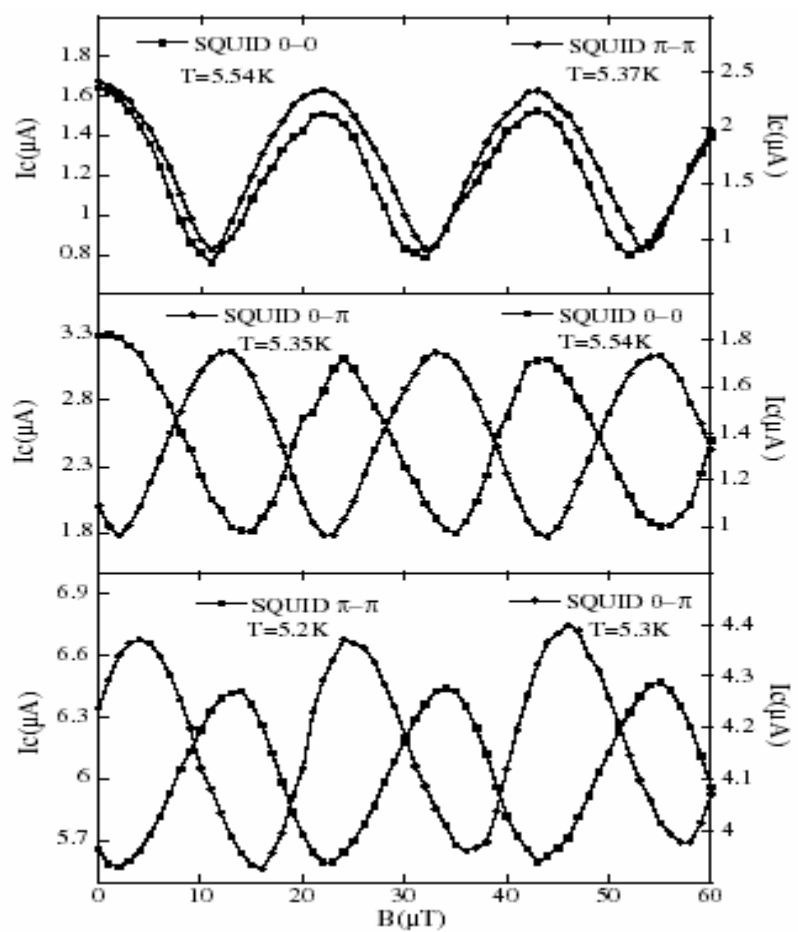
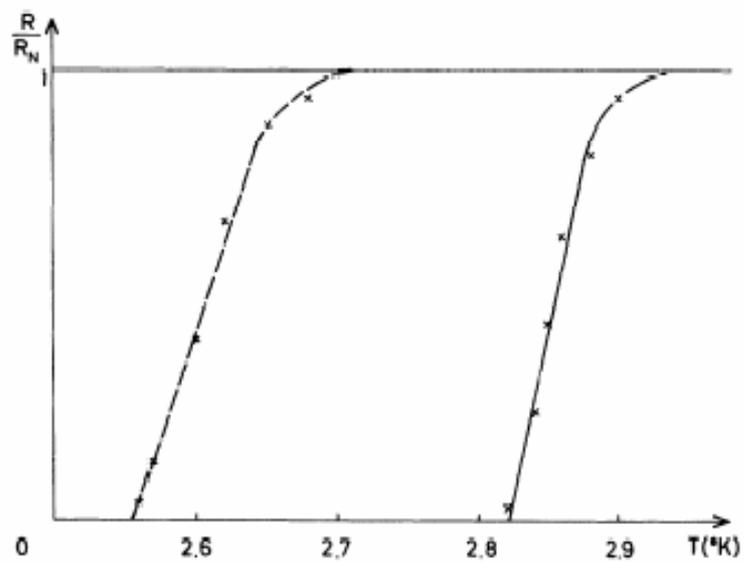
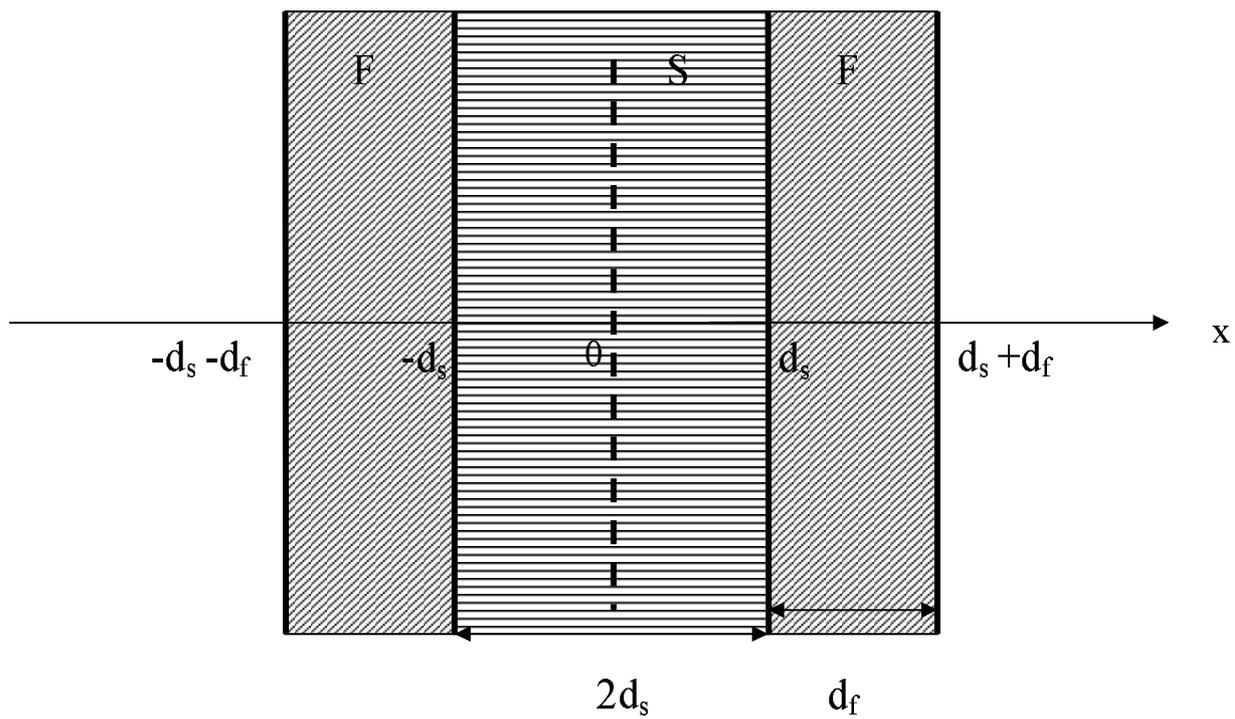
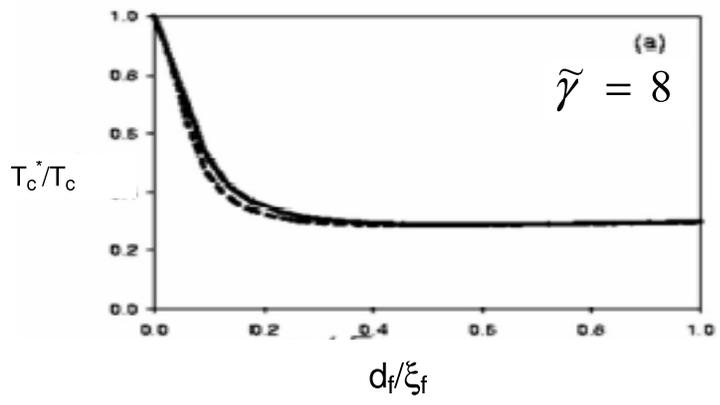
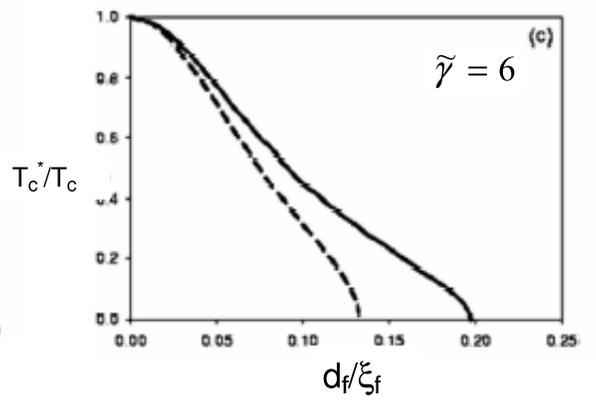
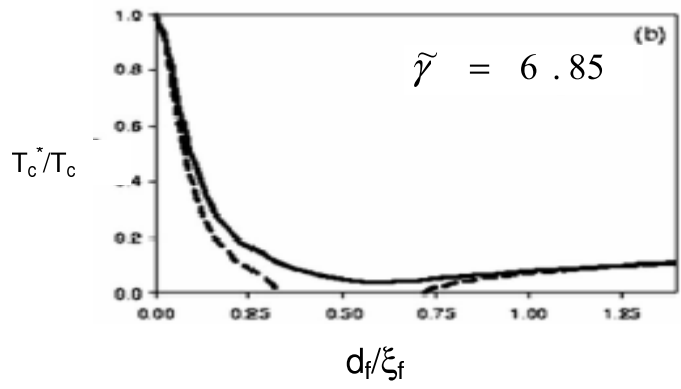
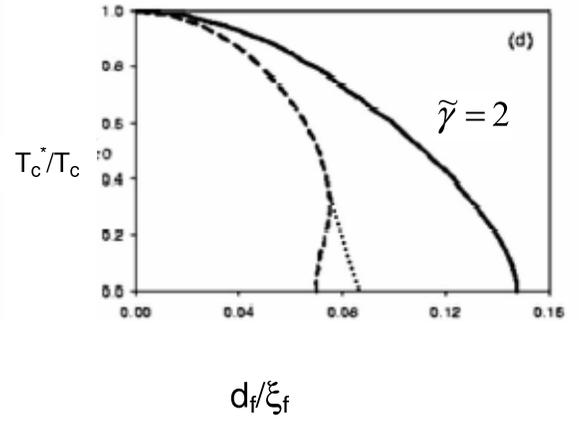
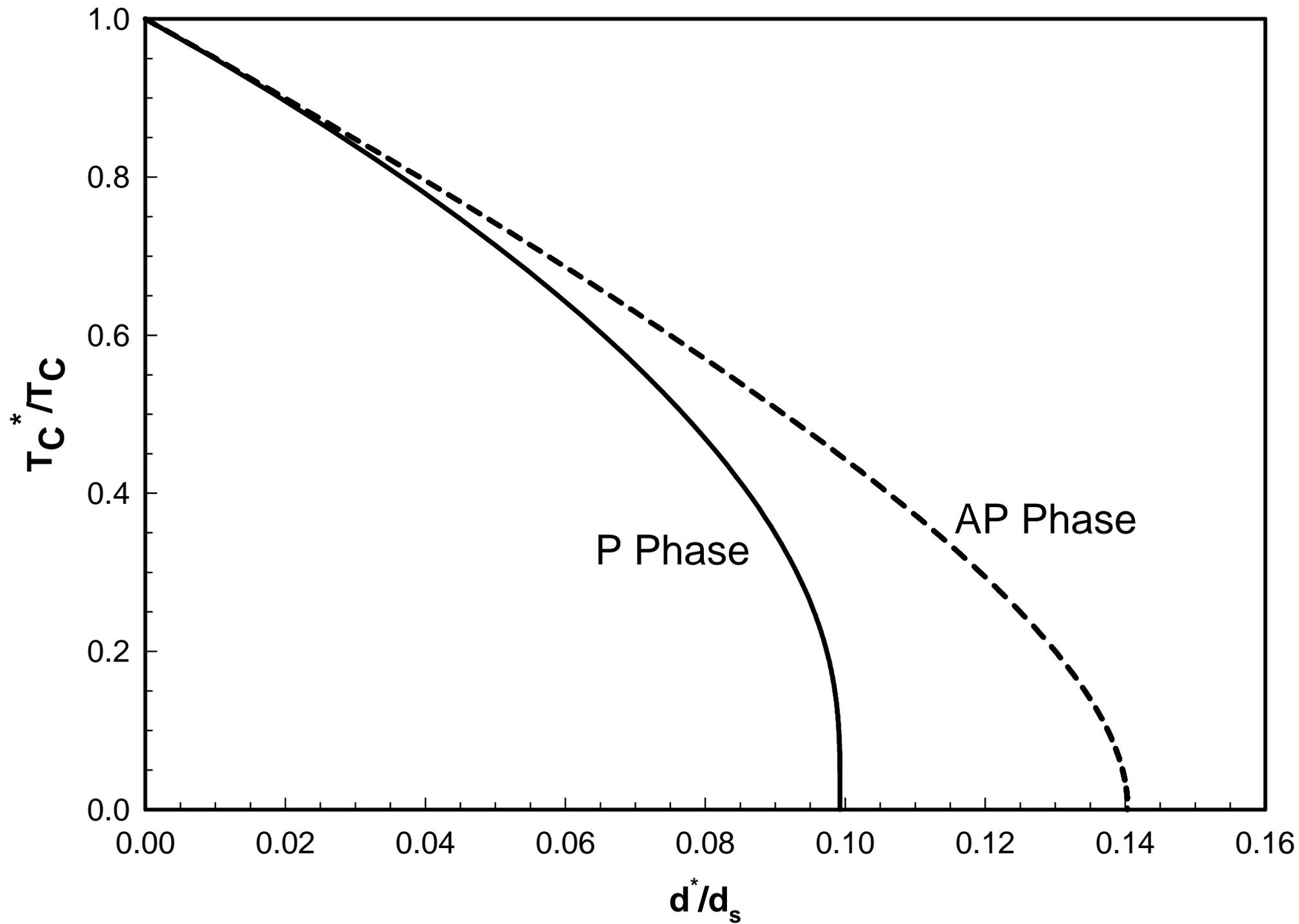
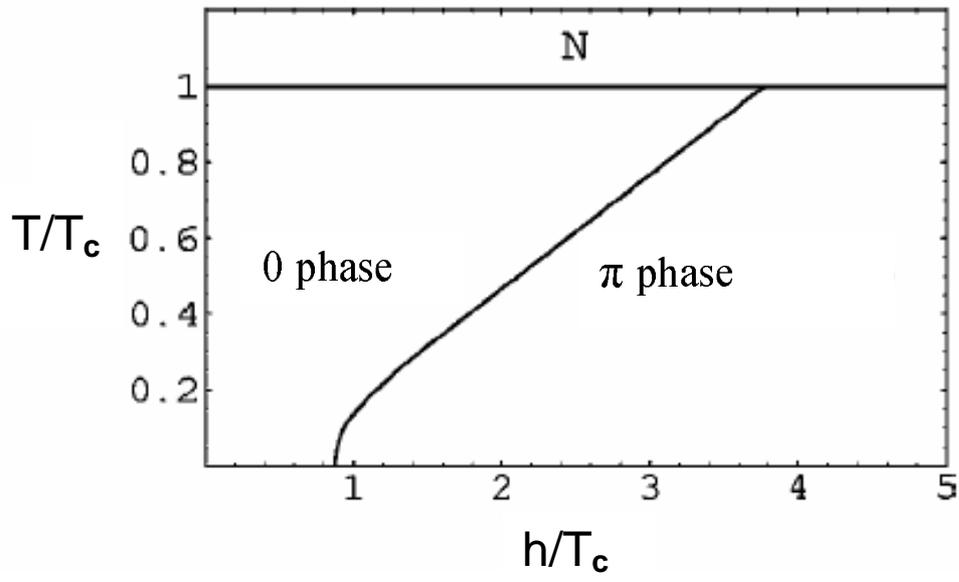
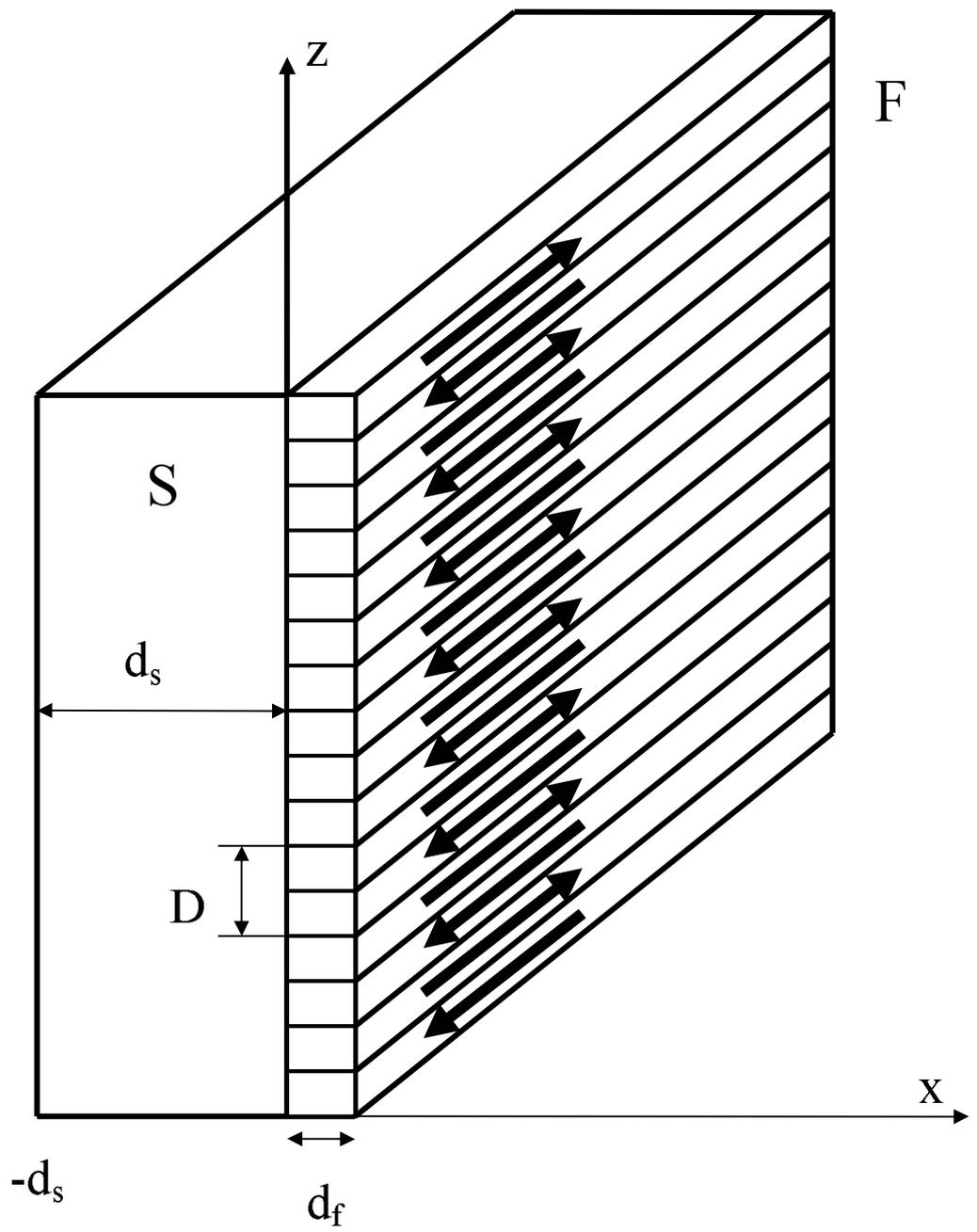
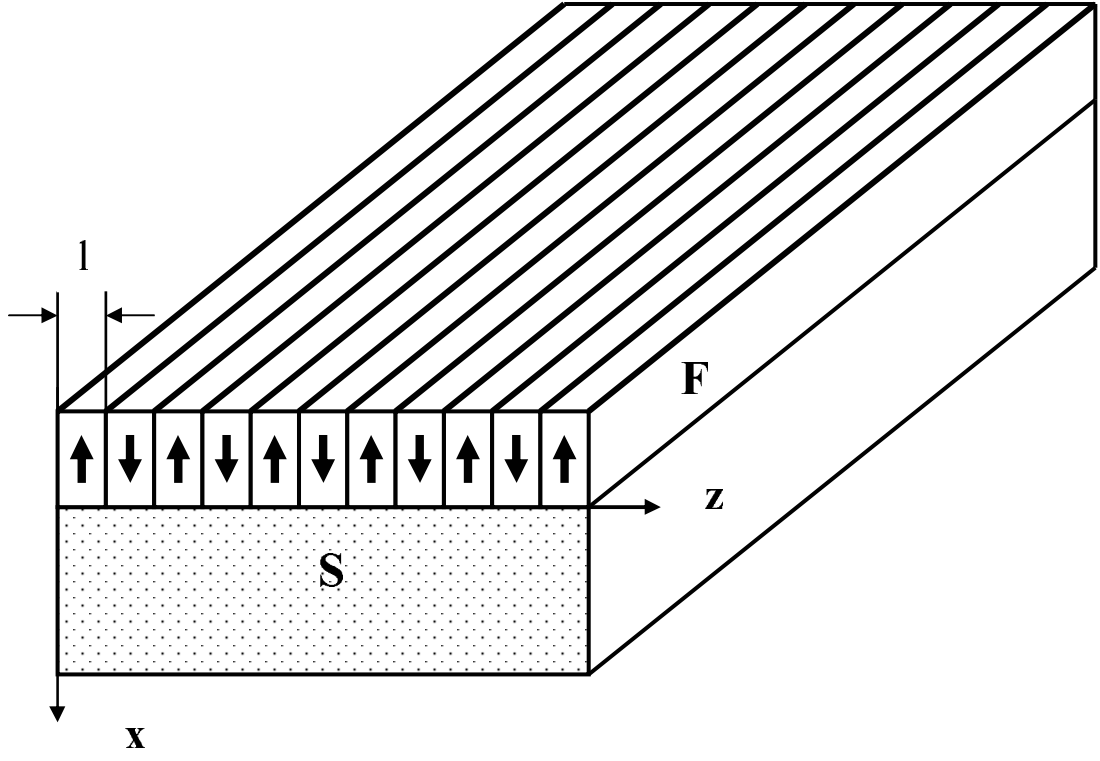
\end{document}